\documentclass[aps,floats,amssymb,amsmath,prb,nofootinbib,twocolumn,superscriptaddress]{revtex4-1}

\usepackage{calc}
\usepackage{psfrag}
\usepackage{graphicx}
\usepackage{color}
\usepackage[dvipsnames]{xcolor}
\usepackage[unicode=true,pdfusetitle,
 bookmarks=true,bookmarksnumbered=false,bookmarksopen=false,
 breaklinks=false,pdfborder={0 0 0},backref=false,colorlinks=true]
 {hyperref}
\usepackage{geometry}
\pdfpageattr {/Group << /S /Transparency /I true /CS /DeviceRGB>>} %necessary for transparent plots

\newcommand{\up}{\uparrow}
\newcommand{\dn}{\downarrow}
\newcommand{\loc}{\mathrm{loc}}
\newcommand{\imp}{\mathrm{imp}}
\newcommand{\latt}{\mathrm{latt}}

\newcommand{\per}{\mathrm{per}}
\newcommand{\sgn}{\mathrm{sgn}}
\newcommand{\NCS}{NCS}


\begin{document}

\title{Practical consequences of Luttinger-Ward functional multivaluedness for cluster DMFT methods}
\author{J. Vu\v ci\v cevi\'c}
\affiliation{Institut de Physique Th\'eorique (IPhT), CEA, CNRS, UMR 3681, 91191 Gif-sur-Yvette, France}
\affiliation{Scientific Computing Laboratory, Center for the Study of Complex Systems, Institute of Physics Belgrade,
University of Belgrade, Pregrevica 118, 11080 Belgrade, Serbia}
\author{N. Wentzell}
\affiliation{Institut de Physique Th\'eorique (IPhT), CEA, CNRS, UMR 3681, 91191 Gif-sur-Yvette, France}
\author{M. Ferrero}
\affiliation{Centre de Physique Th\'eorique, Ecole Polytechnique,CNRS, Universit\'e Paris-Saclay, 91128 Palaiseau, France}
\affiliation{Coll\`ege de France, 11 place Marcelin Berthelot, 75005 Paris, France}
\author{O. Parcollet}
\affiliation{Institut de Physique Th\'eorique (IPhT), CEA, CNRS, UMR 3681, 91191 Gif-sur-Yvette, France}

\begin{abstract}
The Luttinger-Ward functional (LWF) has been a starting point
for conserving approximations in many-body physics for 50 years. The recent
discoveries of its multivaluedness and the associated divergence of
the two-particle irreducible vertex function $\Gamma$ have revealed an inherent limitation
of this approach. Here we demonstrate how these undesirable properties of the
LWF can lead to a failure of computational methods based on an
approximation of the LWF. We apply the Nested Cluster Scheme (NCS) 
to the Hubbard model and observe 
the existence of an additional stationary point of the self-consistent
equations, associated with an unphysical branch of the LWF. In the strongly
correlated regime, starting with the first divergence of $\Gamma$, this
unphysical stationary point becomes attractive in the standard iterative technique used to solve DMFT.
This leads to an incorrect solution, even in the large cluster size limit, for which we discuss diagnostics.
\end{abstract}

\maketitle

The Luttinger-Ward functional (LWF) $\Phi$
is a  central object in the quantum many-body theory of strongly-correlated fermionic systems.
$\Phi$~\cite{BaymKadanoff1961} is defined as the interacting part of 
the Legendre transform of the free energy
with respect to the bare propagator $G_0$~\cite{Dominicis1964}. 
It is a functional of the full propagator $G$, formally equal to the sum of all vacuum skeleton
diagrams~\cite{Dominicis1964a,Nozieres1997}.
$\Phi$ has been the basis of many approximations in the field over the last decades.

Dynamical mean-field theory (DMFT)\cite{Georges1996,Kotliar2006} and its
cluster extensions \cite{Hettler1998,Lichtenstein2000,Kotliar2001,Maier2005a}
are a class of $\Phi$-derivable approximations with a systematic control parameter: the size $N_c$
of the cluster. 
They interpolate between DMFT ($N_c =1$) and the exact solution of the lattice model
for $N_c=\infty$.
Cluster methods allow to treat the Mott physics \`a la DMFT and to include
short-range spatial correlations.
They have led to significant progress in recent years, in particular on the Hubbard model.
\cite{Civelli2009,
Civelli2008,Civelli2005,deLeo2008,TremblayPRB2008,Kyung2006a,Kyung2006,Okamoto2010,Osolin2017,Parcollet2004,Park2008,CivelliImadaPRB2016,Sordi2011,Sordi2010,Sordi2013,
Zhang2007,Aryanpour2002,Dang2015,Ferrero2009,Ferrero2008,Ferrero2010,Gull2010,
Hettler2000,Huscroft2001,Jarrell2001,Kent2005,Leblanc2015,Macridin2004,Macridin2006,Macridin2007,Maier2004,
MaierPRL2005,MaierPRL2006,Maier2011,MaierArXiv2015,Staar2016,MaierPRB2014,
Staar2013,Li2015,Stanescu2006,Biroli2002,Biroli2004,Leblanc2015,Sakai2012,
Ayral2015c,Ayral2016,Ayral2015,Ayral2017a,Vucicevic2017}. 
Cluster DMFT methods are formulated in terms of one (or a few) auxiliary
quantum impurity models in a non-interacting bath encoded in the bare propagator $\cal G$. 
The bath is determined self-consistently in such a way that the impurity Green function $G^\text{imp}$
coincides with some (local) components of the Green function of the lattice model $G^\text{latt}$.
This \emph{representability} property, i.e.~the possibility to find $\cal G$ for
a given $G^\text{imp}$ in a quantum impurity model lies at the very heart of DMFT
methods\cite{Georges2004, Kotliar2006}.

Surprisingly, it was recently
discovered~\cite{Kozik2015,Gunnarsson2017,Rohringer2017} in simple
strongly-correlated models that the functional $\Phi[G]$ is in fact
multivalued, i.e.~has multiple branches.  
As a consequence, the relation $G[{\cal G}]$ cannot always be
inverted in quantum impurity models as several ${\cal G}$ yield the same Green
function $G$.
%Such a {\it reverse impurity solver} can also become unstable at
%strong coupling \cite{Kozik2015}. 
%The existence of several branches of $\Phi$
%manifest itself by some divergences in the two-particle irreducible vertex $\Gamma$
%. 
This has deep consequences for numerical methods in some parameter regimes.  
The crossing of two branches of $\Phi$ leads to divergence of the two-particle irreducible
vertex $\Gamma$\cite{Schafer2013,Schafer2016,Gunnarsson2016,Ribic2016,Gunnarsson2017} and
therefore the breakdown of the parquet decomposition\cite{Gunnarsson2016,Gunnarsson2015,Wu2017}.
Moreover, at strong coupling, the bold diagrammatic series can converge to an incorrect
result, as was checked explicitly using a Bold Quantum Monte Carlo algorithm\cite{Kozik2015}.
Similar pathological behavior was observed in the context of $GW$-like approximations of $\Phi$\cite{Tandetzky2015}.

In this Letter, we show that the multivaluedness of $\Phi$ 
has unexpected and severe consequences in certain cluster DMFT methods,
and can potentially lead to incorrect results.  Concretely, we
study the \emph{nested cluster} DMFT scheme (\NCS)~\cite{Schiller1995, Georges1996, Biroli2004} which is
an early example of the recently introduced Self-Energy Embedding Theory (SEET)%
\cite{ZgidPRBR2015,Lan2017b,Lan2017a,ZgidCollaboration2017,Zgid2017}. \NCS\ is a particularly interesting scheme since it
addresses the main drawbacks of the most widely used cluster methods: cellular DMFT (CDMFT)
\cite{Kotliar2001} and the Dynamical Cluster Approximation (DCA)\cite{Hettler1998}.  It is a real
space cluster method which is translationally invariant (unlike CDMFT) and
yields a continuous self-energy in reciprocal space (unlike DCA).  In the classical
limit, it reduces to the well-known Bethe-Kikuchi method of classical
statistical physics~\cite{Biroli2004,Georges1996}.

%It is expressed in terms of two or three coupled quantum impurity models.

We solve the \NCS\ for the Hubbard model and compare it to benchmarks
established with converged large DCA clusters. At weak to moderate couplings
the scheme is stable and performs very well. Even at strong coupling,
there is a physical solution which is very close to the benchmarks already
at moderate cluster size. However, {\it i)} in the standard iterative method used to
solve the DMFT equations, this solution is unstable towards an unphysical
solution characterized by a non-causal Weiss field; 
{\it ii)} as the cluster size increases, this stable unphysical solution converges to an
incorrect result; {\it iii)}
this occurs in the strong-coupling regime as delimited by the generalization of the
divergences of the irreducible vertex observed in Refs.~\onlinecite{Schafer2013,Janivs2014,Stan2015,Gunnarsson2016,Schafer2016,Gunnarsson2017,Rohringer2017}.

We consider the Hubbard model on a square lattice:
\begin{equation}
  H=-t\sum_{\langle ij\rangle\sigma} c_{i\sigma}^{\dagger}c_{j\sigma}
    -\mu\sum_{i\sigma}n_{i\sigma}+U\sum_{i}n_{i\uparrow}n_{i\downarrow},\label{eq:Hubbard}
\end{equation}      
where $c_{i\sigma}^\dagger$ creates a fermion with spin
$\sigma$ at site $i$. The density operator is $n_{i\sigma} =
c_{i\sigma}^\dagger c_{i\sigma}$. The nearest-neighbor hopping amplitude is
$t$, the on-site interaction $U$ and the chemical potential $\mu$. 
$D=4t$ is the unit of energy.
We use the CT-INT algorithm to solve the quantum impurity model~\cite{Rubtsov2004,GullRMP2011}.

Let us first address the representability issue of the Green function $G$
by a Weiss field ${\cal G}$ in a cluster impurity model.
We consider a $2 \times 2$ CDMFT
calculation for $T/D = 0.125$ and various $U$ and dopings $\delta$,
where it yields a quantitatively good solution as compared to
converged large cluster DCA benchmarks (see Fig.\ref{fig:bench_full}).
The CDMFT self-consistency equation reads \cite{Kotliar2001}
$G^{\text{imp}}[{\cal G}] = G^{\text{loc}}[{\cal G}]$ with 
\begin{equation*}\label{eq:CDMFTselfconsistency}
G^{\text{loc}}[{\cal G}] (i \omega_n) \equiv \!\! 
\sum_{k \in \mathrm{RBZ}}
\Bigl(
i\omega_n + \mu - \hat \epsilon_k-\Sigma^{\text{imp}}[{\cal G}](i \omega_n) 
\Bigr)^{\smash{-1}},
\end{equation*}
where $\hat \epsilon_k$ is the dispersion over the superlattice of clusters, RBZ is the reduced Brillouin zone
and $\Sigma^\mathrm{imp}$ (resp. $G^\mathrm{imp}$) is the impurity cluster self-energy (resp. Green function).
The CDMFT equations are solved with the usual iterative technique for
DMFT: given ${\cal G}^{(i)}$ at iteration $i$,
the impurity model yields $\Sigma^{\text{imp}}[{\cal G}^{(i)}]$ and the next iteration 
${\cal G}^{(i+1)}$ is given by
\begin{align}\label{eq:StdIterationMethod}
 {\cal G}^{(i+1)} = 
\left(
 G^{\text{loc}}\bigl[{\cal G}^{(i)}\bigr] ^{-1}
+ \Sigma^\text{imp}\bigl[{\cal G}^{(i)}\bigr]
\right)^{-1}.
\end{align}
Starting from the converged CDMFT solution $G^\mathrm{cdmft}$ we then implement a
{\it reverse quantum impurity solver}~\cite{Kozik2015}:
we seek a bare propagator ${\cal G}^\text{rev}$ of the cluster model 
such that $G^\text{imp}[{\cal G}^\text{rev}] = G^\text{cdmft}$,
with a similar iterative method as in Eq.~(\ref{eq:StdIterationMethod})
but with $G^{\text{loc}}[{\cal G}^{(i)}]$ replaced by $G^\text{cdmft}$, which remains fixed in the calculation.

\begin{figure*}[!ht]
 %\centering{}
 \includegraphics[width=6.5in,trim=1cm 0cm 1cm 0cm]{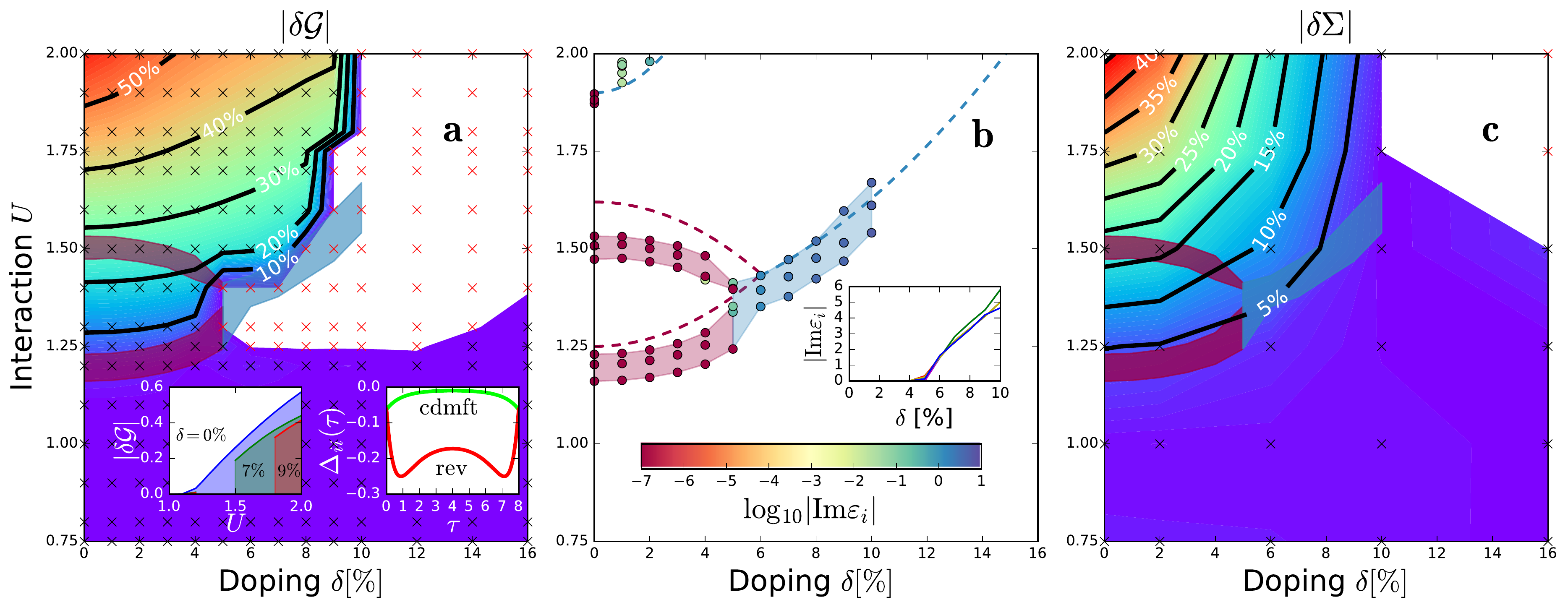}
 \caption{
For all panels, the temperature is $T/D=0.125$.
$\mathbf{a)}$
Color plot of $|\delta{\cal G}| \equiv |{\cal G}_{00}^\mathrm{cdmft}(i\omega_0)-{\cal G}_{00}^\mathrm{rev}(i\omega_0)|/|{\cal G}_{00}^\mathrm{cdmft}(i\omega_0)|$;
black crosses are data points, red crosses are points where the reverse impurity solver does not converge.
Left inset: Slices for fixed doping, showing discontinuity vs. $U$ for $\delta = 7\%, 9 \%$.
Right inset: Hybridization (local) of ${\cal G}^\mathrm{cdmft}$ (top, green) and ${\cal G}^\mathrm{rev}$ (bottom, red), for $U=2, \delta=0$ 
showing its causality violation.
$\mathbf{b)}$ 
Vertex divergences, where the real part of an eigenvalue $\varepsilon_i$ of $\tilde\chi_\mathrm{c}^{\Omega=0}$ crosses zero
for single-site DMFT (dashed line) and $2\times 2$ CDMFT (colored circles).
Color encodes $\text{Im}\, \varepsilon_i$ at the given point; colored stripes are guides for the eyes.
Inset: $\text{Im}\,  \varepsilon_i$ vs. doping for the bottom two groups of circles for $2\times 2$ CDMFT.
$\mathbf{c)}$ 
Color plot of 
$|\delta\Sigma| \equiv |\mathrm{Im} \Sigma_{00}^\mathrm{cdmft}(i\omega_0)-\mathrm{Im}\Sigma_{00}^\mathrm{nested}(i\omega_0)|/|\mathrm{Im}\Sigma_{00}^\mathrm{cdmft}(i\omega_0)| $,
i.e.~the difference between the imaginary part of the local self-energy for $2\times 2$ \NCS\ and the $2\times 2$ CDMFT  (the latter is close to the exact solution, see Appendix~\ref{subsec:comparison}, \ref{fig:bench_full}).
}
 \label{fig:div_and_rev} 
\end{figure*}

In Fig.~\ref{fig:div_and_rev}a, we present the relative difference between the local
component of the converged CDMFT Weiss field ${\cal G}^\text{cdmft}$ and the result of
the reverse impurity solver ${\cal G}^\text{rev}$. We observe three regions.
At weak coupling, the reverse impurity solver yields ${\cal G}^\text{cdmft}$ as
naively expected. At strong coupling and high doping, the reverse solver does
not converge.  At strong coupling and low doping, ${\cal G}^\text{rev}$
progressively deviates from ${\cal G}^\text{cdmft}$, even though they both yield the exact same Green function $G^\text{cdmft}$.  
As soon as ${\cal G}^\text{rev}$ is different from ${\cal G}^\text{cdmft}$ it acquires a
non-causal hybridization function $\Delta$%
\footnote{{The hybridization is $\Delta$ is defined by $\Delta(i\omega_n) \equiv i\omega_n + C - {\cal G}^{-1}$
where $C$ is a constant such that $ \Delta(i\omega_n) \xrightarrow{n\rightarrow\infty} 0$}}
as shown in the inset of Fig.~\ref{fig:div_and_rev}a. Indeed, $\Delta(\tau)$ is not concave over the
full $[0,\beta]$ interval and therefore has a corresponding spectral function
with negative parts.
This calculation demonstrates the existence of multiple branches of
$\Phi$ for the $2 \times 2$ impurity problem by exhibiting explicitly two
${\cal G}$ (and hence $\Sigma$) giving the same $G$, see also Refs.~\onlinecite{Schafer2013,Kozik2015,Stan2015,Schafer2016,Gunnarsson2016,Gunnarsson2017}.  
We will see below that a similar phenomenon occurs in \NCS.

%%%%   VERTEX DIVERGENCE

It is interesting to note that in the reverse impurity calculation at low
doping $\delta < 5\%$, one first finds ${\cal G}^\mathrm{rev} = {\cal G}^\text{cdmft}$
for small interactions $U < 1.25$ and then \emph{continuously} switches to an
unphysical solution for $\cal G$ as $U$ is increased. This means that the
physical branch of $\Phi$ crosses the unphysical branch.
As has been discussed in the particle-hole symmetric case~\cite{Gunnarsson2017},
this crossing has to be accompanied by a divergence of the corresponding
two-particle irreducible vertex function $\Gamma$, since it is the second derivative of
$\Phi$ with respect to $G$.  We generalize the results of Refs.~\onlinecite{Schafer2013,Schafer2016,Gunnarsson2016} to the doped case
and map these divergences of $\Gamma$ in
the $2\times 2$ CDMFT case, to obtain a characterization of the strong-coupling
region which is not linked to the details of an iterative algorithm.
Given the two-particle propagator
$G_{2,\sigma\sigma',ijkl}^{\omega\omega'\Omega} = \frac{1}{\beta}\langle
c^\dagger_{i,\sigma}(\omega)c_{j,\sigma}(\omega +
\Omega)c^\dagger_{k,\sigma'}(\omega' + \Omega)c_{l,\sigma'}(\omega')\rangle $
and the single-particle Green function $G$,
%$G_{\sigma,ij}(i\omega) =-\frac{1}{\beta}\langle c_{i\sigma}(i\omega)c^\dagger_{i\sigma}(i\omega)\rangle$,
$\Gamma$ can be calculated with the inverse Bethe-Salpeter equation
\begin{equation}
  \Gamma_{\mathrm{c},ijkl}^{\omega,\omega',\Omega} = \beta^2\big[[\chi_{0}^{\Omega}]^{-1} - [\tilde\chi_\mathrm{c}^{\Omega}]^{-1}\big]_{ij\omega,kl\omega'},
\end{equation}
where $\chi_{0,ijkl}^{\omega\omega'\Omega} = - G_{li}(i\omega)G_{jk}(i\omega +
i\Omega)\beta \delta_{\omega,\omega'}$, and
$\tilde\chi^{\omega\omega'\Omega}_{\mathrm{c},ijkl} =
G^{\omega\omega'\Omega}_{2,\up\up,ijkl}+G^{\omega\omega'\Omega}_{2,\up\dn,ijkl}
- 2G_{ji}(\omega)G_{lk}(\omega') \beta\delta_{\Omega,0}$. The inverse is
assumed to be in combined indices $(ij\omega)$ and $(lk\omega')$, where $\omega,\omega'$ 
denote fermionic and $\Omega$ bosonic Matsubara frequencies. If
$\tilde\chi_c(i\Omega)$ as a matrix has an eigenvalue $\varepsilon_i=0$, it is
singular and $\Gamma$ diverges at the given $i\Omega$. While in single-site
DMFT at ph-symmetry the eigenvalues of
$\tilde\chi_\mathrm{c}(i\Omega=0)$ are purely real by symmetry, it is no longer
necessarily true here\cite{Gunnarsson2016}.

Fig.~\ref{fig:div_and_rev}b shows trajectories in the $(\delta,U)$-plane where
the real part of an eigenvalue of $\tilde{\chi}_c$ crosses zero for 
single-site DMFT and $2\times 2$ CDMFT. 
In single-site DMFT, at half-filling, there are three $\Gamma$
divergences in the examined range of interaction, in agreement with
Ref.~\onlinecite{Schafer2016}. As we go to finite doping, the divergence close to $U=1.8$
disappears immediately as the corresponding eigenvalue acquires an imaginary
part.  However, the divergences close to $U=1.2$ and $U= 1.5$ extend up to
$\delta \simeq 5\%$ where they merge. For higher doping the divergences
disappear because the corresponding eigenvalues acquire an imaginary part.
In CDMFT, the behavior is very similar except that each divergence
is split into four, the two middle ones occurring simultaneously. 
Hence, we conjecture (see also Refs.~\onlinecite{Schafer2013,Gunnarsson2016,Schafer2016,Gunnarsson2017}) 
that the divergences in $\Gamma$ are not an artifact of
the single-site model but rather survive and multiply in the cluster impurity model. % (the large cluster limit).
Finally, in the left inset of Fig.~\ref{fig:div_and_rev}a, we see that for $\delta \geq 6\%$,
the unphysical solution appears discontinuously when $U$ is increased, in agreement with the absence of a divergence in $\Gamma$. 

Let us now turn to the \NCS. It approximates $\Phi$ by $\Phi^{(L)}$, defined as its restriction 
to the set of real-space two particle irreducible (2PI)
diagrams that involve lattice points lying within a box of 
shape $L\times L$.
$\Phi^{(L)}$ can be expressed as a
linear combination of the LWFs $\Phi_{L\times L}$ of a $L\times L$ cluster and the LWF of its subclusters,
with appropriate weights that eliminate the double counting of diagrams. 
Each cluster LWF is associated to an impurity model, via the representability property.
The lattice self-energy $\Sigma^\latt$ is therefore a linear combination of the self-energies of the impurities.
This couples the impurity models together and
the baths adjust so that e.g.~the impurity Green function 
is the same for every site of every cluster.
This method was introduced for a two site cluster (a dimer) in Ref.~\onlinecite{Schiller1995}, see also Refs.~\onlinecite{Georges1996,Biroli2004,Jabben2012}.

A priori, solving large nested clusters seems like a daunting task, 
requiring to solve a large number of coupled impurity problems, one for every
subcluster of the $L\times L$ cluster.
However, as shown in Appendix \ref{sec:square_clusters}, it is sufficient to solve only {\it four} coupled clusters
since 
\begin{multline}\label{eq:def_nested} 
\Phi^{(L)}[G] = \sum_i 
\Phi_{L\times L}\bigl[G|_{{C}^{L\times L}_i}\bigr] 
-\Phi_{L-1 \times L}\bigl[ G|_{{C}^{L-1 \times L}_i}\bigr] 
\\
-\Phi_{L\times L-1}\bigl[ G|_{{C}^{L\times L-1}_i}\bigr] 
+\Phi_{L-1\times L-1}\bigl[G|_{{C}^{L-1\times L-1}_i}\bigr]
\end{multline}
where ${C}^{n\times p}_i$ is the cluster of shape $n \times p$ whose
bottom-left lattice point is $i$, and $G|_{{C}^{n\times p}_i}$ the
restriction of the Green function to this cluster (i.e.~the set 
$\{ G_{lm} \}_{l,m \in {C}^{n \times p}_i}$).
If we assume rotational invariance, the last two terms give the same contribution 
and the method can be solved using three coupled cluster impurity models.
We present the full formalism for the \NCS\ with several examples in Appendix~\ref{sec:ncs}.

\begin{figure*}[!ht]
 %\centering{}
 \includegraphics[width=7in,trim=3cm 0cm 3cm 0cm]{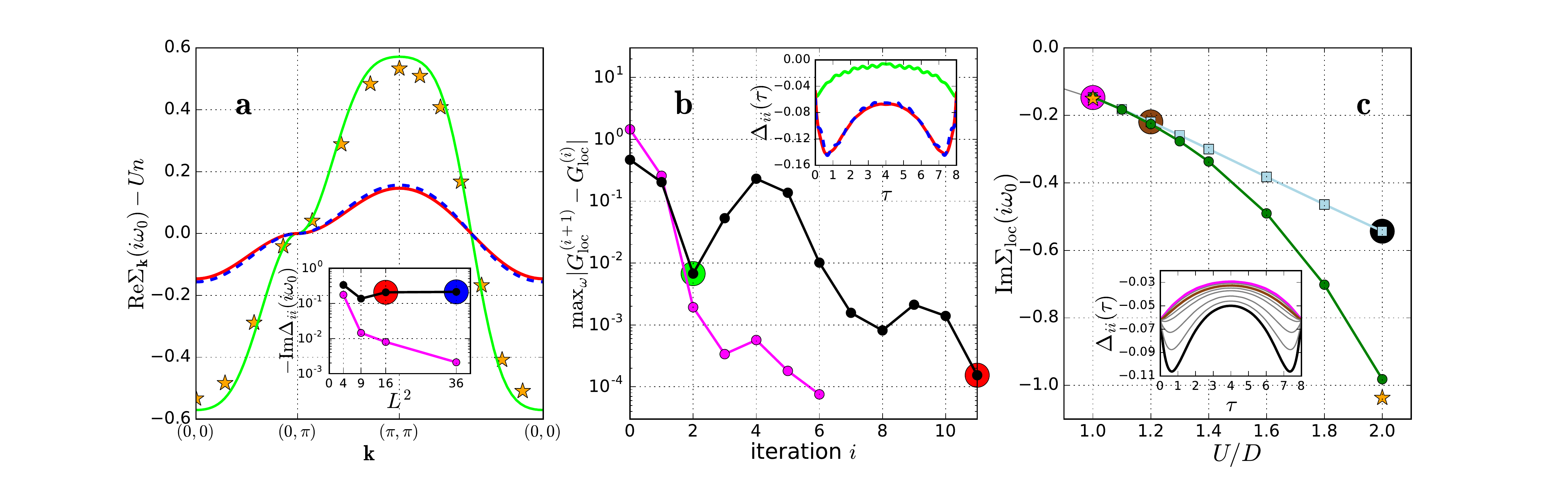}
 \caption{%
$\mathbf{a)}$ 
Real part of $\Sigma_\mathbf{k}(i\omega_0) -Un$ vs. momentum $\mathbf{k}$,
in \NCS\ for $L=4$ (solid line, red), $L=6$ (dashed line, blue);
unstable solution at the third iteration for $L=4$ (solid line, green); the stars denote 98 sites DCA results.
Inset:  Hybridization at the center of the $L\times L$ cluster 
at the first Matsubara frequency vs. $L$ for $U/D=2$ (circles, black), and $U/D=1$ (circles, magenta).
$\mathbf{b)}$ 
Norm of the difference of $G_{loc}$ between iterations vs. the iteration number $i$, for $L=4$, $U/D=2$ (circles, black), and $U/D=1$ (circles, magenta).
Green (resp.~red) dot corresponds to the unstable (resp.~stable) solution, Cf text.
Inset: Hybridization at the corner of the $L\times L$ cluster vs. $\tau$ for the unstable and the stable non-causal solution,
with same convention as in $\mathbf{a)}$.
$\mathbf{c)}$
$\text{Im}\, \Sigma_\text{loc} (i\omega_0)$ vs. $U/D$ for \NCS\ $L=2$ (square, blue online) 
and $2\times 2$ CDMFT, (circle, green online). The star is 98 sites DCA.
Inset: Local hybridization for $L=2$ vs. $\tau$.
}
\label{fig:nested_failure} 
\end{figure*}

We solve the \NCS\ using the standard iterative method of solution for DMFT equations as in Eq.~(\ref{eq:StdIterationMethod}).
At weak coupling, the NCS yields a solution in excellent
agreement with large DCA cluster benchmarks (see Fig.~\ref{fig:bench_full}).
However, at strong coupling the situation is more complex.
First, in Fig.~\ref{fig:div_and_rev}c, we observe that the $2\times2$ \NCS\ gives a poor result compared to CDMFT 
in the strong-coupling region delimited by the divergences of $\Gamma$ discussed above.
We then solve larger clusters $L = 2,3,4,6$ to examine the convergence of the
method with the cluster size. We observe an
unexpected and severe problem: \emph{the nested cluster scheme converges for
$L\rightarrow \infty$ but to an incorrect solution} even though formally
$\Phi^{(L)}\xrightarrow{L\rightarrow\infty} \Phi$.
In Fig.~\ref{fig:nested_failure}a and Appendix~\ref{subsec:two_solutions} we show the momentum dependent self-energies
obtained for $L=4, 6$: 
they are very close to each other, indicating convergence, 
but quite far from the benchmarks.

To gain further insight, we study the convergence of the $L=4$ case
at strong coupling iteration by iteration. In
Fig.~\ref{fig:nested_failure}b, we plot the difference between successive $G_\text{loc}$
for $U/D=1$ and $U/D=2$, $\delta=0$.  At $U/D=1$ convergence is
roughly exponential until the level of Monte-Carlo noise is reached.  However,
for $U/D=2$, we observe that the self-consistency is almost converged after 3
iterations (the green point on Fig.~\ref{fig:nested_failure}b) to an 
unstable solution before finally converging to another solution of the equation (red
point).
Remarkably, in Fig.~\ref{fig:nested_failure}a, we see that 
this unstable (green) solution is almost perfectly on top of the benchmark,
contrary to the stable (red) one.

Furthermore, we observe two pathologies of the stable (red) solution, 
which can be used as diagnostics in the absence of benchmarks.
First, the inset of Fig.~\ref{fig:nested_failure}b shows the local hybridization
function of both solutions (at the corner of the $L\times L$ cluster), 
$\Delta_\text{stable}$ and $\Delta_\text{unstable}$.
$\Delta_\text{stable}$ clearly violates causality at $U/D=2$,
similarly to the reverse impurity solver studied above, while $\Delta_\text{unstable}$ is fine.
Moreover, we see in Fig.~\ref{fig:nested_failure}c that this effect appears as a function of $U$ for $U > 1.2$, {\it i.e.} exactly when
the solution deviates from the benchmark (or CDMFT in this case).
Second, the $\Delta_\text{stable}$ bath does not decay in the large $L$ limit at strong coupling ($U/D=2$), contrary to $U/D=1$,
as illustrated in the inset of Fig.~\ref{fig:nested_failure}a.
Contrary to CDMFT or DCA, the \NCS\ does not impose $\Delta=0$ for every converged solution for $L\rightarrow\infty$, 
but only the weaker condition $ \Sigma^\latt = \Sigma^\imp+ \Delta$ (see Appendix~\ref{subsec:two_solutions} and Fig.~\ref{fig:latt_vs_imp_delta} for further discussion).
For the physical solution, we conjecture that $\Delta\rightarrow 0$ for $L\rightarrow\infty$:
the large cluster will be a Hubbard model with no bath.
The unphysical solution converges on the other hand to a certain resummation of the bold diagrams series.

For CDMFT and DCA, the standard iterative method of solution is {\it
iteratively causal}~\cite{Kotliar2001,Hettler1998}, {\it i.e.} one
can prove that the bath stays causal at each iteration (and therefore at
convergence).  Hence the causality violation of the bath cannot occur and the
solution stays on the physical branch.  The \NCS\ does not have this property, 
which, as we have seen, has drastic consequences on the stability of the physical solution in the iterative procedure. 
In the dimer case, \NCS\ was already known to yield non-causal self-energies at low
temperatures and strong-coupling~\cite{Schiller1995, Georges1996}.  But in
previous works~\cite{Biroli2004, OkamotoFictiveImpurity2003}, this was simply interpreted as the
signature of an insufficiently large cluster, {\it i.e.} a defect that
the large $L$ should cure.

To summarize, the nested cluster is a translationally invariant, real-space cluster method with a
physical solution very close to numerically exact benchmarks already at
moderate cluster sizes, both at weak and strong coupling. However, the
multivaluedness of the LWF leads to an instability of the standard iterative
procedure of solution in the strong-coupling region (as delimited by the
divergence of the irreducible vertex $\Gamma$) towards an unphysical solution,
even in the infinite cluster limit. This failure is signaled by causality violations of the hybridization function.
All this points to the importance of distinguishing between a
cluster method and the iterative procedure used to solve its equations. 
An important challenge is therefore
to design new ways of solving the cluster DMFT equations that are guaranteed to
stay on the physical branch of the LWF and stabilize the "hidden" physical
solution, e.g.~by implementing the "shifted-action"\cite{Rossi2016} proposal in this context.
%In a wider class of SEET methods it is essential to verify the results by
%checking the causality and convergence properties of the hybridization function.
%Alternatively, one can use cluster methods based on higher order functionals,
%which we believe are less likely to be multivalued (TRILEX\cite{Ayral2015, Ayral2015c, Ayral2017a}, QUADRILEX\cite{Ayral2016}).
Alternatively, one can use cluster methods based on higher order functionals
(TRILEX\cite{Ayral2015, Ayral2015c, Ayral2017a}, QUADRILEX\cite{Ayral2016}).
We believe these are less likely to be multivalued, as it would require the existence
of two systems with identical single-particle, but also higher-order correlation 
functions, which is a priori harder to achieve.
Moreover, going to higher-order functionals would correspond to adding more degrees of freedom to the
solution, which in itself could remove the multivaluedness.

\begin{acknowledgments}
We are grateful to A.~Georges and P.~Thunstr\"om for useful insights and discussion.
We further thank T.~Sch\"afer and A.~Toschi for critical reading of the manuscript.
The DCA 98A and 50A data was provided by J.~Leblanc.
We thank Z.~Mitrovi\'c Vu\v ci\v cevi\'c for figure editing.
This work is supported by the FP7/ERC, under Grant Agreement No.~278472-MottMetals. Part
of this work was performed using HPC resources from GENCI-TGCC (Grant
No.~2016-t2016056112). The CT-INT algorithm has been implemented using the TRIQS toolbox\cite{Parcollet2014}.
\end{acknowledgments}

%%%%%%%%%%%%%%%%%%%%%%%%%%%%%%%%%%%%%%%%%%%%%%%%%%%%%%%%%%%%%%%%%%%%%%%%%%%%%%%%%%%%%%%%%%%%%%%%
%%%%%%%%%%%%%%%%%%%%%%%%%%%%%%%%%%%%%%%%%%%%%%%%%%%%%%%%%%%%%%%%%%%%%%%%%%%%%%%%%%%%%%%%%%%%%%%%
%%%%%%%%%%%%%%%%%%%%%%%%%%%%%%%%%%%%%%%%%%%%%%%%%%%%%%%%%%%%%%%%%%%%%%%%%%%%%%%%%%%%%%%%%%%%%%%%
%%%%%%%%%%%%%%%%%%%%%%%%%%%%%%%%%%%%%%%%%%%%%%%%%%%%%%%%%%%%%%%%%%%%%%%%%%%%%%%%%%%%%%%%%%%%%%%%
%%%%%%%%%%%%%%%%%%%%%%%%%%%%%%%%%%%%%%%%%%%%%%%%%%%%%%%%%%%%%%%%%%%%%%%%%%%%%%%%%%%%%%%%%%%%%%%%
%%%%%%%%%%%%%%%%%%%%%%%%%%%%%%%%%%%%%%%%%%%%%%%%%%%%%%%%%%%%%%%%%%%%%%%%%%%%%%%%%%%%%%%%%%%%%%%%
%%%%%%%%%%%%%%%%%%%%%%%%%%%%%%%%%%%%%%%%%%%%%%%%%%%%%%%%%%%%%%%%%%%%%%%%%%%%%%%%%%%%%%%%%%%%%%%%
%%%%%%%%%%%%%%%%%%%%%%%%%%%%%%%%%%%%%%%%%%%%%%%%%%%%%%%%%%%%%%%%%%%%%%%%%%%%%%%%%%%%%%%%%%%%%%%%
%%%%%%%%%%%%%%%%%%%%%%%%%%%%%%%%%%%%%%%%%%%%%%%%%%%%%%%%%%%%%%%%%%%%%%%%%%%%%%%%%%%%%%%%%%%%%%%%

\appendix

\section{Benchmarks} \label{sec:benchmark}

In this section we present results for various cluster DMFT methods applied to the two-dimensional square-lattice Hubbard model as introduced in the main text. 
We pay special attention to the nested cluster scheme (NCS), which is discussed in detail in Section \ref{sec:ncs}.
Detailed summary of other cluster DMFT methods is provided in Appendix \ref{app:cluster_dmft}. 

We first present an extensive benchmark against exact results (subsection \ref{subsec:comparison}), which we use in the main text to determine the quality of solutions and to identify problematic regimes. We then adress in particular the causality violations in the problematic region (subsection \ref{subsec:causality}). In subsection \ref{subsec:cumulant}, we provide a comparison between two variants of the nested cluster scheme, differing in the nested quantity (self-energy vs. cumulant).
In subsection \ref{subsec:two_solutions} we discuss the stable and unstable solution of the nested equations.

\subsection{Comparison against exact results} \label{subsec:comparison}

\begin{figure*}[!ht]
 %\centering{}
 \includegraphics[width=6.4in,trim=1cm 0cm 1cm 0cm]{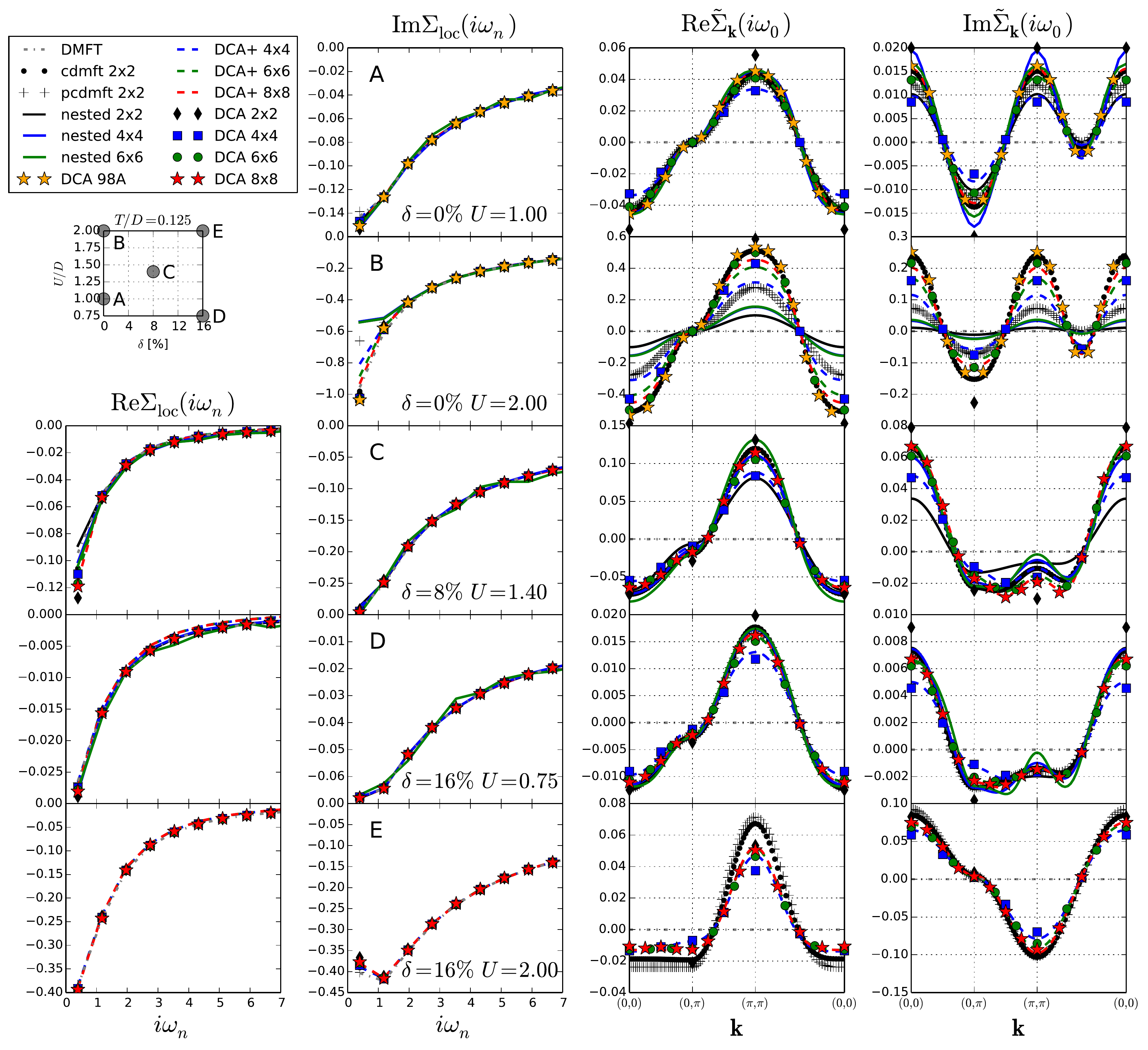}
 \caption{ Benchmark of DMFT and various cluster DMFT methods. Hubbard model square lattice, temperature $T/D=0.125$. We present separately the local and the non-local parts of self-energy, $\tilde\Sigma_\mathbf{k}=\Sigma_\mathbf{k}-\Sigma_\loc$.
 Color denotes cluster size, symbols/linestyles different methods. Stars denote the best available result.
 Agreement is excellent in all points except pt.B: NCS converges to a wrong solution, PCDMFT $2\times 2$ is considerably worse than in other points, and DCA$^+$ converges very slowly with cluster size, while being a poor approximation at small cluster size. In other points, NCS $4\times 4$ performs well, but at $6\times 6$ amplification of statistical noise becomes a problem (see text for details).
 }
 \label{fig:bench_full} 
\end{figure*}

In Fig.~\ref{fig:bench_full} we show the results of cluster DMFT methods for the Hubbard model, at various cluster sizes, in the four corners and the center of the phase diagram examined in the main text. The temperature is $T/D=0.125$. At half-filling, $\mathrm{Re}\Sigma_\loc(i\omega_n)=U/2$ by symmetry, so we omit this data. The non-local part $\tilde\Sigma_\mathbf{k}(i\omega_n)=\Sigma_\mathbf{k}(i\omega_n)-\Sigma_\loc(i\omega_n)$ we present at the lowest Matsubara frequency, along a triangular path enclosing the irreducible Brillouin zone. With stars we denote the best available result: at half-filling, we have DCA $N_c=98$, and away from half-filling, the biggest cluster is $8\times 8$ ($N_c=64$). These results are converged with respect to cluster size, and can be considered exact solutions of the Hubbard model.

The presented CDMFT result is the self-energy periodized by Eq.~(\ref{eq:cdmft_periodization}) (in appendix \ref{sec:cdmft} below). In DCA we are showing only the values at coarse-grained wave-vectors $\mathbf{K}$ (see appendix \ref{sec:dca}).

We first concentrate on the points other than pt.~B. We see excellent agreement of all methods. The local part is captured correctly already at $2\times 2$ cluster size.
DCA typically overestimates the amount of $\mathbf{k}$-dependence at $2\times 2$, then underestimates it at $4\times 4$, and is mostly converged at $6\times 6$. DCA$^+$ has a similar behavior ($2\times 2$ not shown for the sake of clarity). CDMFT and PCDMFT give almost the same result, and are on top of the benchmark except for the real non-local part in pt.~E, where the overall shape is correct, but the amplitude is overestimated slightly; PCDMFT also noticeably misses the local imaginary part in pt.~A. 
Nested cluster performs well, and at $4\times 4$ cluster size is even more accurate than DCA around $\mathbf{k}=(0,0)$. In pt.~E, it doesn't converge at any cluster size. 
Away from half-filling and at cluster size $6\times 6$, statistical noise amplification in nested cluster becomes significant (see Section \ref{sec:nested_noise} for details). It is particularly noticeable in the local part of self-energy at high Matsubara frequencies, in points C and D. Also in these points, there is a peak-like feature around $\mathbf{k}=(\pi,\pi)$ in the non-local imaginary part. It comes from the numerous long distance self-energy components which are small and comparable to the statistical error bar. These fine details of the solution can not be perfectly converged due to the statistical noise. 
%The amplification of noise renders nested cluster impractical for very big cluster calculations, unless non-stochastic solvers are used.

Now we turn to pt.B. CDMFT $2\times 2$ is, again, on top of the benchmark. DCA behaves no differently than in other points, and is almost converged at $6\times 6$; the local part is correct already at $2\times 2$. On the other hand, DCA$^+$ is not on top of the benchmark even at $8\times 8$, and especially the local part is strongly underestimated: at $8\times 8$ it is still worse than single-site DMFT. The non-local part is underestimated as well: the $6\times 6$ calculation is comparable to DCA $4\times 4$. PCDMFT, similarly, underestimates both the local and non-local part. Nested cluster converges to a wrong solution with respect to $N_c$: the local part is indistinguishable already between $2\times 2$ and $4\times 4$, and the non-local part between $4\times 4$ and $6\times 6$. The local part is underestimated by about $50\%$, and imaginary non-local part by almost an order of magnitude. The failure of PCDMFT, NCS and DCA$^+$ in this particular point is strongly reminiscent of the failure of bold-diagrammatic QMC presented in Ref.\onlinecite{Kozik2015}, for the same model parameters: the self-energy obtained in these methods is more metallic and much more local than the exact solution. We note that the similar phenomenon can also be observed in the original DCA$^+$ paper\cite{Staar2013} - in the strongly coupled regime, the $N_c=16$ DCA$^+$ self-energy result is much more local and metallic than that of the DCA at the same cluster size.

In conclusion, in this phase diagram, the best performing $2\times 2$ method is CDMFT. We take it as a reference method for benchmarking on a denser $(\delta,U)$-grid (Fig.~\ref{fig:div_and_rev}c in the main text, and Fig.~\ref{fig:scan_DMFT_and_DCA} below). At $4\times 4$ cluster size, in the points where it works, NCS does have an advantage over CDMFT and DCA. DCA $4\times 4$ coarse-graining is still quite crude - due to symmetries of the lattice, it yields only 6 independent self-energy components; NCS at the same size yields 10 independent self-energy components, and captures longer distance processes. In DCA, interpretation of the results in real-space is problematic; NCS results can be looked at in both $\mathbf{r}$ and $\mathbf{k}$-space. CDMFT is also problematic at $4\times 4$. At this size both the translational symmetry and the homogeneity within a supercell are broken, and the periodization becomes even less straight-forward. Finally, we note that in pt.B, even though NCS fails with forward substitution algorithm, there still appears to be a stationary point of the NCS equations (Fig.~\ref{fig:nested_failure} in main text) which is in better agreement with the exact result than DCA at the same cluster size.

\begin{figure*}[!ht]
 %\centering{}
 \includegraphics[width=3.0in,trim=0cm 0cm 0cm 0cm]{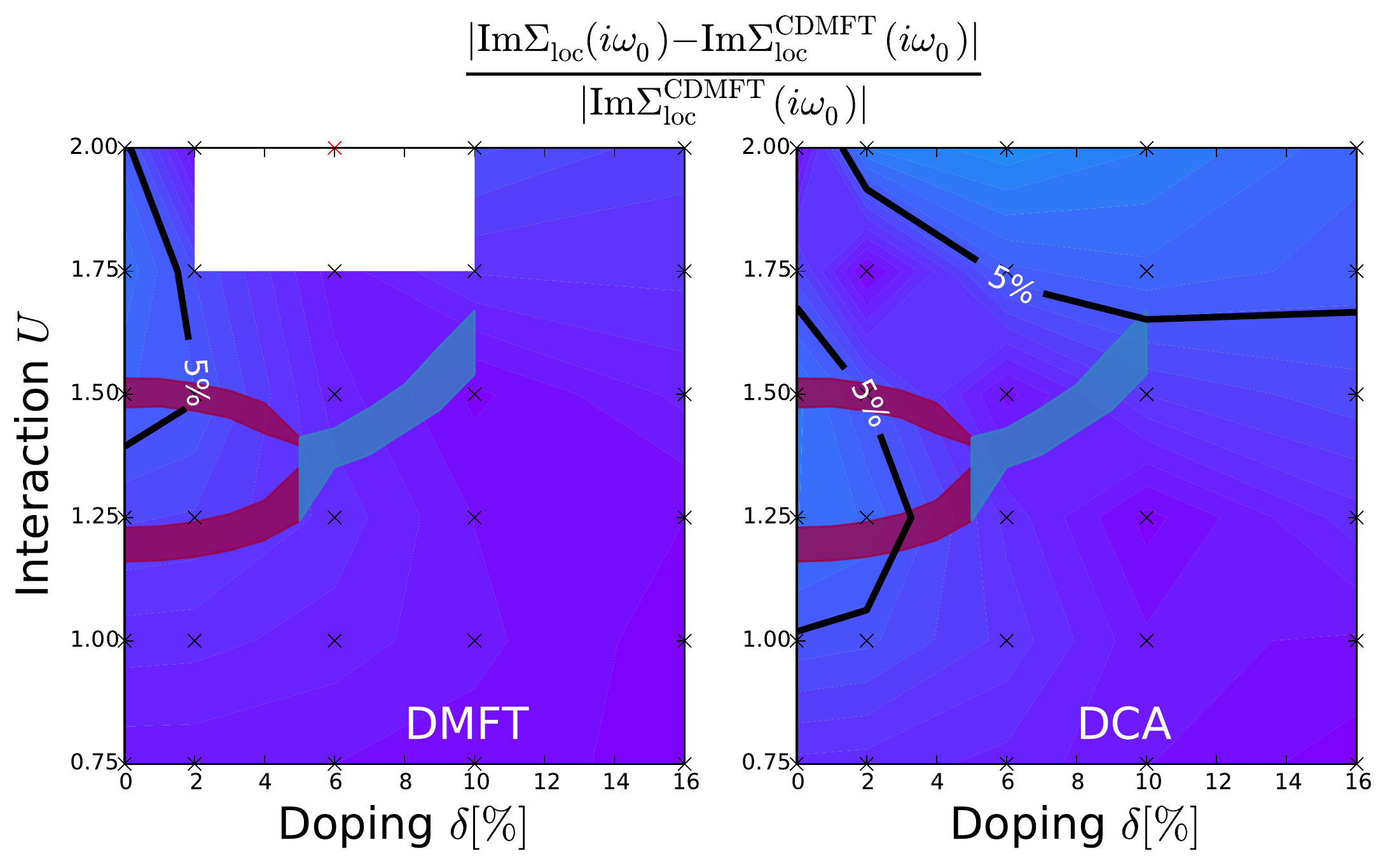}
 \includegraphics[width=3.0in,trim=0cm 0cm 0cm 0cm]{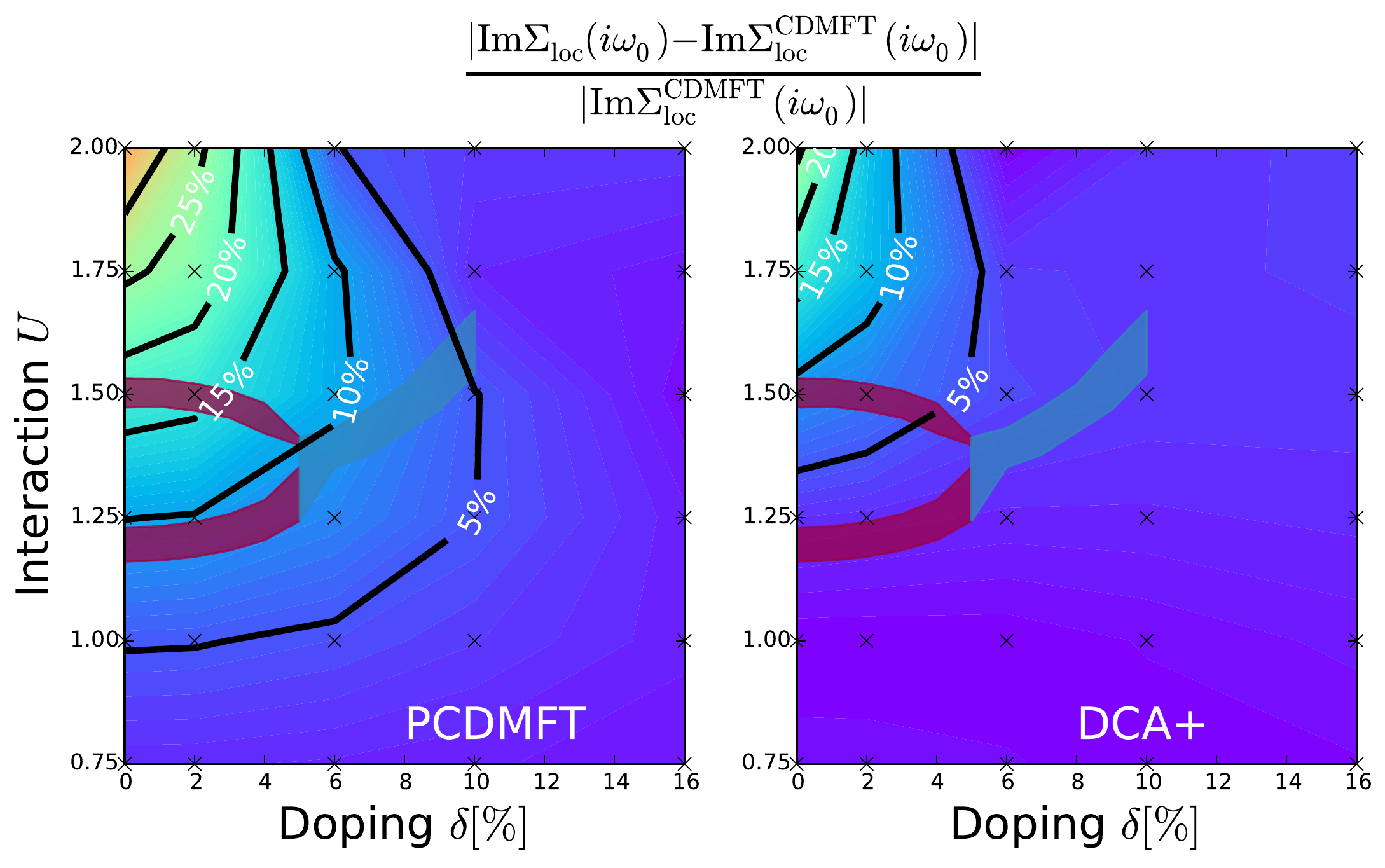}
 \caption{ Hubbard model, square lattice, temperature $T/D=0.125$.
 Relative difference in $\mathrm{Im}\Sigma_\loc$ from the reference method (CDMFT $2\times 2$). 
 DCA, PCDMFT and DCA$^+$ clusters are $2\times 2$.
 The red and blue stripes are the $\Gamma$ divergence trajectories from CDMFT $2\times 2$ calculation (Fig.~\ref{fig:div_and_rev}b in main text). 
 In DCA and DMFT performance is uniform across the phase diagram and appears unrelated to the divergence trajectories.
 In PCDMFT and DCA$^+$ the result is considerably poorer in the region roughly bounded by the divergence trajectories.
 }
 \label{fig:scan_DMFT_and_DCA} 
\end{figure*}

We finalize our analysis by a high resolution benchmark of DMFT and $2\times 2$ cluster methods (DCA, DCA$^+$ and PCDMFT), analogous to Fig.~\ref{fig:div_and_rev}c in the main text.
In Fig.~\ref{fig:scan_DMFT_and_DCA} we present the deviation from the exact result of these methods. DMFT and DCA perform uniformly well across the phase diagram, and are at most $\approx 5\%$ away from the correct result. No features can be associated with the $\Gamma$ divergence trajectories. On the other hand, NCS, DCA$^+$ and PCDMFT all fail in similarly shaped regions around pt.B, but give good results in other regimes. In DCA$^+$ and PCDMFT the coincidence of the problematic region with the $\Gamma$-divergence trajectories is less conclusive, but we can similarly connect the failure with the non-causality of the hybridization function. It is however unclear whether a correct stationary point is present in these methods at all.

\subsection{Causality properties} \label{subsec:causality}

In this section we analyze the causality properties of various quantities in the cluster methods presented above.

Quantities like Green's functions, self-energies and hybridization baths should have Lehmann spectral representation. The diagonal components of these quantities, should satisfy in real-frequency
\begin{equation}
 \mathrm{Im}X_{ii}(\omega) < 0
\end{equation}
where $X$ stands for $G$, $\Sigma$, $\cal G$ or $\Delta$. This has implications for the shape of these objects in imaginary time.
\begin{subequations}
\begin{eqnarray}
 X_{ii}(\tau) &=& \frac{1}{\pi}\int d\omega \frac{e^{-\tau\omega} }{1+e^{-\beta\omega}} \mathrm{Im}X_{ii}(\omega) \\ \nonumber
 \partial^{2n}_\tau X_{ii}(\tau) &=& \frac{1}{\pi}\int d\omega \frac{\omega^{2n} e^{-\tau\omega} }{1+e^{-\beta\omega}} \mathrm{Im}X_{ii}(\omega) \\ \label{eq:causality_condition}
                                 &<& 0, \;\;\; n\in \mathbb{N}
\end{eqnarray}
\end{subequations}
All even-order derivatives with respect to $\tau$ must be negative. This rules out the appearance of inflection points in $X_{ii}(\tau)$ and any of its even-order derivatives.

\begin{figure*}[!ht]
 %\centering{}
 \includegraphics[width=6.4in,trim=-1cm 0cm 1cm 0cm]{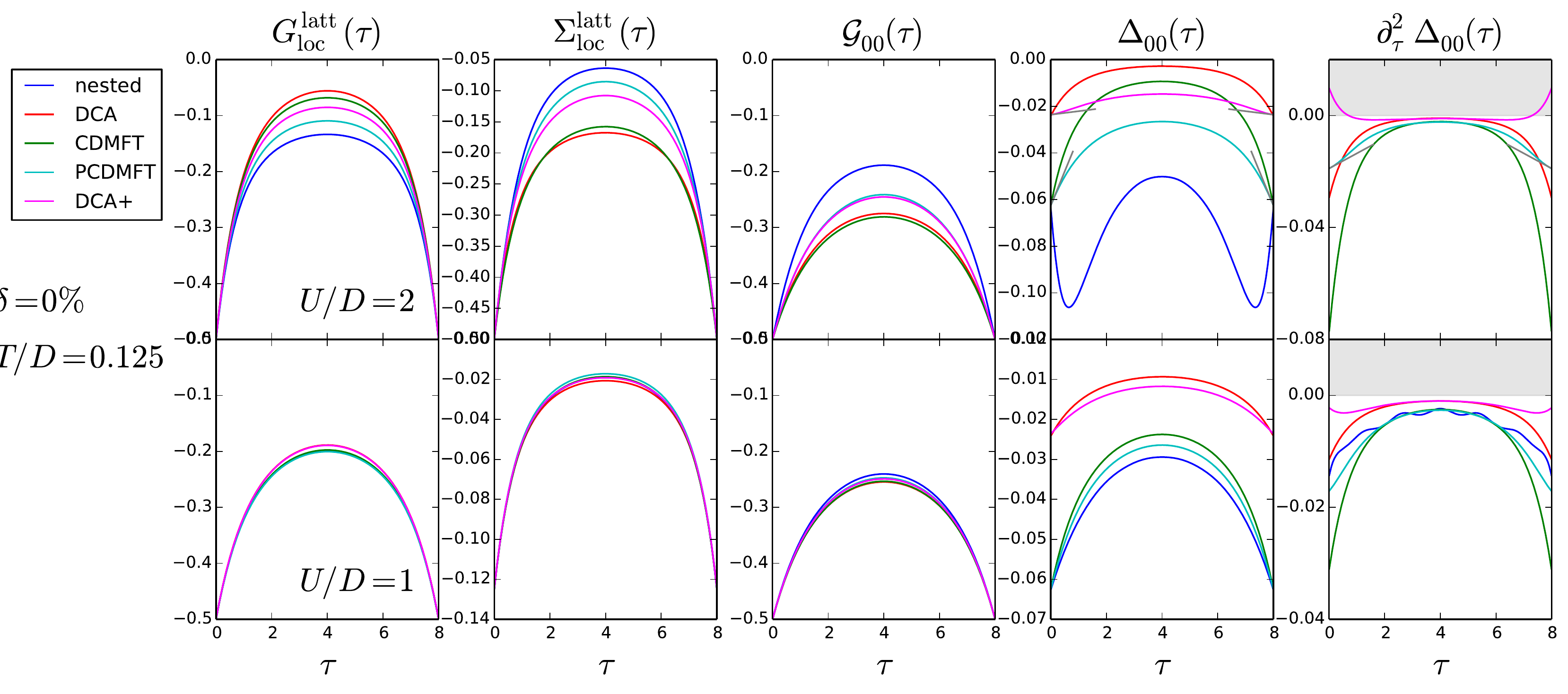}
 \caption{ Causality analysis of relevant quantities, in various cluster methods at $2\times 2$ cluster size. Temperature $T/D=0.125$, square lattice. 
   At strong coupling, DCA$^+$ and NCS have pronounced inflection points in $\Delta(\tau)$, PCDMFT in the second-derivative of $\Delta(\tau)$.
   Gray lines extrapolate the linear component close to $\tau=0$ and $\beta$. Other quantities are all causal, including the bare propagator on the impurity.
   Increasing the cluster size in DCA$^+$ improves the causality in $\Delta$, but not in NCS. In the upper right panel, NCS result is omitted for the sake of clarity (the non-causality is already obvious in $\Delta_{00}(\tau)$).
 }
 \label{fig:causality} 
\end{figure*}

In Figure \ref{fig:causality} we present the results for the local $G$, $\Sigma$ on the lattice, as well as the diagonal components of the bare propagator $\cal G$ and the hybridization function $\Delta$ on the impurity, all in imaginary time. All methods used are at $2\times 2$ cluster size. In NCS we present the impurity quantities only for the biggest cluster. In all methods at $2\times 2$, all the diagonal components of $\cal G$ and $\Delta$ are the same by symmetry (in DCA/DCA$^+$ this holds at any cluster size).

We see that all the quantities except the hybridization bath are causal. At $U/D=1$, there is a slight violation of \eqref{eq:causality_condition} in the second derivative of $\Delta$ in NCS, PCDMFT and DCA$^+$, but it is a tiny effect.
%one can connect with statistical noise from the impurity solver. 
In this regime, small fluctuations in the non-causal direction do not cause problems for these methods and the result is correct. However, it is clear that these methods do not impose causality on the hybridization function strictly, which then leads to problems at strong coupling. At $U/D=2$ we see a strong violation of \eqref{eq:causality_condition} in NCS, a clear inflection point in $\Delta(\tau)$ in DCA$^+$, and in PCDMFT there is an inflection point in $\partial^2_\tau \Delta(\tau)$. 
Here we observe a similar trend in DCA$^+$, NCS and PCDMFT: $\Delta_{00}(\tau\sim\beta/2)$ is generically overestimated (by absolute value) with respect to DCA and CDMFT, respectively (note that the difference in the bath between DCA/DCA$^+$ on one side and CDMFT/PCDMFT/NCS on the other is due to a different way of closing self-consistency in these two groups of methods: $\mathbf{k}$-space vs. $\mathbf{r}$-space clusters; see Appendix \ref{app:cluster_dmft}). The bigger $\Delta_{00}(\tau\sim\beta/2)$ translates to having a bigger bath at the low frequency - the observed non-causal bath is also bigger, and as we see in Fig.~\ref{fig:nested_failure} in the main text, in NCS it does not even decay with increasing cluster size.

\subsection{Cumulant vs. self-energy nesting} \label{subsec:cumulant}

\begin{figure*}[!ht]
 %\centering{}
 \includegraphics[width=6.4in,trim=1cm 0cm 1cm 0cm]{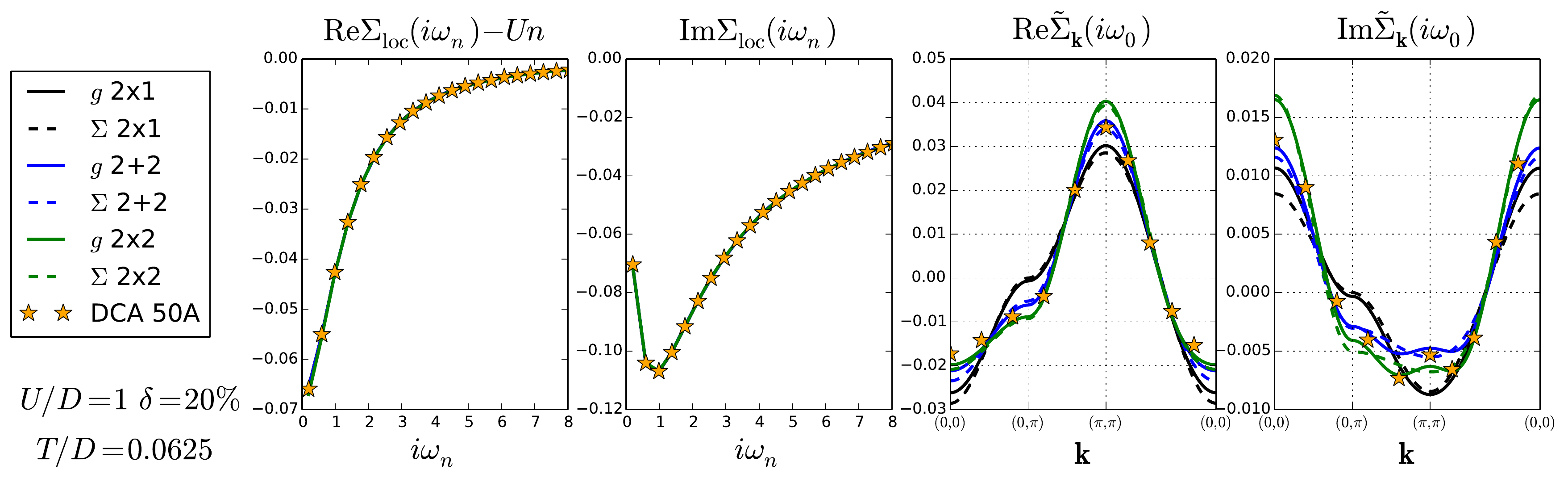}
 \caption{ Benchmark of the two variants of NCS (cumulant vs. self-energy embedding) at small-cluster sizes. Stars denote the best available result (50-site DCA). The cumulant variant performs only slightly better. $2\times 2$ calculation is already very close to the benchmark data.
 }
 \label{fig:cumul_vs_sigma} 
\end{figure*}

In Fig.~\ref{fig:cumul_vs_sigma} we compare the two variants of the NCS: one embeds either the cumulant $g$, or the self-energy $\Sigma$ (for details see Section \ref{sec:cumul_nested}). The results are compared to a 50-site DCA calculation. The temperature is $T/D=0.0625$ and the (hole) doping is $20\%$.

We present results for the simplest dimer calculation ($2\times 1$, see Section \ref{sec:2x1}), the double dimer $2+2$ (see Eq.~(\ref{eq:2+2}) and the corresponding section), and the $2\times 2$ calculation (Section \ref{sec:2x2}). We see that the result is solid already at $2+2$, and is overall improved at $2\times 2$. However, it is clearly not yet converged, and looking at the non-local part, the convergence is not monotonic. This is clearly expected at such small cluster size.

We observe that the cumulant variant performs slightly better, but the difference is almost negligible. We have checked that none of the features of the failure of NCS depend on the choice of the nested quantity ($g$ or $\Sigma$). In the problematic region, the cumulant variant converges to almost exactly the same wrong solution as the self-energy variant.

\subsection{Unstable and unphysical solutions} \label{subsec:two_solutions}

\begin{figure*}[!ht]
 %\centering{}
 \includegraphics[width=5.5in,trim=2cm 0cm 2cm 0cm]{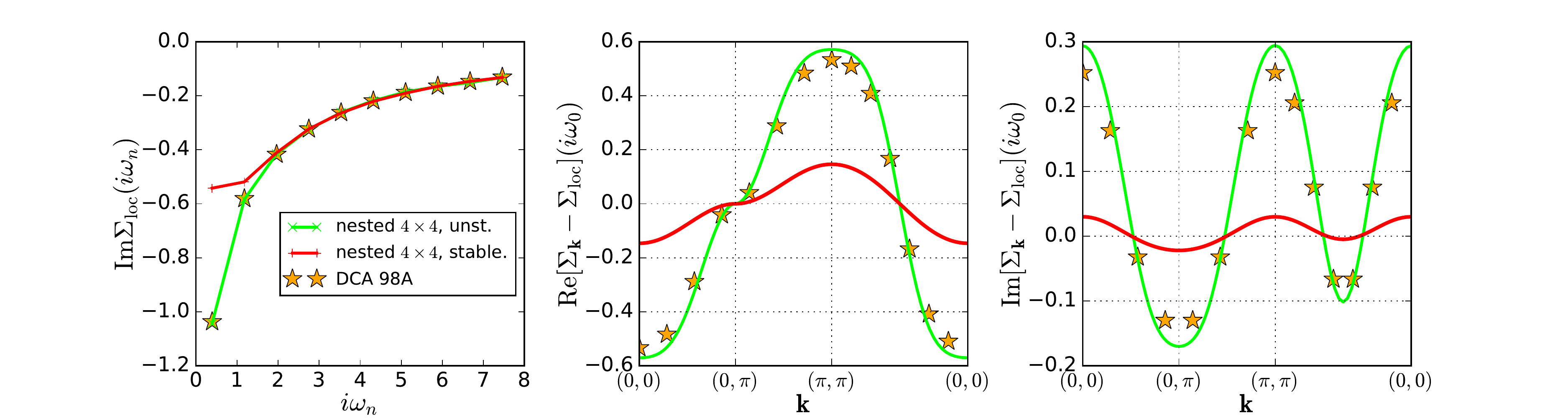}
  \caption{ Temperature $T/D=0.125$, $U/D=2$, $\delta=0$, NCS $4\times 4$. Red line: converged solution (stable, unphysical); green line: solution aftrer 3 iterations (almost converged, physical, unstable); stars: DCA 98A (exact benchmark).
 }
 \label{fig:unstable_solution} 
\end{figure*}

In Fig.~\ref{fig:unstable_solution} we present the self-energy for the apparently unstable (green line) and the stable solution (red line) in NCS $4\times 4$, compared to the exact benchmark (from Fig.~\ref{fig:bench_full}). We observe that the unstable solution is in excellent agreement with the exact benchmark, even better than DCA of the same size cluster. The stable solution on the other hand, is much more metallic and much more local. However, it does have the correct asymptotics and is aparently causal (see subsection \ref{subsec:causality}).

\begin{figure}[!ht]
 %\centering{}
 \includegraphics[width=2.0in,trim=0cm 0cm 0cm 0cm]{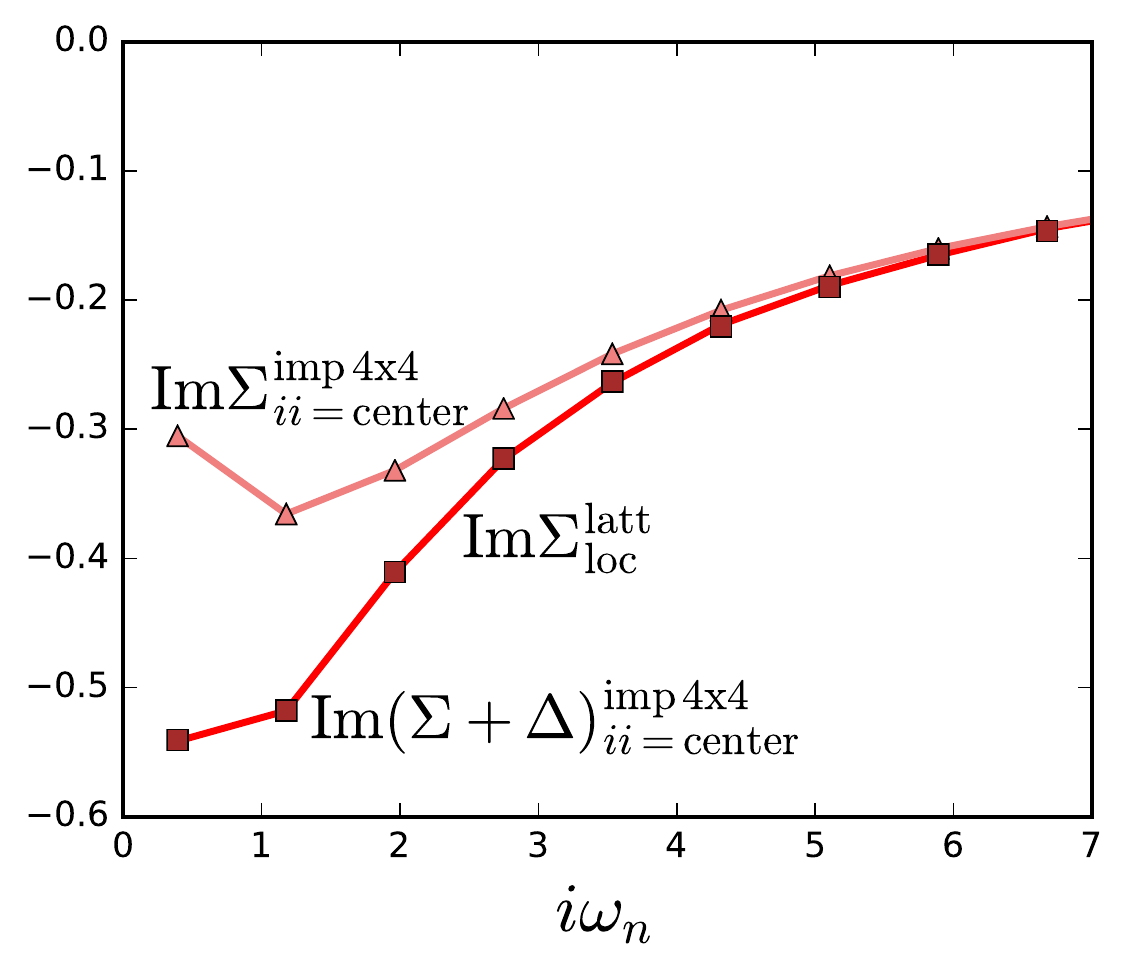}
  \caption{ Temperature $T/D=0.125$, $U/D=2$, $\delta=0$, NCS $4\times 4$. The unphysical solution retains a finite hybridization function even in the large cluster limit, such that $\Sigma^\latt = \Sigma^\imp + \Delta$. Quantities are presented at the center of the biggest cluster.
 }
 \label{fig:latt_vs_imp_delta} 
\end{figure}

Even in the large cluster limit, NCS does not guarantee $\Sigma^\imp\rightarrow \Sigma^\latt$, and therefore at large cluster size, a principal solution is possible
\begin{equation}
 \Sigma^{\imp,C}[\Delta^C]+\Delta^C \approx {\cal F}^C\Sigma^\latt[\Sigma^\imp]
\end{equation}
where ${\cal F}^C$ projects a lattice quantity onto impurity degrees of freedom of the cluster $C$. We check this explicitly in our unphysical solution in Fig.~\ref{fig:latt_vs_imp_delta} and find excellent agreement.

\section{Vertex divergences} \label{sec:vertex}

\begin{figure}[!ht]
 \includegraphics[width=0.4\columnwidth]{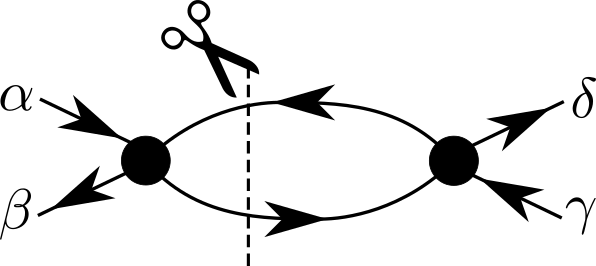}
 \caption{ Scattering diagrams can be classified according to their two-particle reducibility. If after cutting two Green's function lines, the diagram is separated into two vertex diagrams, with the external indices $\alpha$ and $\beta$ in one and $\gamma$ and $\delta$ in the other, the diagram is reducible in the $\mathrm{ph}$ channel. }
 \label{fig:PH_reduce} 
\end{figure}

%The irreducible vertex function $\Gamma$ is inherently linked to the LWF\cite{Kozik2015}
The irreducible vertex function $\Gamma_r$ contains all possible two-particle scattering processes that are two-particle irreducible\cite{Rohringer2012,Dominicis1964,Dominicis1964a} in the given channel $r$ (see Fig.~\ref{fig:PH_reduce} for an illustration of the two-particle reducibility concept). The reducibility channels are particle-hole ($\mathrm{ph}$), transverse particle-hole $\overline{\mathrm{ph}}$ and particle-particle $\mathrm{pp}$, depending on which of the external indices remain connected after cutting two propagator lines\cite{Rohringer2012}.
$\Gamma_{\mathrm{ph}}$ in particular corresponds to the second-order functional derivative of the LWF% $\Phi$
\begin{eqnarray}
   \Gamma_{\mathrm{ph},\alpha\beta\gamma\delta} &=&\left. \frac{\delta \Sigma_{\delta\gamma}[G]}{\delta G_{\alpha\beta}}\right|_{G=G[G_0,\Sigma]}\\ \nonumber
   &=& \left.\frac{ \delta^2 \Phi[G] }{\delta G_{\alpha\beta} \delta G_{\gamma\delta}}\right|_{G=G[G_0,\Sigma]} \label{eq:gamma_lwf}
\end{eqnarray}
\begin{figure}[!ht]
 \includegraphics[width=\columnwidth]{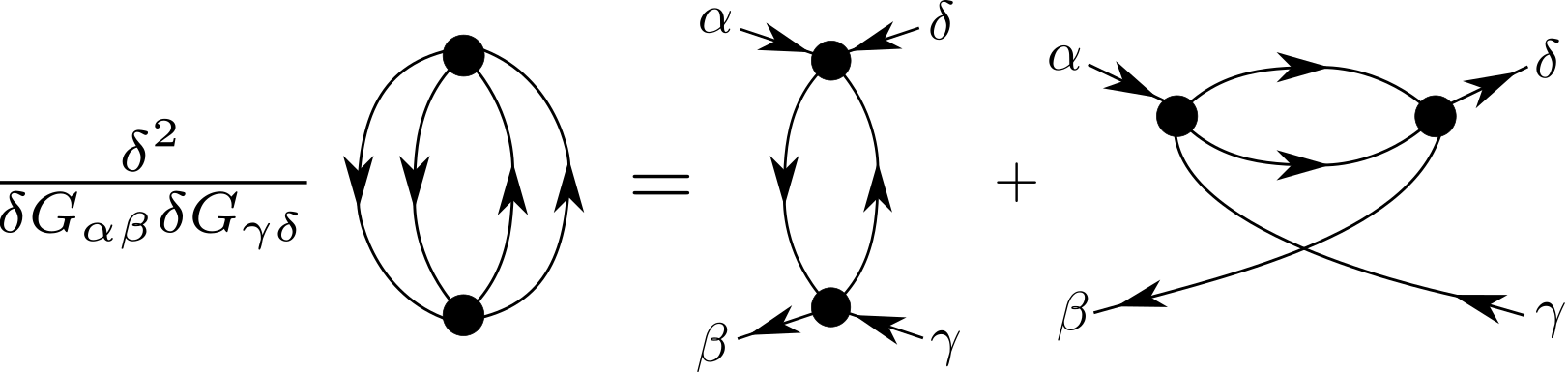}
 \vspace{0.1cm}
 
 \includegraphics[width=\columnwidth]{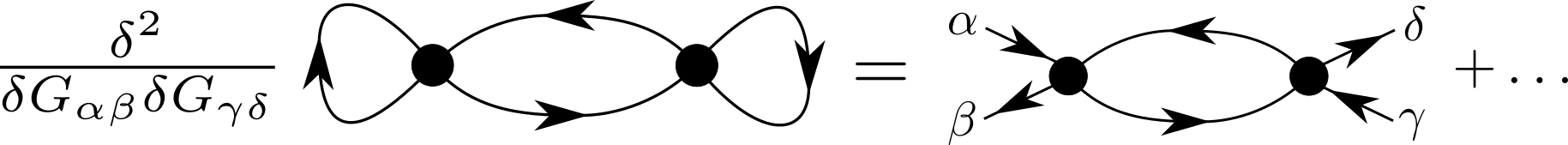}
 \caption{ Upper panel: Second-order functional derivative of the second order contribution to $\Phi$ generates diagrams reducible in the $\mathrm{pp}$ and $\overline{\mathrm{ph}}$ channel. Lower panel: The $\mathrm{ph}$-reducible diagrams can only be obtained by functional derivatives of non-skeleton vacuum diagrams which are not found in $\Phi$. }
 \label{fig:dG_PHI} 
\end{figure}
The Greek indices combine the orbital and spin index and the imaginary time, e.g. $\alpha=(i_\alpha, \sigma_\alpha, \tau_\alpha)$. 
%This quantity contains all the diagrams 
%This connection allows to identify the multi-valued domains in the phase diagram by looking at the divergencies of $\Gamma$ appearing as lines in the $U-\delta$ phase diagram (see left panel of Fig.~1 in main part). 
This relation is illustrated in Fig.~\ref{fig:dG_PHI} for diagrams of the second order.

The connection between $\Gamma$ and $\Phi$ is the reason why $\Gamma$ is sensitive to the multivaluedness of the LWF: it
diverges along the lines in the phase diagram where two branches of the LWF cross\cite{Gunnarsson2017} (see Fig.~\ref{fig:div_and_rev}a in main part). However, note that also $\Gamma_{\mathrm{pp}}$ can diverge in some cases \cite{Schafer2013,Schafer2016}.%, although it is not directly related to the LWF.

One can define the irreducible vertex function in the ``charge'' channel as
$\Gamma_{\mathrm{c}}=\Gamma_\mathrm{ph,\up\up\up\up}+\Gamma_\mathrm{ph,\up\up\dn\dn}$
where we have omitted the time/frequency and orbital indices for clarity.

In this paper we are interested in identifying divergences of $\Gamma_\mathrm{c}$. It doesn't appear explicitly in the cluster DMFT equations, so we only need to calculate it at the end of the self-consistency loop. Note that due to the LWF construction of the methods, we calculate it only from the correlation functions on the impurity.

%To perform this analysis for our cluster DMFT methods we have to calculate the irreducible vertex functions of the effective impurity model obtained in the end of the self-consistency loop.
%This is achieved, however, not via Eq.~\ref{eq:gamma_lwf}, but rather on the basis of the so-called Bethe-Salpeter equation (in the charge channel), that allows for a direct calculation of $\Gamma_{\mathrm{c}}$ from the two- and four-point correlator of the impurity.

\subsection{The Bethe-Salpeter equation}

\begin{figure}[!ht]
 \includegraphics[width=3.2in, trim=0cm 0cm 0cm 0.0cm]{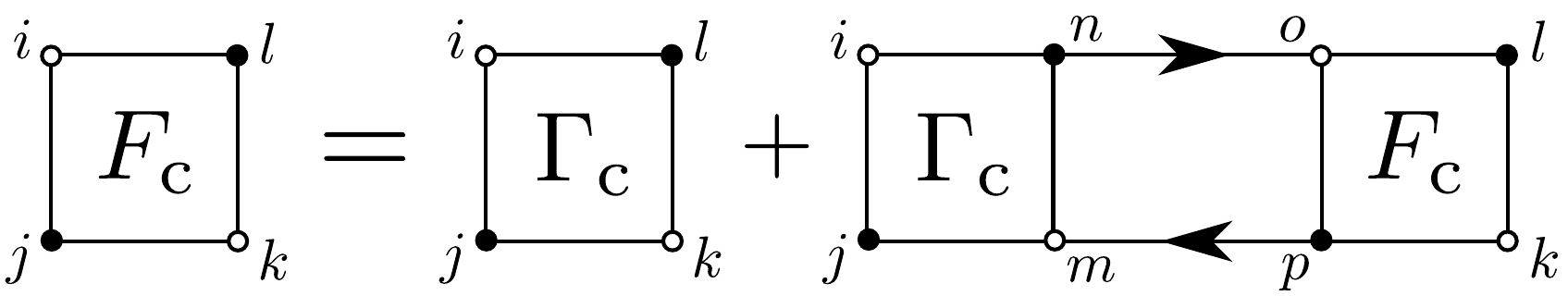}
 \caption{ Diagrammatic representation of the Bethe-Salpeter equation in the charge channel. Empty circles denote an ingoing connector of a vertex function, while black dots correspond to outgoing ones. }
 \label{fig:BSE} 
\end{figure}

In general, $\Gamma_\mathrm{ph}$ can be calculated from the Green's function $G$ and the four-point correlation function
\begin{equation}
   \begin{split}
 &\chi_{4,\sigma\sigma',ijkl}^{\omega\omega'\Omega} = \frac{1}{\beta}\langle c^\dagger_{i,\sigma}(\omega)c_{j,\sigma}(\omega + \Omega)c^\dagger_{k,\sigma'}(\omega' + \Omega)c_{l,\sigma'}(\omega')\rangle \\
 &\quad - G_{ji}(\omega)G_{lk}(\omega') \beta\delta_{\Omega,0}+ \delta_{\sigma,\sigma'} G_{li}(\omega)G_{jk}(\omega + \Omega)\beta\delta_{\omega,\omega'} 
 \end{split}
 \label{eq:chi4_spin}
\end{equation}
where we have assumed SU(2) symmetry and absence of spin-orbit interactions.
First we calculate the general $\chi_4$, and then calculate the charge channel simply via
\begin{equation}
 \chi_{4,\mathrm{c}} = \chi_{4,\uparrow\uparrow} + \chi_{4,\uparrow\downarrow}\label{eq:chi4_charge}
\end{equation}
From this object one can obtain the full vertex function $F_{\mathrm{c}}$, which contains \emph{all} the possible two-particle scattering processes (including the reducible ones). It is identical to the four-point correlation function with amputated incoming/outgoing two-point propagators
\begin{equation}
   \begin{split}
 &F^{\omega\omega'\Omega}_{\mathrm{c},ijkl} = \\
 &\sum_{mnop} G^{-1}_{mi}(\omega) G^{-1}_{ok}(\omega'+\Omega) \chi^{\omega\omega'\Omega}_{4,\mathrm{c},mnop}  G^{-1}_{jn}(\omega+\Omega) G^{-1}_{lp}(\omega')
   \end{split}
\end{equation}
%On the other hand $\Gamma_\mathrm{c}$ contains only the diagrams two-particle-irreducible\cite{Rohringer2012} in the charge scattering channel.
$\Gamma_\mathrm{c}$ is linked to $F$ by the corresponding Bethe-Salpeter equation (BSE).
The BSE can be understood as a Dyson Equation at the two-particle level\cite{Bickers2004}, and it reads 
%By raising the notion of one-particle irreducibility as required by the self-energy diagrams to the so-called two-particle irreducibility (in a particular scattering channel) , one can derive the Bethe-Salpeter equation
%
\begin{equation}
\begin{split}
   F^{\omega\omega'\Omega}_{\mathrm{c},ijkl} &=  \Gamma^{\omega\omega'\Omega}_{\mathrm{c},ijkl} \\
   &- \frac{1}{\beta} \sum_{\omega''} \sum_{mnop} \Gamma^{\omega\omega''\Omega}_{\mathrm{c},ijmn} \, G_{on}(\omega'') \, G_{mp}(\Omega + \omega'') \, F^{\omega''\omega'\Omega}_{\mathrm{c},opkl}\label{eq:BSE}
\end{split}
\end{equation}
The diagrammatic representation of BSE is presented in Fig.~\ref{fig:BSE}.

One can invert the BSE to obtain a closed expression for $\Gamma$.
%For practical purposes we invert Eq.~\ref{eq:BSE} to get a direct equation for $\Gamma$.
%In this process it proves useful to introduce a matrix notation for all four-point quantities (vertices and correlators) 
After rewriting the vertex quantities as matrices w.r.t. the properly grouped indices for a given transfer frequency $\Omega$, $\hat{V}^{\Omega}_{(i,j,\omega),(l,k,\omega')} = V^{\omega\omega'\Omega}_{ijkl}$, the BSE becomes a matrix equation
\begin{equation}
\begin{split}
   \hat{F}_{\mathrm{c}}^{\Omega} = \hat{\Gamma}^{\Omega}_{\mathrm{c}} + \frac{1}{\beta^2} \hat{\Gamma}^{\Omega}_{\mathrm{c}} \hat{\chi}_0^{\Omega} \hat{F}^{\Omega}_{\mathrm{c}}\label{eq:BSE_compact}
\end{split}
\end{equation}
where
\begin{align}
   \chi_{0,ijkl}^{\omega\omega'\Omega} = - G_{li}(i\omega)G_{jk}(i\omega + i\Omega)\beta \delta_{\omega,\omega'}
\end{align}
A few algebraic steps then yield
\begin{align}
   \hat{\Gamma}^\Omega_{\mathrm{c}} = \beta^2\left[\left[\hat{\chi}_{0}^\Omega\right]^{-1} - \left[\tilde{\chi}^\Omega_{\mathrm{c}}\right]^{-1}\right]
   \label{eq:BSE_mat_inv}
\end{align}
where we have defined the so-called generalized susceptibility\cite{Rohringer2012}
\begin{align}
   \tilde{\chi}_{\mathrm{c}} %= \hat{\chi}^{0} + \hat{\chi}^{0} \hat{F}_{\mathrm{c}} \hat{\chi}^{0} 
   = \hat{\chi}_{0} + \hat{\chi}_{4,\mathrm{c}}.
   \label{eq:gen_susc}
\end{align}
The matrix $\hat{\chi}_0^\Omega$ is always invertible. This does not necessarily hold for the generalized susceptibility $\tilde{\chi}_{\mathrm{c}}$. As it approaches a singular matrix, $\Gamma_\mathrm{c}$ diverges.
%This quantity is responsible for the divergencies in $\Gamma$, as the matrix can become singular in certain trajectories of the phase diagram.
%This is typically the case as one approaches the strong-coupling regime or lower temperatures, where they first appear for $\Omega=0$, to then turn up also at finite $\Omega$.

While the analysis of $\Gamma$ divergences can be performed for an arbitrary transfer frequency $\Omega$, we here focus only on the $\Omega=0$ case.
$\tilde{\chi}_{\mathrm{c}}^{\Omega=0}$ is a symmetric matrix. In a single-site model at particle-hole symmetry, it is also purely real, which makes it Hermitian, and its eigenvalues purely real. In cluster-impurity models, and/or away from ph-symmetry, it can have complex elements, and its eigenvalues are no longer necessarily real\cite{Gunnarsson2016}.
%We note that both, $\hat{\chi}_0$ and $\tilde{\chi}$, are real-valued only for the particle-hole symmetric case. Further, both matrices are symmetric at $\Omega=0$.

\subsection{Eigenvalues and divergences}

In this part we present the procedure for determining the divergence trajectories of the irreducible vertex function, $\Gamma_{\mathrm{c}}$. At a fixed temperature $T/D=0.125$, in the $(\delta,U)$ phase diagram discussed in the main text, we determine trajectories $U^{d=1,2,3,...}(\delta)$ where $d$ indexes different divergences, counting from the low $U$ ($U^d<U^{d'},\;d<d'$).

From Eq.~(\ref{eq:BSE_mat_inv}) it is clear that $\Gamma_{\mathrm{c}}^{\Omega=0}$ diverges when an eigenvalue of $\tilde{\chi}_\mathrm{c}^{\Omega=0}$ goes through zero.
The dimension of the matrix is $M = N_\omega \times N^2_c$, where $N_\omega$ is the number of fermionic frequencies stored, and $N_c$ is the number of sites in the cluster.
We start by solving the eigenproblem for $\tilde{\chi}_\mathrm{c}^{\Omega=0}$. We fully diagonalize this matrix at each discrete value of $U$ ($U_l$) at a fixed doping $\delta$, and obtain a set of $M$ eigenvectors and eigenvalues $\{(\mathbf{v}^l_i,\varepsilon^l_i)\}_{i\in[0,M)}$.
In single-site DMFT, at the lowest $U$, the real part of all eigenvalues is positive ($\mathrm{Re}\varepsilon_i>0$). Therefore, as we iterate over the interaction values $U_l$, it is straightforward to detect when the real part of an eigenvalue crosses zero - it is whenever a new eigenvalue with the negative real part appears. However, with this simple method, the error bar for $U^d(\delta)$ is given by the interaction step $U_{l+1}-U_l$. Furthermore, this method could potentially miss an event where between two $U_l$'s two eigenvalues cross zero, one becoming negative, the other one positive. This is particularly important in CDMFT $2\times 2$ where there are many negative eigenvalues present already at the lowest $U$. Furthermore, we would like to know the exact value of the imaginary part of the eigenvalue ($\mathrm{Im}\varepsilon_i$) when its real part is crossing zero - if it's non zero ($\mathrm{Im}\varepsilon_i\neq0$), $U^d(\delta)$ at that point does not correspond to an actual divergence of $\Gamma$.

One can do better by connecting the eigenvalues $\varepsilon^l_i$ according to matching eigenvectors and then interpolating $\varepsilon^l_i\rightarrow \varepsilon_i(U)$. $U^d(\delta)$ is then defined by $\mathrm{Re}\varepsilon_i(U^d)=0$. We start from the lowest $U$ ($l=0$), and for each eigenvector $\mathbf{v}^l_i$ we search for an eigenvector $\mathbf{v}^{l+1}_j$, such that $|\mathbf{v}^{l}_i\cdot\mathbf{v}^{l+1}_j|$ is maximal. After this is done for all eigenvectors $\mathbf{v}^l_i$, one proceeds with the next $l$ until all the eigenvector/eigenvalue pairs are connected across the entire range of $U$. This procedure is, however, not entirely straightforward, especially when the step in $U$ is big. The eigenbasis rotates with changing $U$, and in a given $U$ step, different eigenvectors may ``exchange''. In the single-site DMFT calculation, we had to additionally require that $\epsilon^l_i$ is smooth to avoid getting eigenvalues mixed up. In CDMFT $2\times 2$, the vector space is much bigger and we encountered no such problems. Note also that, as doping is changed, the eigenvectors change considerably, and we were unable to reliably connect the eigenvalues at the same $U$, but different values of doping.

\begin{figure}[!ht]
 %\centering{}
 \includegraphics[width=3.2in,trim=1cm 0cm 1cm 0cm]{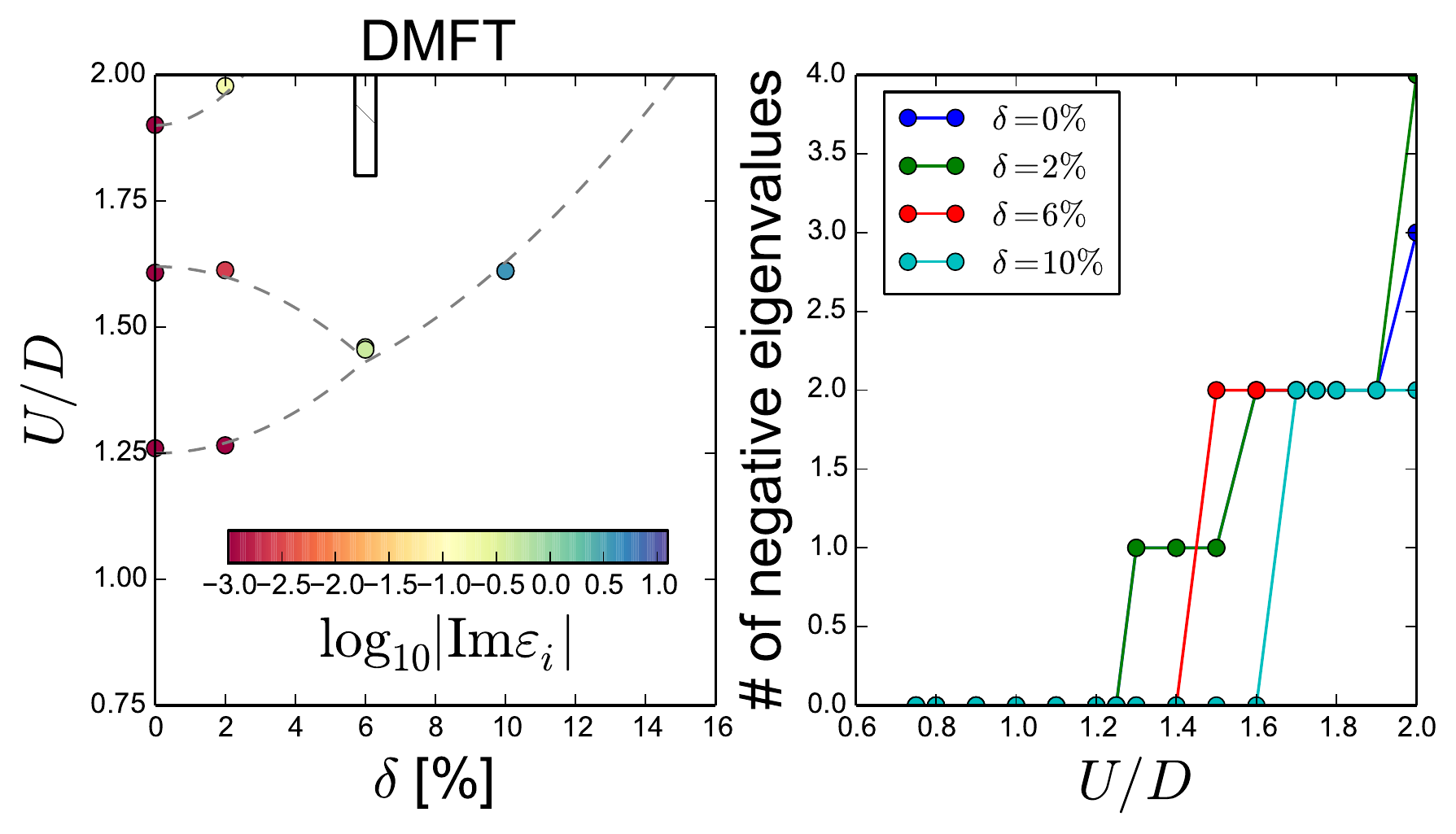}
 \includegraphics[width=3.2in,trim=1cm 0cm 1cm 0cm]{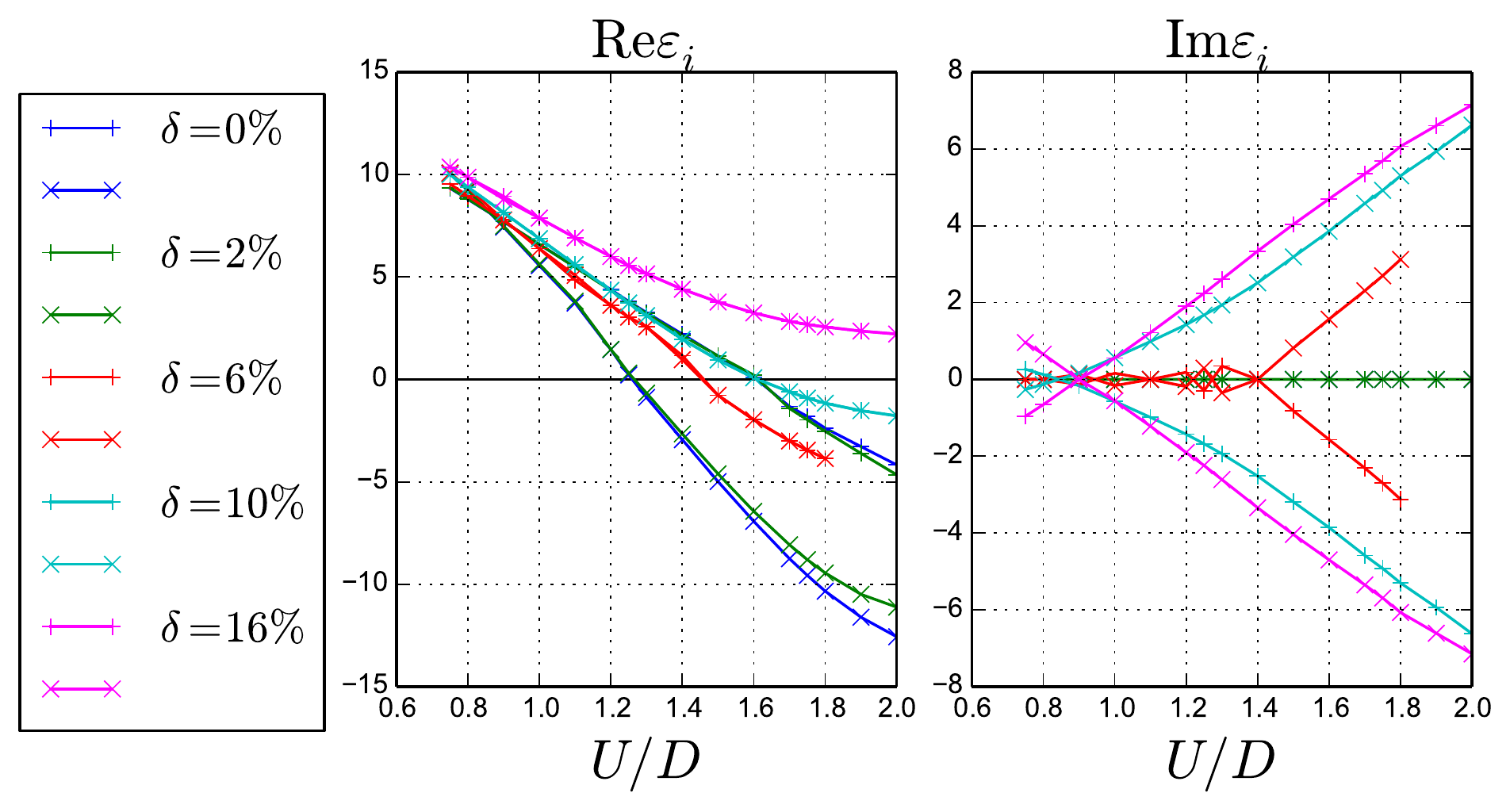}
 \caption{ Single-site DMFT calculation. Temperature $T/D=0.125$, square lattice. 
   Upper left panel: the colored points denote where the real part of an eigenvalue of $\tilde{\chi}_\mathrm{c}(i\Omega=0)$ crosses zero; the color denotes the imaginary part at that point, with respect to the colorbar in the inset; gray dashed lines are eye-guides, used also in the Fig.~\ref{fig:div_and_rev} in the main text. Upper right panel: the number of negative eigenvalues; at $\delta>5\%$ two eigenvalues cross zero at the same time. Bottom panels: evolution of the first two eigenvalues crossing zero, with $U$ and $\delta$. They remain purely real before becoming mutually complex conjugate. No eigenvalues cross zero at $\delta=16\%$.
 }
 \label{fig:eigenvalues_DMFT} 
\end{figure}

In Fig.~\ref{fig:eigenvalues_DMFT} we present the results from the single-site DMFT calculation. Here we have data at $\delta=0,2,6,10,16\%$. On the top left panel, results for $U^d(\delta)$ are presented with colored circles; the color represents the imaginary part of the eigenvalue crossing zero (color code is in the inset). The dashed lines are guides for the eye, and are also presented on Fig.~\ref{fig:div_and_rev}a in the main text. The total count of negative eigenvalues as a function of $U_l$ is given on the top right panel. We see that at $\delta=0,2\%$, eigenvalues cross zero one by one. Then at $\delta = 6,10\%$, we see that two eigenvalues cross zero in the same $U$-step. In the bottom panels we plot the interpolation $\epsilon_i(U)$ obtained after connecting the eigenvalues at different values of $U$. We present only the first two eigenvalues to cross zero in the examined range of $U$. We note that these eigenvalues are the highest valued ones at the lowest $U$. At $\delta=16\%$ no eigenvalues have the real part cross zero, and instead we present the two mutually complex conjugate eigenvalues which are the biggest ones at the lowest $U$, and thus apparently correspond to the 2 eigenvalues crossing zero at the lower $\delta$'s. We see that at low $\delta$ we have two separate eigenvalues which are purely real and cross zero at different values of $U$. Then at $6\%$ doping, the two eigenvalues crossing zero are mutually complex conjugate, and cross zero at the same time, but with finite imaginary parts of opposite signs. As doping is further increased, the two eigenvalues remain mutually complex conjugate and have the real part grow towards positive values such that at $\delta=16\%$ they no longer cross zero. The imaginary part grows with both doping and interaction.

\begin{figure}[!ht]
 %\centering{}
 \includegraphics[width=3.2in,trim=1cm 0cm 1cm 0cm, page=1]{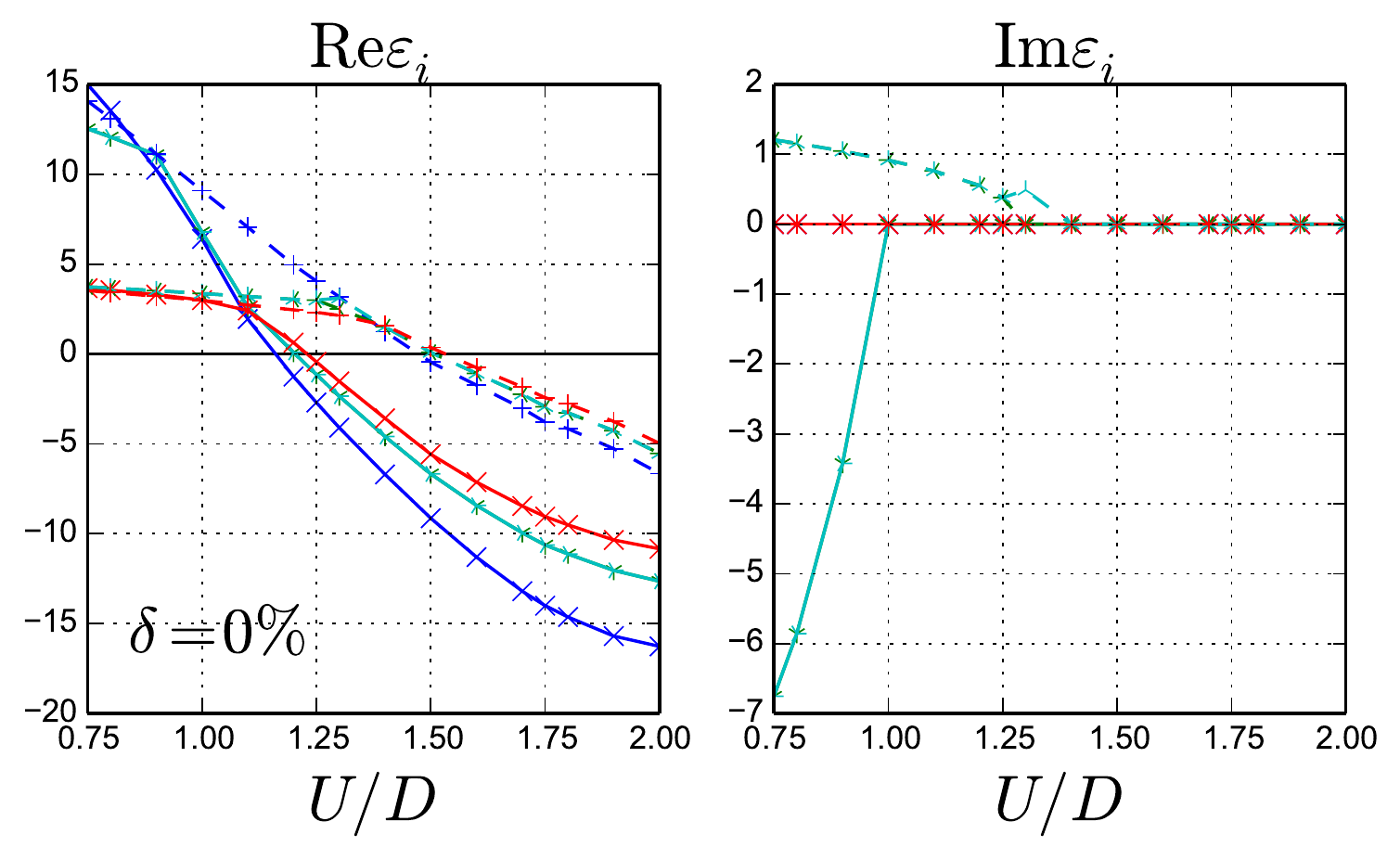}
 \includegraphics[width=3.2in,trim=1cm 0cm 1cm 0cm, page=2]{CDMFT_eigenvalues_figure}
 \includegraphics[width=3.2in,trim=1cm 0cm 1cm 0cm, page=3]{CDMFT_eigenvalues_figure} 
 \caption{ 2x2 CDMFT calculation. Temperature $T/D=0.125$, square lattice. Evolution of first 8 eigenvalues to cross zero, with $U$ and $\delta$.   
 }
 \label{fig:eigenvalues_CDMFT} 
\end{figure}

In Fig.~\ref{fig:eigenvalues_CDMFT} we present the result from CDMFT $2\times 2$. We show the result for the first eight eigenvalues to cross zero. These are separated in two groups of four (yielding $U^d(\delta)$ with $d=1-4$ and $d=5-8$), and each group apparently corresponds to one of the two eigenvalues crossing zero in single-site DMFT. At higher $U$ there is another group of four eigenvalues crossing zero ($d=8-11$, not shown), apparently corresponding to the 3rd divergence in single-site DMFT. The two middle eigenvalues in all groups are mutually complex conjugate (the ones yielding $U^d(\delta)$ with $d=2,3$, $d=6,7$ and $d=10,11$) . At $\delta=3\%$, we see that the first two groups merge at around $U=1.8$ ($d=1$ with $d=5$, $d=2$ with $d=6$, and so on). This point is denoted with the vertical gray dashed line. The merging of eigenvalues occurs at different $U$ for various dopings, along the gray dashed line on the phase diagram in the inset. At $\delta=8\%$, there are still 8 eigenvalues crossing zero, but they have only 3 distinct real parts - first and last doubly degenerate, the middle one 4-times degenerate. After merging, the imaginary part of the eigenvalues grows from zero, both with $U$ and $\delta$, similarly to the single-site DMFT case.

Note that we have performed the analogous analysis also in DCA, DCA$^+$ and PCDMFT. The overall picture is very similar. The only qualitative difference is the presence of additional crossings of zero at low $U$ in DCA/DCA$^+$. These however occur with a very big imaginary part and do not correspond to singular behavior of $\Gamma$.

\section{Nested cluster scheme} \label{sec:ncs}

%Small intro, history, connection with SEET, more about what has been done, our contribution (3 clusters enough!).

In this section we present the fully general formalism of the self-energy embedding theory (SEET) and then focus on its application to infinite lattice systems (NCS).
The main idea is to approximate the Luttinger-Ward functional $\Phi$ (LWF) by a sum of functionals, including counter terms to cancel double counting of diagrams.
By now it is clear that combining different LW functionals is a very general approach, and can lead to a great variety of approximations. For example, one can rederive within the SEET framework also the GW+EDMFT method\cite{Zgid2017,Sun2002}.  Moreover, CDMFT can be viewed as a special case of NCS, where no counter terms are needed in the construction of the LWF.

We develop a general algorithm to obtain NCS based LWF approximations and the corresponding self-energy expressions with no doubly counted diagrams, given a set of independent clusters one wishes to solve.  Also, it was not clear previously whether pushing the cluster size will also increase the number of impurity problems one needs to solve. Here we prove that in the simplest scheme (square clusters), one needs to solve only 3 impurity problems, regardless of the cluster size.

\subsection{General formulation}

Consider a system with single-particle degrees of freedom $i \in {\cal L}$. At this point these may be lattice sites, or more general orbitals, and the system may or may not be infinite.
The exact Luttinger-Ward functional depends on all the components of the Green's function
\begin{equation}
 \Phi[G] \equiv \Phi[\{G_{ij}\}_{i,j\in{\cal L}}]
\end{equation}
Consider now an approximation of the Luttinger-Ward functional, such that it is a sum of functionals, each depending on components of $G$ that connect only a certain subset $C$ of orbitals $i$, i.e. components of $G$ within a ``cluster'' $C\subset{\cal L}$
\begin{equation}
 \Phi \approx \sum_{C\in{\cal C}} \Phi_C[G|_C]
\end{equation}
with 
\begin{equation} \label{eq:bar_notation}
G|_C \equiv \{ G_{ij} \}_{ij\in C}
\end{equation}
where $|_C$ denotes the restriction of the orbital-space domain of the Green's function to the cluster $C$.
It is assumed that the clusters are mutually independent
$$C \nsubseteq C', \forall C,C'\in{\cal C}$$
and cover the entire system
$$\bigcup_{C\in{\cal C}} C = {\cal L}$$
However, if any of the clusters are overlapping
\begin{equation}
 \exists C,C' \in {\cal C}: C \cap C' \neq \{\}
\end{equation}
then we are double-counting diagrams constructed entirely from $G$ components connecting the orbitals present in both $C$ and $C'$. To avoid this, we need to add functional counter terms, each dependent only on $G_{ij}$ within an overlap of clusters in $\cal C$. In general
\begin{equation}
\Phi[G] \approx \sum_{C\in{\cal C}} \Phi_C[G|_C] + \sum_{C\in{\cal O}}\, p_C\, \Phi_C[G|_C]
\end{equation}
where $\cal O$ is the set of all possible overlaps between any number of non-identical clusters in set $\cal C$, i.e.
$${\cal O} = \bigcup_{n\in[2,N_{\cal C}]} \Bigg\{\bigcap_{a=1}^n C_a\Bigg\}_{C_a\in{\cal C}} \setminus {\cal C} $$
$N_{\cal C}$ is the size of the set ${\cal C}$.
% $${\cal O} = \bigcup_{n\in[2,N_{\cal C}]} \bigcup_{1\leq a_1< N_{\cal C}}\bigcup_{a_1< a_2\leq N_{\cal C}}..\bigcup_{a_{n-1}< a_n\leq N_{\cal C}}
%  \Bigg\{\bigcap_{a=1}^n C_a\Bigg\} $$
$p_C$ are appropriately chosen integer prefactors, possibly negative or even zero. 
%{\red If the set $\cal O$ does not have the property $C \cap C' \in {\cal O}, \forall C,C\in{\cal O}$, further terms must be added to account for overlaps of overlaps (in the schemes discussed here  that is not necessary) GIVE ME ONE EXAMPLE OTHERWISE ERASE THIS}.

\begin{figure}[!ht]
 %\centering{}
 \includegraphics[width=1.8in,trim=0cm 0cm 0cm 0cm]{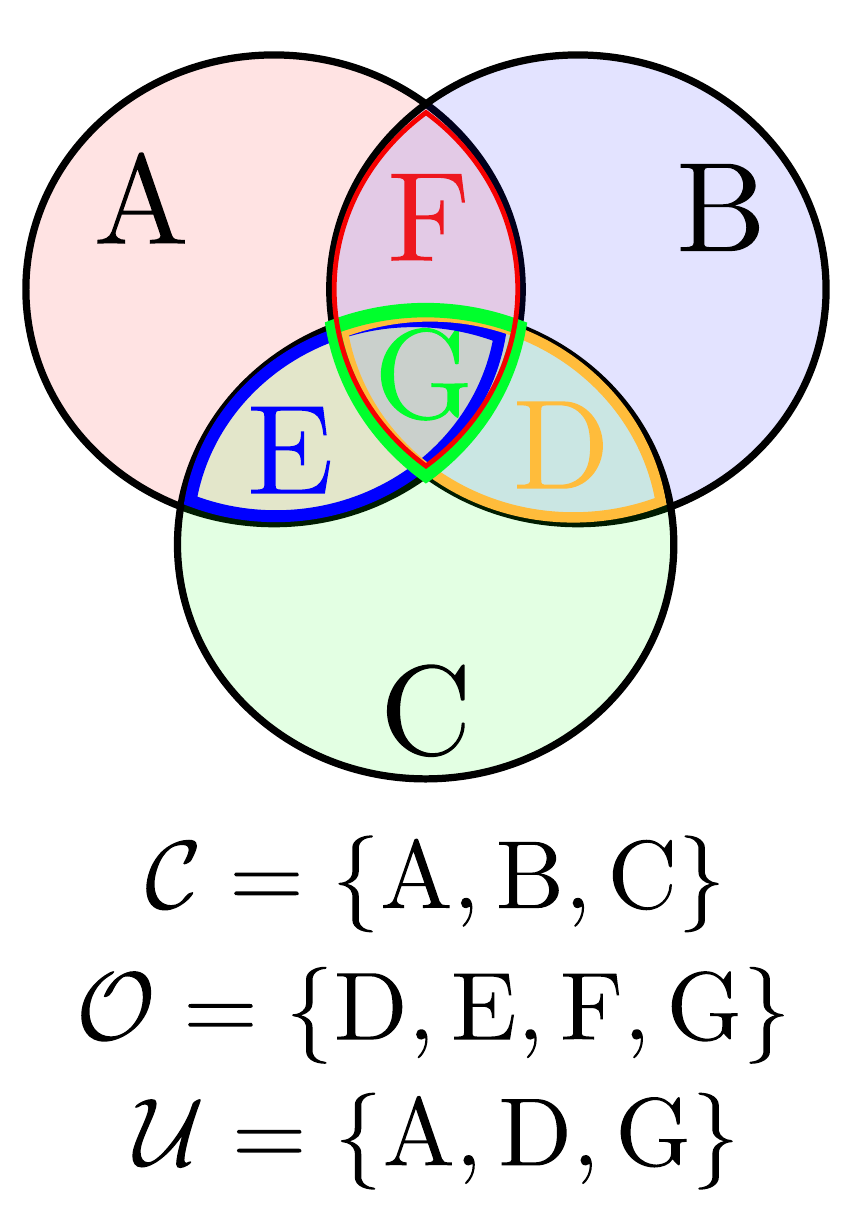} 
 \caption{ Illustration of reasoning in SEET. $\cal C$ sets of orbitals are chosen. $\cal O$ are all possible overlaps. $\cal U$ are sets independent by symmetry that one needs to solve in practice.
 }
 \label{fig:venn_diagram} 
\end{figure}

We emphasize that NCS is defined only by the choice of $\cal C$; the set $\cal O$ and prefactors $p_C$ are then determined uniquely by the requirement that no diagrams are counted more than once.
We can rewrite more simply
\begin{equation}
\Phi[G] \approx \sum_{C\in{\cal C} \cup {\cal O}} p_C\Phi_C[G|_C]
\end{equation}
where $p_{C\in{\cal C}}=1$. Hereinafter, summation $\sum_C$ is assumed to go over ${\cal C}\cup{\cal O}$ unless stated differently.
Finally, the prefactors $p_C$ must satisfy for each $C$
\begin{equation} \label{eq:nested_pC_condition}
 \sum_{\substack{C' \in {\cal C}\cup{\cal O} \\ C'\supseteq C}} p_{C'} = 1
\end{equation}
which means that the contribution of diagrams that involve orbitals from a given cluster $C$ are taken into account exactly once. In Section \ref{sec:algorithm} we present an algorithm to find $p_C$ which satisfy this requirement.

\subsubsection{Obtaining self-energy from the functional}

Anticipating that the formalism will be used for the lattice systems below, and to avoid introducing new notation, here we assume that the system is a lattice, with lattice sites $\{i\}$. Note, however, that the considerations presented here are still fully general. 

The self-energy is given by the functional derivative with respect to the Green's function
\begin{eqnarray} \label{nc_sigma_approx}
 \Sigma_{\alpha\beta}  &=& \sum_C\, p_C\, \left. \frac{ \partial \Phi[\{ G_{\gamma\delta} \}_{i_\gamma,i_\delta\in C}] }
                                            { \partial G_{\beta\alpha}} \right|_{G=G[G_0,\Sigma]}  
\end{eqnarray}
where $\alpha,\beta...$ are combined indices defined in Section \ref{sec:vertex}.
We can obtain it by solving a set of impurity problems
\begin{eqnarray} \nonumber
 S^C &=& \sum_{IJ\in C,\sigma} \iint d\tau d\tau' c^+_{\sigma,I}(\tau) [-(\hat{\cal G}^C)^{-1}_\sigma]_{IJ}(\tau-\tau') c_{\sigma,J} (\tau') \\ 
   &&  + U \sum_{I\in C} \int d\tau c^+_{\up,I}(\tau) c^+_{\dn,I}(\tau) c_{\dn,I}(\tau) c_{\up,I}(\tau)
\end{eqnarray}
corresponding to each cluster $C$, under the condition that
\begin{equation} \label{eq:nested_sc}
 G^{\imp\,C}_{IJ}(i\omega_n) = G^\latt_{i_{C,I}j_{C,J}}(i\omega_n)
\end{equation}
where $i_{C,I}$ is the mapping between the index $I$ of a site within a cluster $C$ and its index $i$ within the lattice. The Green's function on the lattice is a matrix in site-indices
\begin{equation}
 \hat{G}^\latt(i\omega_n) = [(i\omega_n+\mu)\hat{I} - \hat{\Sigma}^\latt]^{-1}
\end{equation}
and the self-energy approximation on the lattice is given by
\begin{eqnarray} \label{eq:general_nested_sigma}
  \Sigma^\latt_{ij}  &=& \sum_{C\supseteq\{i,j\}}\, p_C\, \Sigma^{\imp\,C}_{I_{C,i}J_{C,j}}                               
\end{eqnarray}
where $I_{C,i}$ is the mapping between the site index on the lattice and in the cluster $C$, inverse of the previously defined $i_{C,I}$.
Note that up to now we have not used any lattice symmetries. Therefore, this prescription can be used for solving (finite-size) disordered and inhomogenous lattice models (e.g. one could write down a cluster extension of real-space DMFT (ref) ).

\subsubsection{Application to lattice models}

When there are symmetries in the system Hamiltonian, one should choose $\cal C$ in a way that does not artificially break those symmetries.
For example, if there is translational symmetry on the lattice, clusters must be arranged uniformly across the entire lattice; if there is rotational symmetry, the arrangement must be the same along equivalent directions. A simple realization of a translationally and rotationally invariant set $\cal C$ for a square lattice would include $2\times 2$ plaquettes on all possible positions on the lattice. On the contrary, if the plaquettes are only tiled over the system, with no overlaps (as is the case in CDMFT), the translational symmetry is artificially broken.

If translational, rotational and mirror symmetry are present, the number of clusters one actually needs to solve is reduced - one solves only one cluster of each different shape and/or size. Due to translational invariance, the position of the cluster on the lattice does not make a difference, just its shape/size.
Due to rotational symmetry, quantities on clusters of the same (non-square) shape, but different orientation, can be inferred one from another. 

Translational symmetry also allows to rewrite the lattice quantities as functions of the real-space vector rather than matrices in site-index. The self-consistency condition can be rewritten as
\begin{equation} \label{eq:nested_sc_lattice}
 G^{\mathrm{imp}\,C}_{IJ}(i\omega_n) = G^\latt_{\mathbf{r}=\mathbf{r}_{i_{C,I}}-\mathbf{r}_{j_{C,J}}}(i\omega_n)
\end{equation}
%Now the clusters $C$ do not cover the entire lattice, and the choice of $i_{C,I}$ is not unique.

The Green's function on the lattice, again, is calculated from the approximated self-energy
\begin{equation}
G^\latt_{\mathbf{r}} = \sum_{\mathbf{k}\in BZ} e^{-i\mathbf{k}\cdot\mathbf{r}} G^\latt_\mathbf{k} =  \sum_{\mathbf{k}\in BZ}  \frac{e^{-i\mathbf{k}\cdot\mathbf{r}}}{G_{0,\mathbf{k}}^{-1}(i\omega_n) - \Sigma^\latt_\mathbf{k}(i\omega_n)}
\end{equation}
\begin{eqnarray} 
  \Sigma^\latt_{\mathbf{k}}  &=& \sum_{\mathbf{r}\in BL} e^{i\mathbf{k}\cdot\mathbf{r}} \Sigma^\latt_\mathbf{r}
\end{eqnarray}
which is given by a general expression
\begin{eqnarray} \label{eq:sigma_latt_with_symm}
  \Sigma^\latt_{\mathbf{r}}  &=& \sum_{C\in{\cal U}}\, \sum_{IJ} a_{\mathbf{r},C,I,J}\, \Sigma^{\imp\,C}_{IJ}                               
\end{eqnarray}
The sum runs only over a set of clusters ${\cal U}\subset{\cal C}$ independent by lattice symmetry. If both translational and rotational symmetry are present, $\cal U$ contains a single choice of a cluster, of each size and shape, and the sum over $IJ$ accounts for all the shifts and rotations of the same cluster on the lattice.
% \begin{equation}
%  G^{\mathrm{imp}\;C}_{l_{C,i}m_{C,j}}(i\omega_n) = \Bigg[\bigg(\hat{\cal G}^{C,-1}(i\omega_n) - \hat{\Sigma}^C (i\omega_n)\bigg)^{-1}\Bigg]_{l_{C,i}m_{C,j}}  = G_{ij}(i\omega_n) = \Bigg[\bigg( \hat{G}_{0}^{-1}(i\omega_n) - \hat{\Sigma}(i\omega_n) \bigg)^{-1} \Bigg]_{ij}
% \end{equation}
Note that $\sum_{CIJ} a_{\mathbf{r},C,I,J} = 1$ and $a_{\mathbf{r},C,I,J} \sim \delta_{\mathbf{r},\;\mathbf{r}_{i_{C,I}}-\mathbf{r}_{j_{C,J}} }$. Because some bonds on the cluster correspond to the same real-space vector and can have the same self-energy due to the symmetries of the cluster, one is free to choose which one to use, so $a_{\mathbf{r},C,I,J}$ is not uniquely defined. More importantly, $\sum_{CIJ} |a_{\mathbf{r},C,I,J}| \sim N_c$. This is a problematic property of the method and is the reason why the limit $N_c\rightarrow \infty$ does not guarantee the exact solution, and is the reason for an undesirable amplification of statistical noise when clusters are big.

\paragraph{Large cluster limit.}
As cluster size increases, the difference in self-energy between different clusters becomes smaller, and the self-energy on the clusters becomes more uniform. On the other hand the coefficients $a_{\mathbf{r},C,I,J}$ grow by absolute value roughly proportionally to $N_c$, while their total sum remains 1. This means that in the limit $N_c\rightarrow \infty$, an infinitesimal difference between the self-energies in different clusters and at different positions in the same cluster, all corresponding to the same real-space vector, can in principle be amplified such that
\begin{equation}
\Sigma^\latt_{\mathbf{r}=\mathbf{r}_{i_{C,I}}-\mathbf{r}_{j_{C,J}}} - \Sigma^{\imp\,C}_{IJ} \sim 1 
\end{equation}
Whether this happens or not depends on whether the coefficients $a_{\mathbf{r},C,I,J}$ grow more quickly than do decay the difference between clusters and the inhomogeneity within them.
On the other hand, in the $N_c\rightarrow \infty$ limit we have $G^\imp_0 \rightarrow G_0$, where $G^\imp_0$ denotes the static part of the bare propagator on the impurity (${\cal G} = \big[[G_0^\imp]^{-1} - \Delta\big]^{-1}$). So, if $\Sigma^\latt \neq \Sigma^\imp$, we must have a non-zero $\Delta$ to satisfy the self-consistency condition (recall Eq.~\ref{eq:nested_sc})
\begin{equation}
  \Big[ [G^\imp_0]^{-1} - \Delta - \Sigma^\imp \Big]^{-1} = \Big[ [G_0]^{-1} - \Sigma^\latt \Big]^{-1}.
\end{equation}
Because of this, NCS does not guarantee that in the $N_c\rightarrow \infty$ limit we arrive at the exact solution. A way of checking is to see whether the hybridization function falls off with increasing cluster size.
\paragraph{Amplification of noise.} \label{sec:nested_noise}
Having the property $\sum_{CIJ} a_{\mathbf{r},C,I,J} = 1$, when coefficients are large by absolute value, leads to amplification of QMC statistical error. The problem can be reduced by using symmetries of the clusters, but may prove prohibitive at very large cluster sizes. On the other hand, an approximate solution not involving a stochastic impurity solver, can be safely pushed to bigger cluster sizes.

%In the following we will present a simple general nested cluster scheme for the square lattice, that we use in the main part of the paper, at various cluster sizes. Then we proceed to give a general algorithm to produce self-energy expressions for this one and even more general nested schemes.

\subsection{Square cluster case} \label{sec:square_clusters}

For the special case that $\cal C$ contains all $L \times L$ square clusters of the lattice, 
the nested cluster approximation for $\Phi$ can be written down explicitly for arbitrary size $L$.
It turns out that the only overlaps $C\in \cal O$ with $p_C\neq 0$ are the clusters of shape $L-1\times L$, $L\times L-1$ and $L-1 \times L-1$,
i.e. $\Phi$ is approximated by (Eq.~\ref{eq:def_nested} from the main text)
\begin{multline}\label{eq:def_nested_appendix} 
\Phi^{(L)} = \sum_i 
\,\,\,
\Phi_{L\times L}\bigl[G|_{{C}^{L\times L}_i}\bigr] 
\, + 
\Phi_{L-1\times L-1}\bigl[G|_{{C}^{L-1\times L-1}_i}\bigr]
\\
-\Phi_{L-1 \times L}\bigl[ G|_{{C}^{L-1 \times L}_i}\bigr] 
-\Phi_{L\times L-1}\bigl[ G|_{{C}^{L\times L-1}_i}\bigr] 
\end{multline}
% \begin{multline}\label{eq:def_nested} 
% \Phi^{(L)} \equiv \sum_i 
% \,\,\,
% \Phi_{L^2}\bigl[G|_{{C}^{L\times L}_i}\bigr] 
% \, + 
% \Phi_{(L-1)^2}\bigl[G|_{{C}^{L-1\times L-1}_i}\bigr]
% \\
% -\Phi_{L(L-1)}\bigl[ G|_{{C}^{L-1 \times L}_i}\bigr] 
% -\Phi_{L(L-1)}\bigl[ G|_{{C}^{L\times L-1}_i}\bigr] 
% \end{multline}
Here $G|_C$ denotes the Green's function with the orbital-domain restricted to the sites within cluster $C$ (recall Eq.~(\ref{eq:bar_notation})).
The notation $C_i^{L_x\times L_y}$ denotes a rectangular cluster with width $L_x$ and height $L_y$ with its bottom left site sitting at lattice site $i$.

In the following we prove that $\Phi^{(L)}$ contains only the diagrams which can fit in a cluster $L\times L$, and counts each exactly once.
\subsubsection{Proof of Eq.~\ref{eq:def_nested}}

Let us consider any one diagram of $\Phi^\latt$ in real space.
This defines the (finite) set of lattice sites $D = \{\mathbf{i}\}$ contained in it.
Denoting the coordinate as $\mathbf{i} = (\mathbf{i}_x, \mathbf{i}_y)$, we define 
\begin{align}
 n &= \max_{\mathbf{i} \in D} ( \mathbf{i}_x ) - \min_{\mathbf{i} \in D} ( \mathbf{i}_x ) + 1
 \\ 
 p &= \max_{\mathbf{i} \in D} ( \mathbf{i}_y ) - \min_{\mathbf{i} \in D} ( \mathbf{i}_y ) + 1
\end{align}
Then $(n,p)$ is the shape of the smallest rectangular cluster containing the diagram (with $n = p = 1$ in the local case). 

Let us first count the number of times the diagram appears in 
$\sum_i \Phi_{L_x\times L_y}\bigl[G|_{{C}^{L_x\times L_y}_i}\bigr]$.
This count is identical to the number of ways to place a cluster of shape $(n,p)$ into one of shape $(L_x,L_y)$,
{\it i.e.} $f(L_x +1 - n) f(L_y +1 - p)$ where $f(x) = x \theta(x)$ and $\theta$ is the Heaviside function.
Therefore, the number of times the diagram appears in $\Phi^{(L)}$, with proper weights, 
is given by:
\begin{align*}
R =&\  f(L +1 - n) f(L +1 - p) + f(L - n) f(L - p) \\ 
 &\  - f(L - n) f(L + 1 - p) - f(L + 1 - n) f(L - p)
\end{align*}
Whenever $n\leq L$ and $p\leq L$ we have (denoting $a \equiv L+1-n, b \equiv L+1-p$)
\begin{align*} 
   R =&\  (L+1-n)(L+1-p) + (L-n)(L-p) \\
     &\ - (L-n)(L+1-p) -(L+1-n)(L-p) \\
    =&\  ab + (a-1)(b-1) - (a-1)b - a(b-1) \\
    =&\  1
\end{align*} 
while otherwise $R=0$ by the definition of $f$. \emph{QED.} \\

%Therefore, $\Phi^{(L)}$ contains exactly the diagrams which are contained in a cluster of shape $L \times L$,
%counted once, and only those diagrams.

Note that even with the knowledge of $p_C$ for all subclusters, one still needs to write down the expression for $\Sigma^\latt[\Sigma^\imp]$. We discuss the way this is done in the following sections, including nested schemes more general than the square cluster case discussed here.

\subsection{Algorithm for self-energy coefficients} \label{sec:algorithm}

\begin{figure}[!ht]
 %\centering{}
 \includegraphics[width=3.2in,trim=0cm 0cm 0cm 0cm]{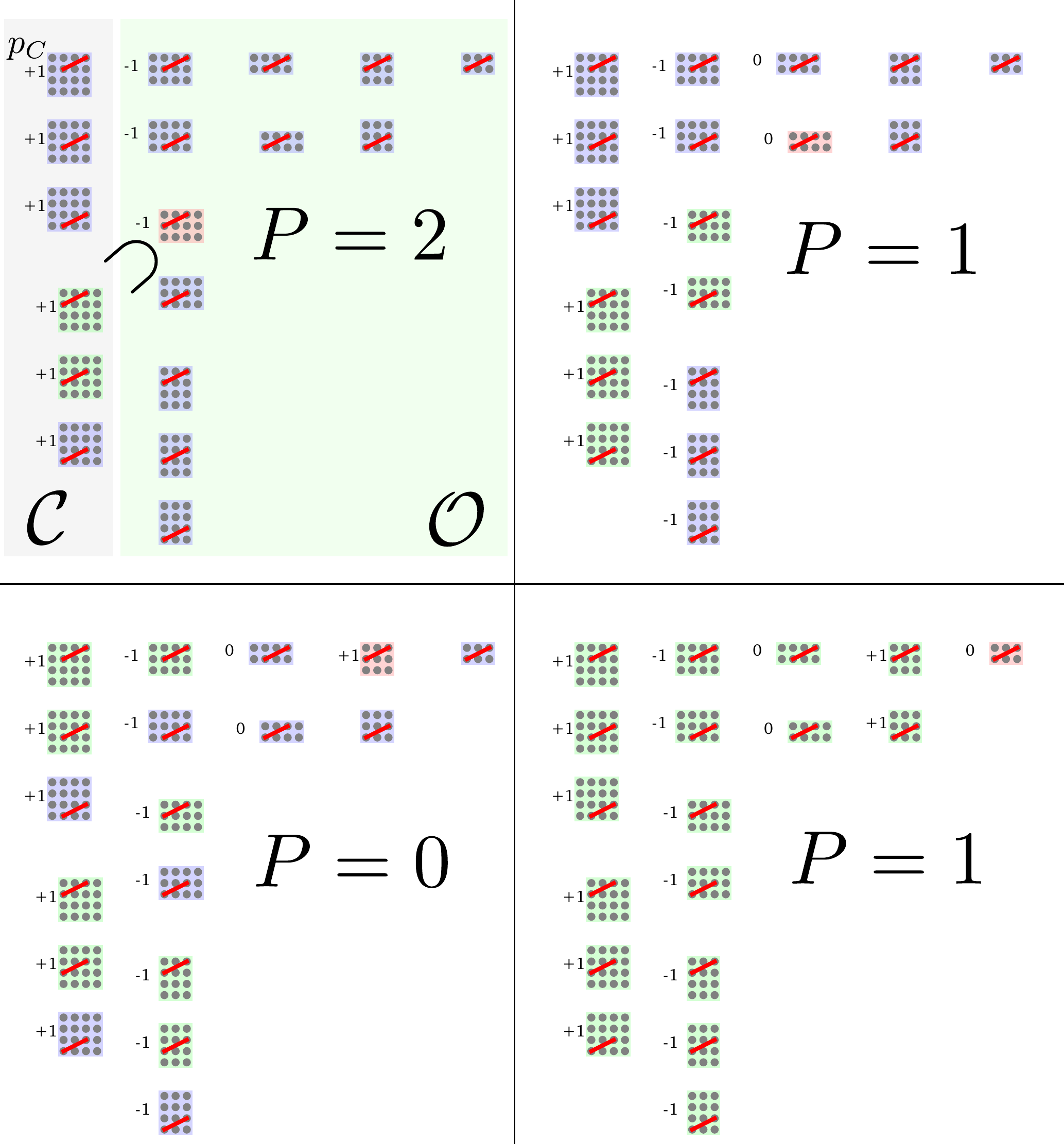} 
 \caption{ Four snapshots of the algorithm for getting the self-energy coefficients; $4\times 4$ nested scheme; coefficients are calculated for an $ij$ bond (red line) corresponding to the real-space vector $\mathbf{r}=(2,1)$; clusters are sorted by size; the prefactors $p_C$ are determined starting from the biggest clusters for which we know $p_C=1$; the red cluster is the one for which $p_C$ is determined at the given step; green clusters are the ones that contain the red cluster. 
 See text for details.
 }
 \label{fig:algorithm} 
\end{figure}

Deriving expressions for the self-energy when $\cal C$ clusters are taken to be bigger than $2\times2$ becomes very cumbersome, and should not be done by hand. Here we present a general algorithm for a uniform and rotationally invariant arrangement of solid rectangular clusters (solid meaning there are no sites missing in the rectangle; a more general algorithm can be devised, but we don't present it here). No symmetries are assumed in the beginning, and the first part of the algorithm gives the fully general expression for $\Sigma^\latt_{ij}$ at a given choice of $ij$. In the second part, the symmetries of the lattice and the clusters are used to fully simplify the expressions.

The algorithm finds the subset of clusters and the corresponding coefficients $p_C$ that appear in the expression Eq.~(\ref{eq:general_nested_sigma}), for a given $ij$ on the lattice.
The prefactors $p_C$ are determined so as to satisfy Eq.~\ref{eq:nested_pC_condition}.
The algorithm finds all the clusters in $\cal C$ and their overlaps $\cal O$ containing the given 2 sites $i$ and $j$ ($i=j$ allowed), orders them by size, and then assigns the prefactors starting from biggest clusters, i.e. the ones in $\cal C$ for which we know $p_{C\in{\cal C}}=1$. For the rest of the clusters $C$, the prefactors $p_{C'}$ of their super-clusters $C'\supset C$ are taken into account to ensure that the contribution of $C$ is taken exactly once. The procedure is ``one-pass'' because the coefficients of smaller clusters cannot affect the coefficients for the bigger ones.

\begin{itemize}
 \item Define the nested-scheme by picking a set of independent rectangular clusters, defined by the size in each direction $(L_x,L_y)$ (independent meaning no cluster can be fit into another). Note that the placement of these clusters on all possible positions on the lattice, with all possible orientations, constructs the set of clusters $\cal C$.
 \item For each pair of the lattice indices $ij$ (``bond'' if $i\neq j$ or ``site'' if $i=j$), perform the following (if you know there are symmetries, this part can be performed for only the independent bonds/sites) 
 \begin{itemize} 
  \item Determine all possible positions of all the clusters such that they contain the bond/site in question $ij$. These form a set of clusters defined by size and position $(x,y,L_x,L_y)$, and the position is assumed to correspond to the left-bottom site of the cluster.
  \item Determine all the overlaps between the clusters obtained in the previous step. Overlaps themselves form a set of clusters defined by size and position $(x,y,L_x,L_y)$. Note that under present assumptions, any overlap of clusters from $\cal O$ is also an overlap of clusters in $\cal C$ ($C \cap C' \in {\cal O}, \forall C,C'\in {\cal O}$)
  \item Group by shape all the clusters obtained in the previous two steps, independently of position and  rotation, i.e. $(x_1,y_1,L_x,L_y)$ goes together with $(x_2,y_2,L_y,L_x)$ 
  \item Order the groups according to $N_c=L_xL_y$ (or $\mathrm{max}(L_x,L_y)$), from biggest to smallest clusters, and place them ``left to right'', so that no cluster contains a cluster to the left of it, but may or may not contain clusters to the right of it. Because clusters in the same group are of the same size, but different position and/or orientation, no cluster can contain a different cluster in the same group.
  \item Assign a prefactor $p(c)$ to each cluster $c$ in each group. 
  \item For each group $g$, starting from biggest clusters (leftmost)
  \begin{itemize} 
   \item For each cluster $c$ in the group $g$, do a weighted count of how many times it is contained in the clusters in the groups left to it. Weighted means to take into account the prefactor of the cluster in which $c$ is found to be contained. In other words, obtain the number $P =\sum_{c'} p(c')$, where the sum goes over all clusters $c'$ which contain $c$. Then set the prefactor of $c$ to be $p(c) = 1-P$. This assures that the total contribution of the cluster $c$ is $1$. The coefficients of the smaller clusters which are yet to change cannot affect this value. By construction, the clusters in ${\cal C}$ have $p(c)=1$: they are not contained in any other clusters, so $P=0$.    
  \end{itemize}
  \item For the bond $ij$, the expression for self energy is now $$\Sigma_{ij} = \sum_c p(c) \Sigma^c_{I_{c,i},J_{c,j}}$$
  where $c$ runs over all clusters in all groups, and $I_{c,i}$ and $J_{c,j}$ are determined trivially for each cluster $c$.
 \end{itemize}
\end{itemize}

The algorithm is visualized in Fig.~\ref{fig:algorithm}.

\paragraph{Use of symmetries}

When there are symmetries, we want to simplify the expression for self-energy by identifying identical contributions in the sum over clusters. First, if there is translational symmetry, clusters of the same shape but different position will have the same self-energy. If there is rotational symmetry, again, clusters of the same shape but different orientation must have the same $\hat\Sigma^C$. We therefore only solve clusters of different shape/size, and the sum over all $C$ is replaced by the sum over independent clusters and a sum over the bonds $IJ$ (recall Eq.~\ref{eq:sigma_latt_with_symm}).
Second, $\Sigma^{\imp\,C}_{IJ}$ may not be the same for every $IJ$ corresponding to the same real-space vector, but clusters will in general have some symmetries, and one is able use them to simplify the expressions further. It is straightforward to identify groups of identical bonds/sites.  Then, the sum over all $IJ$ is replaced by a sum over only the independent bonds/sites $IJ$ on a given cluster, and the prefactors are adjusted accordingly.

Recall now the self-consistency condition in the nested cluster scheme Eq.~\ref{eq:nested_sc_lattice}. Unlike ${\cal G}^C$ and $\Sigma^{\imp\,C}$, when convergence is reached, $G^{\imp\,C}_{IJ}$ will be the same for any choice of $IJ$ corresponding to the same real-space vector.
We find it beneficial for the stability of the loop and the maximal level of convergence reached if this symmetry of $G^{\imp\,C}$ is imposed in each iteration, and if the cluster symmetries are imposed on ${\cal G}^C$ and $\Sigma^{\imp\,C}$.

\begin{figure}[!ht]
 %\centering{}
 \includegraphics[width=3.2in,trim=0cm 0cm 0cm 0cm]{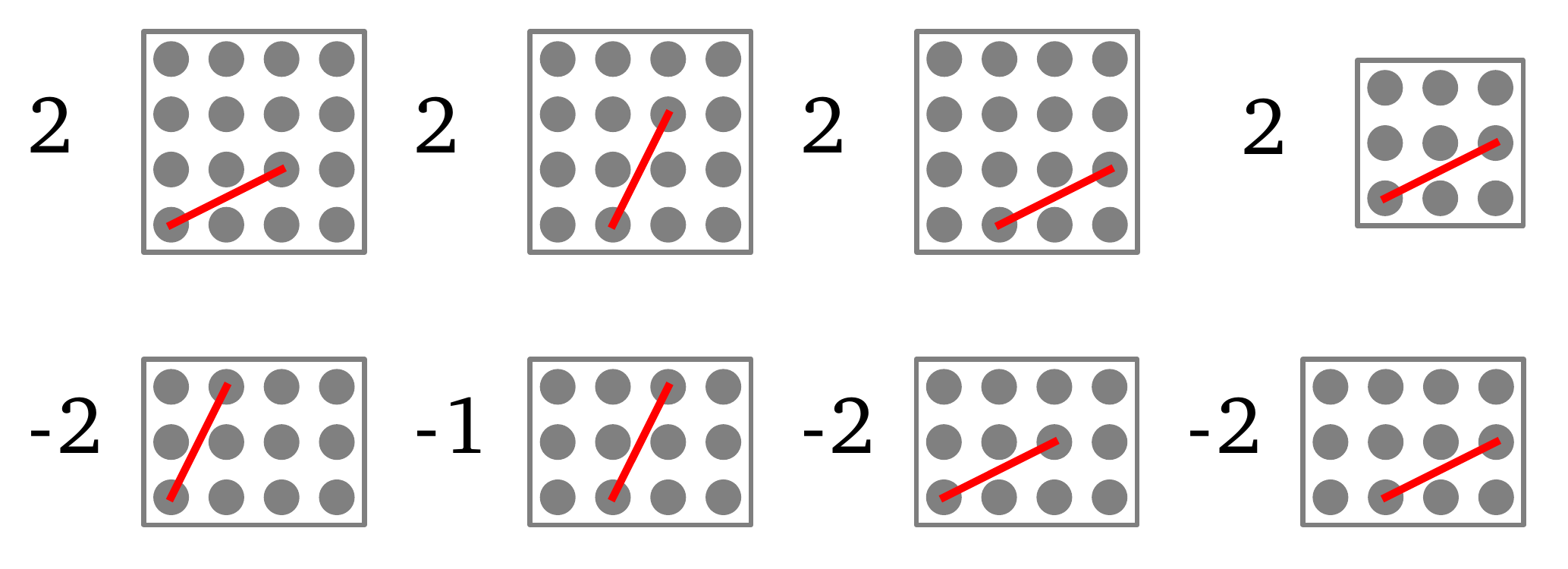} 
 \caption{ After the use of symmetries, the result from the final step in Fig.~\ref{fig:algorithm} simplifies to what is shown in this figure.
    One needs to solve only 3 different clusters, and take into acount a smaller number of bonds on each cluster, for a given real-space vector.
 }
 \label{fig:alg_after_symm} 
\end{figure}

The simplification of the self-energy expression one obtains after using cluster symmetries is visualized in Fig.~\ref{fig:alg_after_symm} (see caption).

\subsection{Nesting the cumulant} \label{sec:cumul_nested}

Here we discuss a different variant of the nested cluster approximation, corresponding to cumulant embedding rather than self-energy embedding theory. The benchmark of this method in comparison with the self-energy nesting variant is presented in Fig.~\ref{fig:cumul_vs_sigma}.

One can in principle define a functional $\Gamma$ of the Green's function such that its derivative yields the cumulant, instead of self energy
\begin{equation}
 \hat{g} = \frac{\partial \Gamma[\hat{G}]}{\partial\hat{G}^\mathsf{T}}
\end{equation}
The cumulant is the full-Green's function stripped of the bare hopping processes, so that
\begin{equation}
 \hat{G}(i\omega_n) = \hat{g}(i\omega_n)+\hat{g}(i\omega_n)\, \hat{t}\, \hat{G}(i\omega_n)
\end{equation}
i.e.
\begin{equation}
 \hat{G}(i\omega_n) = \Big[ \hat{g}^{-1}(i\omega_n) - \hat{t} \Big]^{-1}
\end{equation}
where $\hat{t}$ is the hopping matrix. In $\mathbf{k}$-space
\begin{equation}
 G_\mathbf{k}(i\omega_n) = g_\mathbf{k}(i\omega_n) + g_\mathbf{k}(i\omega_n)\, \varepsilon_\mathbf{k}\, G_\mathbf{k}(i\omega_n)
\end{equation}
\begin{equation}\label{eq:G_from_g}
 G_\mathbf{k}(i\omega_n) = \frac{1}{ g^{-1}_\mathbf{k}(i\omega_n) - \varepsilon_\mathbf{k} }
\end{equation}
which leads to the identity
\begin{equation}
 \hat{g}(i\omega_n)  = \Big[ (i\omega_n+\mu)\hat{I} - \hat{\Sigma}(i\omega_n)\Big]^{-1}
\end{equation}
and the inverse is
\begin{equation} 
 \hat{\Sigma}(i\omega_n)  = (i\omega_n+\mu)\hat{I} - \hat{g}^{-1}(i\omega_n)
\end{equation}

So, we can construct the cumulant on the lattice $g_\mathbf{k}$ from the cumulants on the impurities, the same way we did for the self-energy.
Self-energy on the lattice can be obtained as
\begin{equation}
 \Sigma^\latt_\mathbf{k}(i\omega_n)  = i\omega_n+\mu - \big(g^\latt_\mathbf{k}(i\omega_n)\big)^{-1}
\end{equation}
but this expression is ill-defined at high frequency, so it is important to avoid using it in the DMFT loop.
Therefore, in each iteration, we construct $G^\latt$ directly from the cumulant using Eq.~\ref{eq:G_from_g}, and calculate the self-energy only in the post-processing of the results.

We expect that the cumulant variant works better whenever the cumulant is shorter ranged than the self-energy. In practice we find that the cumulant version does a slightly better job, but the difference is not big (see Section \ref{subsec:cumulant}).

\subsection{Simple examples and summary of equations}

\begin{figure*}[!ht]
 %\centering{}
 \includegraphics[width=6.4in,trim=0cm 0cm 0cm 0cm]{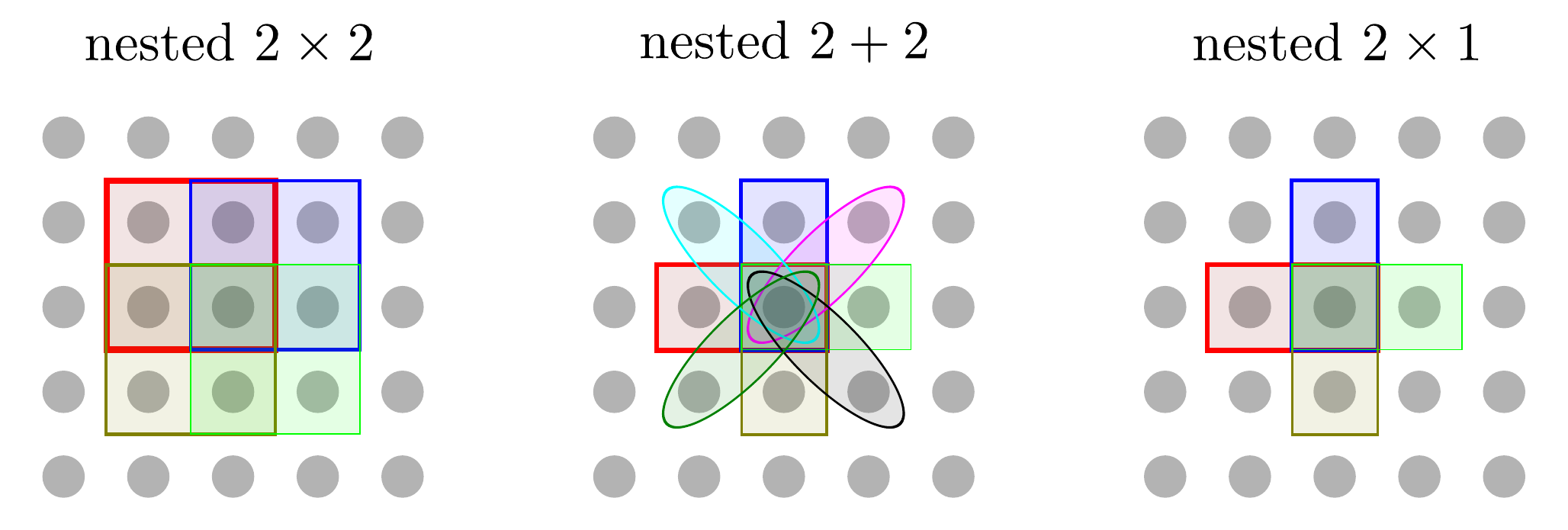} 
 \caption{ Clusters contributing to the local self-energy in different kinds of simple nested schemes.
 }
 \label{fig:plaquette} 
\end{figure*}

\subsubsection{$2\times 1$} \label{sec:2x1}

This example was originally presented in Ref.\onlinecite{Schiller1995}. We rederive it only for pedagogical purposes.

For simplicity, we introduce a shorthand notation $ i+x $ to denote the index of the nearest neighbor of the site $i$ in the $+x$ direction, and similarly $i-x, i+y, ...$.
Recall also the cluster notation $C_i^{2\times 1}\equiv\{i,i+x\}$, $C_i^{1\times 2}\equiv\{i,i+y\}$ and $C_i^{1\times 1}\equiv\{i\}$.

Let's approximate LW functional such that it contains diagrams that involve at most two nearest-neighboring lattice sites
% \begin{equation}
% \Phi = \sum_i\,\sum_{\delta\in\{x,y\}} \Phi[\{ G_{lm} \}_{l,m\in \{i,i+\delta\}}] 
% \end{equation}
%In the present context
\begin{equation}
 %{\cal C} = \{ \{ i,i+\delta\} \}_{\forall i,\;\delta\in\{x,y\}}
 {\cal C} = \{ C_i^{2\times 1} \}_{\forall i} \cup \{ C_i^{1\times 2} \}_{\forall i}
\end{equation}
This means we want to solve at most a 2-site impurity problem.
As for all possible overlaps of the clusters in $\cal C$, one can easily verify
\begin{equation}
% {\cal O} = \{ C \cap C'\}_{\forall C,C'\in{\cal C}} = \{ \{ i \} \}_{\forall i}
 {\cal O} = \{ C \cap C'\}_{\forall C,C'\in{\cal C}} = \{ C_i^{1\times 1} \}_{\forall i}
\end{equation}
which means we will need to take care of double counting. Each overlap cluster is contained in 4 clusters in $\cal C$
% \begin{eqnarray}
%   \{ i \} \subset \{ i, i+x \} \\ \nonumber
%   \{ i \} \subset \{ i-x, i \} \\ \nonumber
%   \{ i \} \subset \{ i, i+y \} \\ \nonumber
%   \{ i \} \subset  \{ i-y, i \}
% \end{eqnarray}
\begin{eqnarray}
  C_i^{1\times 1} \subset C_i^{2\times 1} \\ \nonumber
  C_i^{1\times 1} \subset C_{i-x}^{2\times 1} \\ \nonumber
  C_i^{1\times 1} \subset C_i^{1\times 2} \\ \nonumber
  C_i^{1\times 1} \subset C_{i-y}^{1\times 2}
\end{eqnarray}
which means that we are counting diagrams which involve only the local Green's function 4 times at each site.
To have them taken into account only once, we need to subtract the DMFT functional \eqref{eq:dmft_Phi} 3 times at each site, i.e. $p_{C\in{\cal O}} = -3$
\begin{equation}\label{eq:nested_2x1}
%\Phi = \sum_i\,\sum_{\delta\in\{x,y\}} \Phi[\{ G_{lm} \}_{l,m\in \{i,i+\delta\}}] - 3 \sum_i \Phi[ G_{ii} ]
\Phi \approx \sum_i \Big( \Phi_{2}[ G|_{C_i^{2\times 1}}] + \Phi_{2}[ G|_{C_i^{1\times 2}}] - 3 \Phi_{1}[ G|_{C_i^{1\times 1}} ] \Big)
\end{equation}
Now we write the clusters explicitly to perform the derivatives that yield the self-energy. The local component is given by
\begin{eqnarray} \nonumber
\Sigma_{ii} &=& \frac{\partial}{\partial G_{ii}} \sum_{l}\Bigg(\sum_{\delta\in\{x,y\}} \Phi_2[\{ G_{l'm'} \}_{l'm'\in \{l,l+\delta\}}] - 3 \Phi_1[ G_{ll} ] \Bigg) \\ \label{eq:phi_der_loc}
            &=& \sum_{\delta\in\{x,y,-x,-y\}}\frac{\partial \Phi_2[\{ G_{lm} \}_{lm\in \{i,i+\delta\}}] }{\partial G_{ii}} - 3 \frac{\partial\Phi_1[ G_{ii} ]}{\partial G_{ii}} 
\end{eqnarray}
and the nearest-neighbor components (with $\delta=x,y$)
 \begin{eqnarray} \nonumber
 \Sigma_{i,i+\delta} &=& \frac{\partial}{\partial G_{i+\delta,i}}\Bigg( \sum_{l'}\,\sum_{\delta'\in\{x,y\}} \Phi_2[\{ G_{lm} \}_{lm\in \{l',l'+\delta'\}}]\\ \nonumber
                      && \;\;\;\;- 3 \sum_l \Phi_1[ G_{ll} ] \Bigg) \\ \label{eq:phi_der_nn}
             &=& \frac{\partial \Phi_2[\{ G_{lm} \}_{lm\in \{i,i+\delta\}}] }{\partial G_{i+\delta,i}}
 \end{eqnarray}
When there is translational, mirror and rotational symmetry, the contribution to the local part coming from 4 different nearest-neighbor pairs will be the same, and the self-energy on all n.n. bonds will be the same
\begin{eqnarray} \nonumber
\Sigma^\latt_{\mathbf{r}=(0,0)} &=& 4 \Sigma^{\mathrm{imp}\,2\times 1}_{00} - 3 \Sigma^{\mathrm{imp}\,1\times 1}_{00} \\
\Sigma^\latt_{\mathbf{r}=(0,1)} &=& \Sigma^{\mathrm{imp}\,2\times 1}_{01}
\end{eqnarray}
and the self-consistency is
\begin{eqnarray} \nonumber
 G^{\mathrm{imp}\,2\times 1}_{00/11} = G^{\mathrm{imp}\,1\times 1}_{00} = G^\latt_{\mathbf{r}=(0,0)} \\
 G^{\mathrm{imp}\,2\times 1}_{01/10} = G^\latt_{\mathbf{r}=(1,0)}
\end{eqnarray}

\begin{figure}[!ht]
 %\centering{}
 \includegraphics[width=3.2in,trim=0cm 0cm 0cm 0cm]{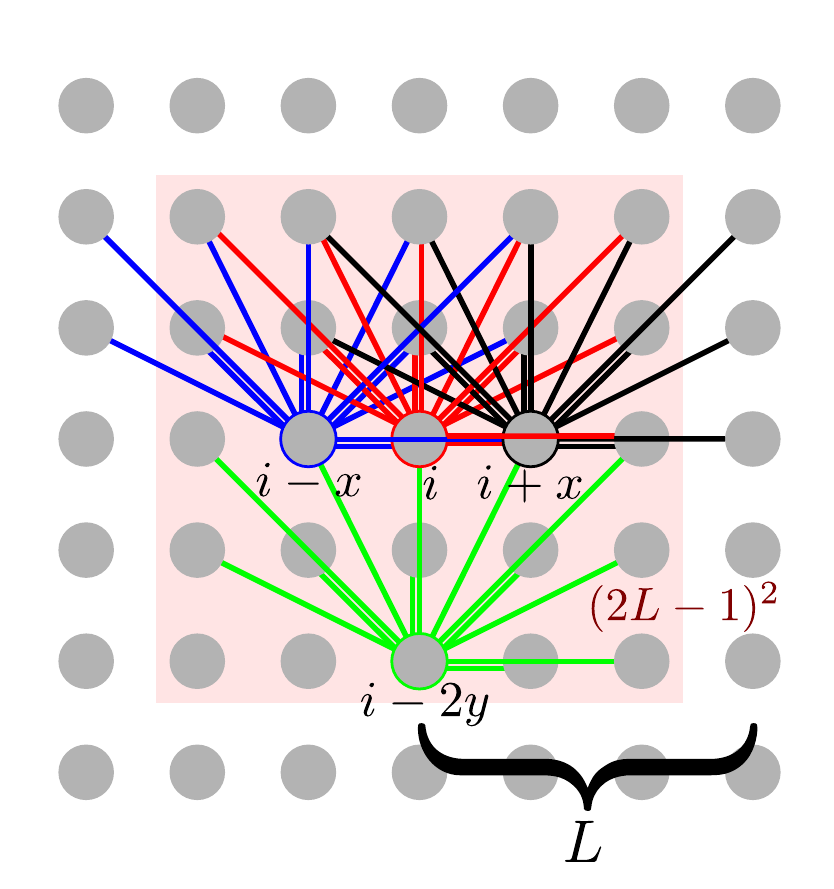} 
 \caption{ Illustration of Eq.~\ref{eq:Ldimers_Phi}. By placing the green set of dimers starting from each site, one covers all the dimers on the lattice, up to Manhattan distance $2(L-1)$. Each site is involved in $(2L-1)^2-1$ different dimers (red site goes with all the sites in the red square).
 }
 \label{fig:L_dimers} 
\end{figure}

\subsubsection{Long distance dimers} 

In this section we present a nested cluster scheme where self-energy at an arbitrary real-space vector is approximated by the self-energy of a corresponding 2-site impurity problem.
The expression for $\Phi$ and $\Sigma^\latt$ can be worked out analytically.

Let's define $i+n_x+n_y$ to be the index of the lattice site at the real-space vector $\mathbf{r}=\mathbf{r}_i + n_x\mathbf{e}_x + n_y\mathbf{e}_y$. We can approximate the LW functional in such a way that it contains diagrams which involve at most 2 sites, but at a distance not bigger than $L-1$ in both directions (maximum Manhattan distance $2L-2$).
We define the cluster notation $C_i^{(n_x,n_y)}\equiv\{i,i+n_x+n_y\}$. Analogously to Eq.~\ref{eq:nested_2x1}, one can prove the following approximation has no double counting
% \begin{eqnarray} \nonumber
% \Phi^{(L)} &=& \sum_i\left(\sum_{\substack{n_x \in (-L,L) \\ n_y \in (0,L)}} + \sum_{\substack{n_x\in(0,L) \\ n_y=0}}\right)\Phi[\{ G_{lm} \}_{lm\in \{i,i+n_x+n_y\}}]\\ \label{eq:Ldimers_Phi}
%       && \;\;\;+ (2 - (2L-1)^2) \sum_i \Phi[G_{ii}]
% \end{eqnarray}
\begin{eqnarray} \nonumber
\Phi &\approx& \sum_i\left(\sum_{\substack{n_x \in (-L,L) \\ n_y \in (0,L)}} + \sum_{\substack{n_x\in(0,L) \\ n_y=0}}\right)\Phi_2[\{ G|_{C_i^{(n_x,n_y)}}]\\ \label{eq:Ldimers_Phi}
      && \;\;\;+ (2 - (2L-1)^2) \sum_i \Phi_{1}[G|_{C_i^{1\times 1}}]
\end{eqnarray}
This is illustrated in Fig.~\ref{fig:L_dimers}. For example, the site with red outline is involved in dimers with all the sites within the red square, of which there are $(2L-1)^2-1$. 

With translational/rotational/mirror symmetry, we get for the self-energy
\begin{eqnarray} \nonumber
\Sigma^\latt_{\mathbf{r}=(0,0)} &=& \sum_{\substack{n_x\in(0,L) \\ n_y \in[0,n_x]}}m_{n_x,n_y}\Sigma^{\mathrm{imp}\;(n_x,n_y)}_{00}\\ \nonumber
  && \;\;\;  + (2 - (2L-1)^2) \Sigma^{\mathrm{imp}\,1\times 1}_{00} \\
\Sigma^\latt_{\mathbf{r}=(n_x,n_y)} &=& \Sigma^{\mathrm{imp}\;(n_x,n_y)}_{01}
\end{eqnarray}
where in the bottom row, $0<n_x<L$, $0 \leq n_y\leq n_x$.
$m_{n_x,n_y}$ is the multiplicity of the (non-zero) real-space vector $\mathbf{r}=n_x\mathbf{e}_x + n_y\mathbf{e}_y$
\begin{equation}
 m_{n_x,n_y} = \left\{ \begin{array}{cc}
                         4, & n_x = 0 \vee n_y = 0 \vee n_x=n_y \\
                         8, & \mathrm{otherwise}
                       \end{array}
               \right.       
\end{equation}
and the self-consistency reads
\begin{eqnarray} \nonumber
 G^{\mathrm{imp}\;(n_x,n_y)}_{00/11} = G^{\mathrm{imp}\,1\times 1}_{00} &=& G^\latt_{\mathbf{r}=(0,0)} \\
 G^{\mathrm{imp}\;(n_x,n_y)}_{01/10} &=& G^\latt_{\mathbf{r}=(n_x,n_y)}
\end{eqnarray}

The simplest example is the $2+2$ scheme corresponding to $L=2$, where we just take the dimer as in the previous example, and add the diagonal one $\{i,i+x+y\}$
\begin{eqnarray} \nonumber  \label{eq:2+2}
\Sigma^\latt_{\mathbf{r}=(0,0)} &=& 4 \Sigma^{\mathrm{imp}\,(0,1)}_{00} + 4 \Sigma^{\mathrm{imp}\,(1,1)}_{00} - 7 \Sigma^{\mathrm{imp}\,1\times 1}_{00} \\ \nonumber
\Sigma^\latt_{\mathbf{r}=(0,1)} &=& \Sigma^{\mathrm{imp}\,(0,1)}_{01} \\
\Sigma^\latt_{\mathbf{r}=(1,1)} &=& \Sigma^{\mathrm{imp}\,(1,1)}_{01}
\end{eqnarray}
Here we are solving 3 impurity problems, 2 of them 2-site, and one single-site.

\subsubsection{$2\times2$} \label{sec:2x2}

Here we discuss the special case of the square clusters scheme presented in subsection \ref{sec:square_clusters}, with $L=2$.
It corresponds to placing a square $2\times 2$ cluster on all possible positions on the lattice. The $\Phi$ approximation is given by Eq.~\ref{eq:def_nested} with $L=2$.

% By the general expression Eq.~\ref{eq:def_nested}, we can write the LW functional approximation more explicitly
% \begin{eqnarray} \nonumber
% \Phi^{(L=2)} &=& \sum_i\Bigg[ \Phi_4[G|_{C_i^{2\times 2}}] + \Phi_1[ G|_{C_i^{1\times 1}}]\\ \nonumber
%      && - \Phi_2[G|_{C_i^{2\times 1}}] - \Phi_2[G|_{C_i^{1\times 2}}] \Bigg]
% \end{eqnarray}
We can write it more explicitly
\begin{eqnarray} \nonumber
\Phi^{(L=2)} &=& \sum_i \Phi_4[\{ G_{lm} \}_{lm\in \{i,i+x,i+y,i+x+y\}}] \\ \nonumber
     && - \sum_{i} \sum_{\delta\in\{x,y\}} \Phi_2[\{ G_{lm} \}_{lm\in \{i,i+\delta\}}] \\
     && + \sum_i \Phi_1[ G_{ii} ]
\end{eqnarray}
Now let's apply the derivative with respect to different components of the Green's function to get the expressions for self-energy
\begin{eqnarray} 
 \Sigma_{ii} &=&   \frac{\partial }{\partial G_{ii}} \Bigg[ \\ \nonumber
             &&  \sum_{l\in\{i,i-x,i-y,i-x-y\}}\Phi_4[\{ G_{l'm'} \}_{l'm'\in \{l,l+x,l+y,l+x+y\}}] \\ \nonumber
             &&  - \sum_{\delta\in\{x,y\}}\sum_{l\in\{i,i-\delta\}} \Phi_2[\{ G_{l'm'} \}_{l'm'\in \{l,l+\delta\}}] \\ \nonumber
             &&  + \Phi_1[\{ G_{ii} \}] \Bigg]
\end{eqnarray}
\begin{eqnarray}
 \Sigma_{i,i+x} &=&   \frac{\partial }{\partial G_{i+x,i}} \Bigg[ \\ \nonumber
                &&  \sum_{l\in\{i,i-y\}}\Phi_4[\{ G_{l'm'} \}_{l'm'\in \{l,l+x,l+y,l+x+y\}}] \\ \nonumber                             
                &&  - \Phi_2[\{ G_{lm} \}_{lm\in \{i,i+x\}}] \Bigg]
\end{eqnarray}
\begin{eqnarray} \nonumber
 \Sigma_{i,i+x+y} &=&   \frac{\partial }{\partial G_{i+x+y,i}} \Phi_4[\{ G_{lm} \}_{lm\in \{i,i+x,i+y,i+x+y\}}]                  
\end{eqnarray}

With full translational/rotational/mirror symmetry, clusters with same size and shape must give identical contributions to the self-energy. Following considerations analogous to Eqs.\ref{eq:phi_der_loc} and \ref{eq:phi_der_nn}, we arrive at the final expression which connects the self energy on the lattice with the one in 3 different impurity problems ($2\times2$, $2\times1$ and $1\times1$)
\begin{eqnarray} \nonumber
 \Sigma^\latt_{\mathbf{r}=(0,0)} &=& 4\Sigma^{\mathrm{imp}\,2\times2}_{00} - 4\Sigma^{\mathrm{imp}\,2\times1}_{00} + \Sigma^{\mathrm{imp}\,1\times1}_{00} \\ \nonumber
 \Sigma^\latt_{\mathbf{r}=(1,0)} &=& 2\Sigma^{\mathrm{imp}\,2\times2}_{01} - \Sigma^{\mathrm{imp}\,2\times1}_{01} \\
 \Sigma^\latt_{\mathbf{r}=(1,1)} &=& \Sigma^{\mathrm{imp}\,2\times2}_{03}
\end{eqnarray}
and the self-consistency condition is given by
\begin{eqnarray} \nonumber
  G^{\mathrm{imp}\,2\times2}_{00/11/22/33} = G^{\mathrm{imp}\,2\times 1}_{00/11} = G^{\mathrm{imp}\,1\times 1}_{00} = G^\latt_{\mathbf{r}=(0,0)} \\ \nonumber
  G^{\mathrm{imp}\,2\times2}_{01/13/32/20/10/...} = G^{\mathrm{imp}\,2\times 1}_{01/10} =  G^\latt_{\mathbf{r}=(0,1)} \\ \nonumber
  G^{\mathrm{imp}\,2\times2}_{03/30/12/21} = G^\latt_{\mathbf{r}=(1,1)} \\
\end{eqnarray}

\subsubsection{$4\times4$}

Using the algorithm \ref{sec:algorithm} and lattice symmetries, we can now automatize the derivation of expressions for the self energy. Here we present as an example the expressions for the $4\times4$ nested-scheme, where $\cal C$ contains all possible positions of a $4\times 4$ cluster, and $\Phi$ approximation is given by Eq.~\ref{eq:def_nested} with $L=4$..
%In the notation below, the first site index is always smaller or equal then the second, so when there are three digits, the last two are the second index.

\begin{eqnarray} \label{eq:4x4_sigma_expr} \\ \nonumber
&&\boxed{
\begin{array}{ccc}
6 & 7 & 8 \\
3 & 4 & 5 \\
0 & 1 & 2 \\
\end{array} 
 } 
\;\;\;\boxed{
\begin{array}{cccc}
12 & 13 & 14 & 15 \\
8 & 9 & 10 & 11 \\
4 & 5 & 6 & 7 \\
0 & 1 & 2 & 3 \\
\end{array} 
 } 
\;\;\;\boxed{
\begin{array}{cccc}
8 & 9 & 10 & 11 \\
4 & 5 & 6 & 7 \\
0 & 1 & 2 & 3 \\
\end{array} 
 } 
\\ \nonumber\Sigma^\mathrm{latt}_{\mathbf{r}=(0,0)} &=& +4\Sigma_{0\,0}^{\mathrm{imp}\,4\times 4}+8\Sigma_{1\,1}^{\mathrm{imp}\,4\times 4}+4\Sigma_{5\,5}^{\mathrm{imp}\,4\times 4}\\ \nonumber &&
-8\Sigma_{0\,0}^{\mathrm{imp}\,4\times 3}-4\Sigma_{4\,4}^{\mathrm{imp}\,4\times 3}+4\Sigma_{0\,0}^{\mathrm{imp}\,3\times 3}\\ \nonumber &&
-8\Sigma_{1\,1}^{\mathrm{imp}\,4\times 3}+4\Sigma_{1\,1}^{\mathrm{imp}\,3\times 3}-4\Sigma_{5\,5}^{\mathrm{imp}\,4\times 3}\\ \nonumber &&
+\Sigma_{4\,4}^{\mathrm{imp}\,3\times 3}\\ \nonumber 
\Sigma^\mathrm{latt}_{\mathbf{r}=(1,0)} &=& +4\Sigma_{0\,1}^{\mathrm{imp}\,4\times 4}+4\Sigma_{1\,5}^{\mathrm{imp}\,4\times 4}+2\Sigma_{1\,2}^{\mathrm{imp}\,4\times 4}\\ \nonumber &&
+2\Sigma_{5\,6}^{\mathrm{imp}\,4\times 4}-4\Sigma_{0\,1}^{\mathrm{imp}\,4\times 3}-2\Sigma_{4\,5}^{\mathrm{imp}\,4\times 3}\\ \nonumber &&
-4\Sigma_{0\,4}^{\mathrm{imp}\,4\times 3}+4\Sigma_{0\,1}^{\mathrm{imp}\,3\times 3}-4\Sigma_{1\,5}^{\mathrm{imp}\,4\times 3}\\ \nonumber &&
+2\Sigma_{1\,4}^{\mathrm{imp}\,3\times 3}-2\Sigma_{1\,2}^{\mathrm{imp}\,4\times 3}-\Sigma_{5\,6}^{\mathrm{imp}\,4\times 3}\\ \nonumber 
\Sigma^\mathrm{latt}_{\mathbf{r}=(1,1)} &=& +2\Sigma_{0\,5}^{\mathrm{imp}\,4\times 4}+4\Sigma_{1\,6}^{\mathrm{imp}\,4\times 4}+2\Sigma_{1\,4}^{\mathrm{imp}\,4\times 4}\\ \nonumber &&
+\Sigma_{5\,10}^{\mathrm{imp}\,4\times 4}-4\Sigma_{0\,5}^{\mathrm{imp}\,4\times 3}-4\Sigma_{1\,4}^{\mathrm{imp}\,4\times 3}\\ \nonumber &&
+2\Sigma_{0\,4}^{\mathrm{imp}\,3\times 3}-4\Sigma_{1\,6}^{\mathrm{imp}\,4\times 3}+2\Sigma_{1\,3}^{\mathrm{imp}\,3\times 3}\\ \nonumber 
\Sigma^\mathrm{latt}_{\mathbf{r}=(2,0)} &=& +4\Sigma_{0\,2}^{\mathrm{imp}\,4\times 4}+4\Sigma_{1\,9}^{\mathrm{imp}\,4\times 4}-4\Sigma_{0\,2}^{\mathrm{imp}\,4\times 3}\\ \nonumber &&
-2\Sigma_{4\,6}^{\mathrm{imp}\,4\times 3}-2\Sigma_{0\,8}^{\mathrm{imp}\,4\times 3}+2\Sigma_{0\,2}^{\mathrm{imp}\,3\times 3}\\ \nonumber &&
-2\Sigma_{1\,9}^{\mathrm{imp}\,4\times 3}+\Sigma_{1\,7}^{\mathrm{imp}\,3\times 3}\\ \nonumber 
\Sigma^\mathrm{latt}_{\mathbf{r}=(2,1)} &=& +2\Sigma_{0\,6}^{\mathrm{imp}\,4\times 4}+2\Sigma_{1\,10}^{\mathrm{imp}\,4\times 4}+2\Sigma_{1\,7}^{\mathrm{imp}\,4\times 4}\\ \nonumber &&
-2\Sigma_{0\,6}^{\mathrm{imp}\,4\times 3}-2\Sigma_{1\,7}^{\mathrm{imp}\,4\times 3}-2\Sigma_{0\,9}^{\mathrm{imp}\,4\times 3}\\ \nonumber &&
+2\Sigma_{0\,5}^{\mathrm{imp}\,3\times 3}-\Sigma_{1\,10}^{\mathrm{imp}\,4\times 3}\\ \nonumber 
\Sigma^\mathrm{latt}_{\mathbf{r}=(2,2)} &=& +2\Sigma_{0\,10}^{\mathrm{imp}\,4\times 4}+2\Sigma_{1\,11}^{\mathrm{imp}\,4\times 4}-4\Sigma_{0\,10}^{\mathrm{imp}\,4\times 3}\\ \nonumber &&
+\Sigma_{0\,8}^{\mathrm{imp}\,3\times 3}\\ \nonumber 
\Sigma^\mathrm{latt}_{\mathbf{r}=(3,0)} &=& +2\Sigma_{0\,3}^{\mathrm{imp}\,4\times 4}+2\Sigma_{1\,13}^{\mathrm{imp}\,4\times 4}-2\Sigma_{0\,3}^{\mathrm{imp}\,4\times 3}\\ \nonumber &&
-\Sigma_{4\,7}^{\mathrm{imp}\,4\times 3}\\ \nonumber 
\Sigma^\mathrm{latt}_{\mathbf{r}=(3,1)} &=& +2\Sigma_{0\,7}^{\mathrm{imp}\,4\times 4}+\Sigma_{1\,14}^{\mathrm{imp}\,4\times 4}-2\Sigma_{0\,7}^{\mathrm{imp}\,4\times 3}\\ \nonumber 
\Sigma^\mathrm{latt}_{\mathbf{r}=(3,2)} &=& +2\Sigma_{0\,11}^{\mathrm{imp}\,4\times 4}-\Sigma_{0\,11}^{\mathrm{imp}\,4\times 3}\\ \nonumber 
\Sigma^\mathrm{latt}_{\mathbf{r}=(3,3)} &=& +\Sigma_{0\,15}^{\mathrm{imp}\,4\times 4}\\ \nonumber 
\end{eqnarray}

% We see here that there are 3 clusters one needs to solve: $4\times4$, $4\times3$ and $3\times3$. It turns out this is a general result, and that if we start by taking all possible sqare clusters of a given size $L$, i.e.  $L \times L$, then we will need to subtract also clusters of the shape $(L-1) \times L$ and add the ones of the shape $(L-1) \times (L-1)$, but none smaller than that. This means that we can do very large clusters while having to solve only 3 cluster-impurity problems.

In practice, we calculate self-energy for all vectors $\mathbf{r}=(x,y)$ such that $y\leq x$, up to $x=\mathrm{max}_C L_x(C)-1$, and the rest is filled by lattice symmetry (eqref).
In Eq.~\ref{eq:4x4_sigma_expr} we have also used the symmetries of the clusters.
The groups of equivalent bonds on all three clusters are given below in curly brackets (inversion symmetry $ij=ji$ is implicit)
\begin{eqnarray}
3 \times 3: \\ \nonumber
&& \boxed{
\begin{array}{ccc}
6 & 7 & 8 \\
3 & 4 & 5 \\
0 & 1 & 2 \\
\end{array} 
} \\ \nonumber 
&\{\;\;& (0,0),\;(2,2),\;(6,6),\;(8,8)\;\;\} \\ \nonumber 
&\{\;\;& (0,1),\;(0,3),\;(2,1),\;(2,5),\;(6,3),\;(6,7),\; \\ \nonumber && 
(8,5),\;(8,7)\;\;\} \\ \nonumber 
&\{\;\;& (0,2),\;(0,6),\;(2,8),\;(6,8)\;\;\} \\ \nonumber 
&\{\;\;& (0,4),\;(2,4),\;(6,4),\;(8,4)\;\;\} \\ \nonumber 
&\{\;\;& (0,5),\;(0,7),\;(2,3),\;(2,7),\;(6,1),\;(6,5),\; \\ \nonumber && 
(8,1),\;(8,3)\;\;\} \\ \nonumber 
&\{\;\;& (0,8),\;(2,6)\;\;\} \\ \nonumber 
&\{\;\;& (1,1),\;(3,3),\;(5,5),\;(7,7)\;\;\} \\ \nonumber 
&\{\;\;& (1,3),\;(1,5),\;(3,7),\;(5,7)\;\;\} \\ \nonumber 
&\{\;\;& (1,4),\;(3,4),\;(5,4),\;(7,4)\;\;\} \\ \nonumber 
&\{\;\;& (1,7),\;(3,5)\;\;\} \\ \nonumber 
&\{\;\;& (4,4)\;\;\} \\ \nonumber 
\end{eqnarray}
\begin{eqnarray}
4 \times 3: \\ \nonumber
&& \boxed{
\begin{array}{cccc}
8 & 9 & 10 & 11 \\
4 & 5 & 6 & 7 \\
0 & 1 & 2 & 3 \\
\end{array} 
} \\ \nonumber 
&\{\;\;& (0,0),\;(3,3),\;(8,8),\;(11,11)\;\;\} \\ \nonumber 
&\{\;\;& (0,1),\;(3,2),\;(8,9),\;(11,10)\;\;\} \\ \nonumber 
&\{\;\;& (0,2),\;(3,1),\;(8,10),\;(11,9)\;\;\} \\ \nonumber 
&\{\;\;& (0,4),\;(3,7),\;(8,4),\;(11,7)\;\;\} \\ \nonumber 
&\{\;\;& (0,5),\;(3,6),\;(8,5),\;(11,6)\;\;\} \\ \nonumber 
&\{\;\;& (0,6),\;(3,5),\;(8,6),\;(11,5)\;\;\} \\ \nonumber 
&\{\;\;& (0,7),\;(3,4),\;(8,7),\;(11,4)\;\;\} \\ \nonumber 
&\{\;\;& (0,8),\;(3,11)\;\;\} \;\;
\{\;\;  (0,3),\;(8,11)\;\;\} \\ \nonumber 
&\{\;\;& (0,9),\;(3,10),\;(8,1),\;(11,2)\;\;\} \\ \nonumber 
&\{\;\;& (0,10),\;(3,9),\;(8,2),\;(11,1)\;\;\} \\ \nonumber 
&\{\;\;& (0,11),\;(3,8)\;\;\} \;\;
 \{\;\;  (1,2),\;(9,10)\;\;\} \\ \nonumber 
&\{\;\;& (1,1),\;(2,2),\;(9,9),\;(10,10)\;\;\} \\ \nonumber 
&\{\;\;& (1,4),\;(2,7),\;(9,4),\;(10,7)\;\;\} \\ \nonumber 
&\{\;\;& (1,5),\;(2,6),\;(9,5),\;(10,6)\;\;\} \\ \nonumber 
&\{\;\;& (1,6),\;(2,5),\;(9,6),\;(10,5)\;\;\} \\ \nonumber 
&\{\;\;& (1,7),\;(2,4),\;(9,7),\;(10,4)\;\;\} \\ \nonumber 
&\{\;\;& (1,9),\;(2,10)\;\;\} \;\;
 \{\;\; (1,10),\;(2,9)\;\;\} \\ \nonumber 
&\{\;\;& (4,4),\;(7,7)\;\;\} \;\;
 \{\;\; (4,5),\;(7,6)\;\;\} \\ \nonumber 
&\{\;\;& (4,6),\;(7,5)\;\;\} \;\;
 \{\;\;  (4,7)\;\;\} \\ \nonumber 
&\{\;\;& (5,5),\;(6,6)\;\;\} \;\;
 \{\;\;  (5,6)\;\;\} \\ \nonumber 
\end{eqnarray}
\vfill
\begin{eqnarray}
4 \times 4: \\ \nonumber
&& \boxed{
\begin{array}{cccc}
12 & 13 & 14 & 15 \\
8 & 9 & 10 & 11 \\
4 & 5 & 6 & 7 \\
0 & 1 & 2 & 3 \\
\end{array} 
} \\ \nonumber 
&\{\;\;& (0,0),\;(3,3),\;(12,12),\;(15,15)\;\;\} \\ \nonumber 
&\{\;\;& (0,1),\;(0,4),\;(3,2),\;(3,7),\;(12,8),\;(12,13),\; \\ \nonumber && 
(15,11),\;(15,14)\;\;\} \\ \nonumber 
&\{\;\;& (0,2),\;(0,8),\;(3,1),\;(3,11),\;(12,4),\;(12,14),\; \\ \nonumber && 
(15,7),\;(15,13)\;\;\} \\ \nonumber 
&\{\;\;& (0,3),\;(0,12),\;(3,15),\;(12,15)\;\;\} \\ \nonumber 
&\{\;\;& (0,5),\;(3,6),\;(12,9),\;(15,10)\;\;\} \\ \nonumber 
&\{\;\;& (0,6),\;(0,9),\;(3,5),\;(3,10),\;(12,5),\;(12,10),\; \\ \nonumber && 
(15,6),\;(15,9)\;\;\} \\ \nonumber 
&\{\;\;& (0,7),\;(0,13),\;(3,4),\;(3,14),\;(12,1),\;(12,11),\; \\ \nonumber && 
(15,2),\;(15,8)\;\;\} \\ \nonumber 
&\{\;\;& (0,10),\;(3,9),\;(12,6),\;(15,5)\;\;\} \\ \nonumber 
&\{\;\;& (0,11),\;(0,14),\;(3,8),\;(3,13),\;(12,2),\;(12,7),\; \\ \nonumber && 
(15,1),\;(15,4)\;\;\} \\ \nonumber 
&\{\;\;& (0,15),\;(3,12)\;\;\} \\ \nonumber 
&\{\;\;& (1,1),\;(2,2),\;(4,4),\;(7,7),\;(8,8),\;(11,11),\; \\ \nonumber && 
(13,13),\;(14,14)\;\;\} \\ \nonumber 
&\{\;\;& (1,2),\;(4,8),\;(7,11),\;(13,14)\;\;\} \\ \nonumber 
&\{\;\;& (1,4),\;(2,7),\;(8,13),\;(11,14)\;\;\} \\ \nonumber 
&\{\;\;& (1,5),\;(2,6),\;(4,5),\;(7,6),\;(8,9),\;(11,10),\; \\ \nonumber && 
(13,9),\;(14,10)\;\;\} \\ \nonumber 
&\{\;\;& (1,6),\;(2,5),\;(4,9),\;(7,10),\;(8,5),\;(11,6),\; \\ \nonumber && 
(13,10),\;(14,9)\;\;\} \\ \nonumber 
&\{\;\;& (1,7),\;(2,4),\;(4,13),\;(7,14),\;(8,1),\;(11,2),\; \\ \nonumber && 
(13,11),\;(14,8)\;\;\} \\ \nonumber 
&\{\;\;& (1,9),\;(2,10),\;(4,6),\;(7,5),\;(8,10),\;(11,9),\; \\ \nonumber && 
(13,5),\;(14,6)\;\;\} \\ \nonumber 
&\{\;\;& (1,10),\;(2,9),\;(4,10),\;(7,9),\;(8,6),\;(11,5),\; \\ \nonumber && 
(13,6),\;(14,5)\;\;\} \\ \nonumber 
&\{\;\;& (1,11),\;(2,8),\;(4,14),\;(7,13)\;\;\} \\ \nonumber 
&\{\;\;& (1,13),\;(2,14),\;(4,7),\;(8,11)\;\;\} \\ \nonumber 
&\{\;\;& (1,14),\;(2,13),\;(4,11),\;(7,8)\;\;\} \\ \nonumber 
&\{\;\;& (5,5),\;(6,6),\;(9,9),\;(10,10)\;\;\} \\ \nonumber 
&\{\;\;& (5,6),\;(5,9),\;(6,10),\;(9,10)\;\;\} \\ \nonumber 
&\{\;\;& (5,10),\;(6,9)\;\;\} \\ \nonumber 
\end{eqnarray}
\vfill\eject
% Then we copy the result by symmetry $\Sigma_{\mathbf{r}=(x,y)} = \Sigma_{\mathbf{r}=(y,x)}$. Then to perform the Fourier tansform to get $\Sigma_\mathbf{k}$, in practice we need to limit the size of the lattice, say $L=64$, i.e. $64\times64$. There we take into account cyclicity to fill in remaining components
% $$\Sigma_{\mathbf{r}=(x,y)} = \Sigma_{\mathbf{r}=(L-x,y)} = \Sigma_{\mathbf{r}=(L-x,L-y)} =  \Sigma_{\mathbf{r}=(x,L-y)}$$

%\newpage
%\appendix

\section{Cluster DMFT methods} \label{app:cluster_dmft}

Here we summarize the (cluster) DMFT methods used in this paper. 

\begin{figure*}[!ht]
 %\centering{}
 \includegraphics[width=7.0in,trim=0cm 1cm 0cm 1cm]{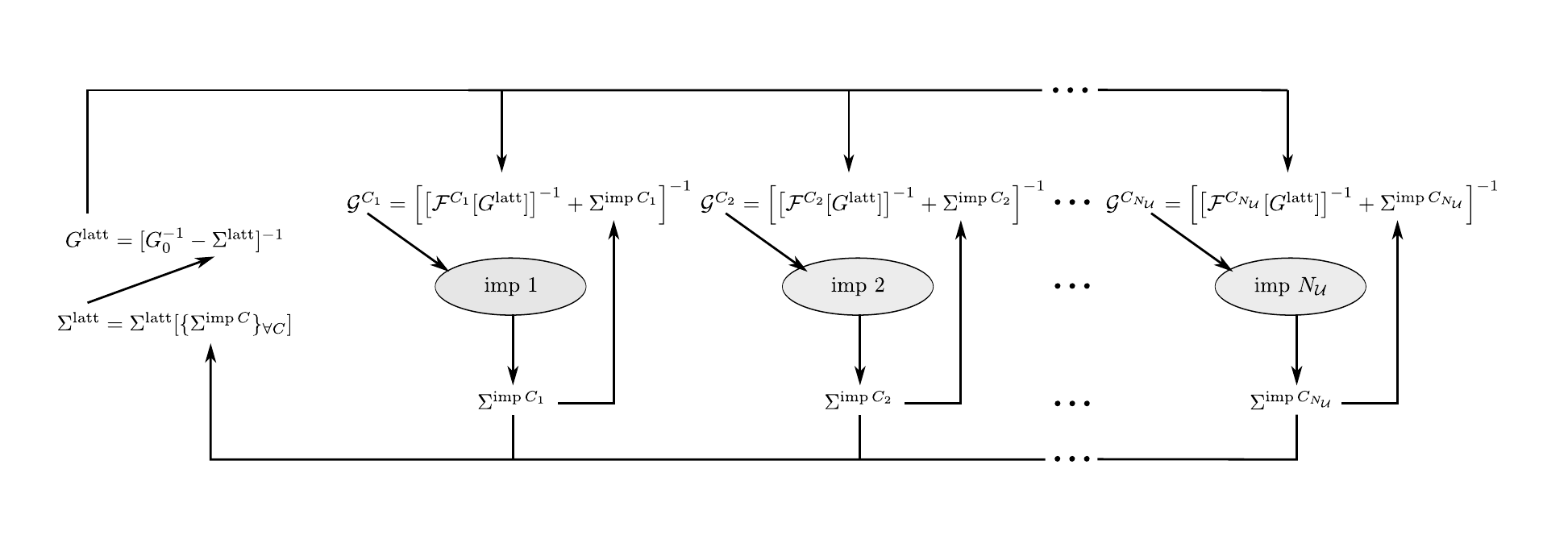} 
 \caption{ Forward substitution solution of a generic cluster DMFT method. $N_{\cal U}$ is the number of independent impurity problems one needs to solve. In all methods except nested cluster, $N_{\cal U}=1$. In nested cluster it can be any number, but in the simplest scheme $N_{\cal U}=3$ independently from the cluster size. The loop starts with an initial guess for $\Sigma^\imp$. ${\cal F}^C[G^\latt]$ project $G^\latt$ onto degrees of freedom of impurity $C$. Convergence is reached when $G^{\imp\,C}={\cal F}^C[G^\latt]$ for each $C$.
 }
 \label{fig:fwd_sub_loop} 
\end{figure*}

The forward-substitution algorithm for the generic cluster DMFT scheme is presented in Fig.~\ref{fig:fwd_sub_loop}. 
Cluster DMFT methods differ in the cluster-impurity action, self-consistency condition, and the self-energy mapping $\Sigma^\latt[\Sigma^\imp]$ - these properties we state for each method in the following sections.
Where possible, we also state the LW functional approximation which leads to the given method.

\subsection{Single-site DMFT}

Single-site DMFT\cite{Georges1996} is the limiting case of all cluster DMFT methods, corresponding to cluster size $N_c=1$.
It can be derived as the local approximation of the LW functional. While the exact LW functional depends on all components of the Green's function, in DMFT it depends only on the local components $G_{ii}$.
\begin{equation} \label{eq:dmft_Phi}
 \Phi[\{G_{ij}\}_{\forall i,j}] \approx \Phi[\{G_{ii}\}_{\forall i}] = \sum_{i}\Phi[G_{ii}]
\end{equation}
The second-step is specific to local interactions, and is crucial to obtain a self-consistent scheme involving a single-site impurity problem.

The impurity action involves degrees of freedom of a single lattice site
\begin{eqnarray}
 S &=& \sum_{\sigma} \iint d\tau d\tau' c^+_{\sigma}(\tau) [-{\cal G}^{-1}_{\sigma}](\tau-\tau') c_{\sigma} (\tau') \\ \nonumber
   &&  + U \int d\tau c^+_{\up}(\tau) c^+_{\dn}(\tau) c_{\dn}(\tau) c_{\up}(\tau)
\end{eqnarray}
The self-consistency condition requires that the local Green's function on the lattice is the same as the one on the impurity
\begin{equation}
 G^\mathrm{imp}(i\omega_n) = G_{ii}^\mathrm{latt}(i\omega_n) \equiv \sum_\mathbf{k \in \mathrm{BZ}} G_\mathbf{k}^\mathrm{latt}(i\omega_n)
\end{equation}
where
\begin{equation}
 G_\mathbf{k}^\mathrm{latt}(i\omega_n) = \Big( G_{0,\mathbf{k}}^{-1}(i\omega_n) - \Sigma^\latt_\mathbf{k}(i\omega_n)\Big)^{-1}
\end{equation}
The self-energy approximation reads
\begin{equation}
 \Sigma^\latt_\mathbf{k}(i\omega_n) \approx \Sigma^\imp(i\omega_n)
\end{equation}
as $\partial \sum_l \Phi [G_{ll}] /\partial G_{ij} \sim \delta_{ij}\partial \Phi [G_{ii}] /\partial G_{ii}$.
The bare-propagator on the lattice
\begin{equation}
 G_{0,\mathbf{k}}(i\omega_n) = \frac{1}{i\omega_n + \mu - \varepsilon_\mathbf{k}}
\end{equation}
is determined by the chemical potential $\mu$ and the bare dispersion $\varepsilon_\mathbf{k}$. On the square-lattice with only nearest-neighbor hopping, it is given by
\begin{equation}
 \varepsilon_\mathbf{k} = -2t(\cos k_x + \cos k_y)
\end{equation}

\begin{figure*}[!ht]
 %\centering{}
 \includegraphics[width=5.2in,trim=0cm 1cm 0cm 0cm]{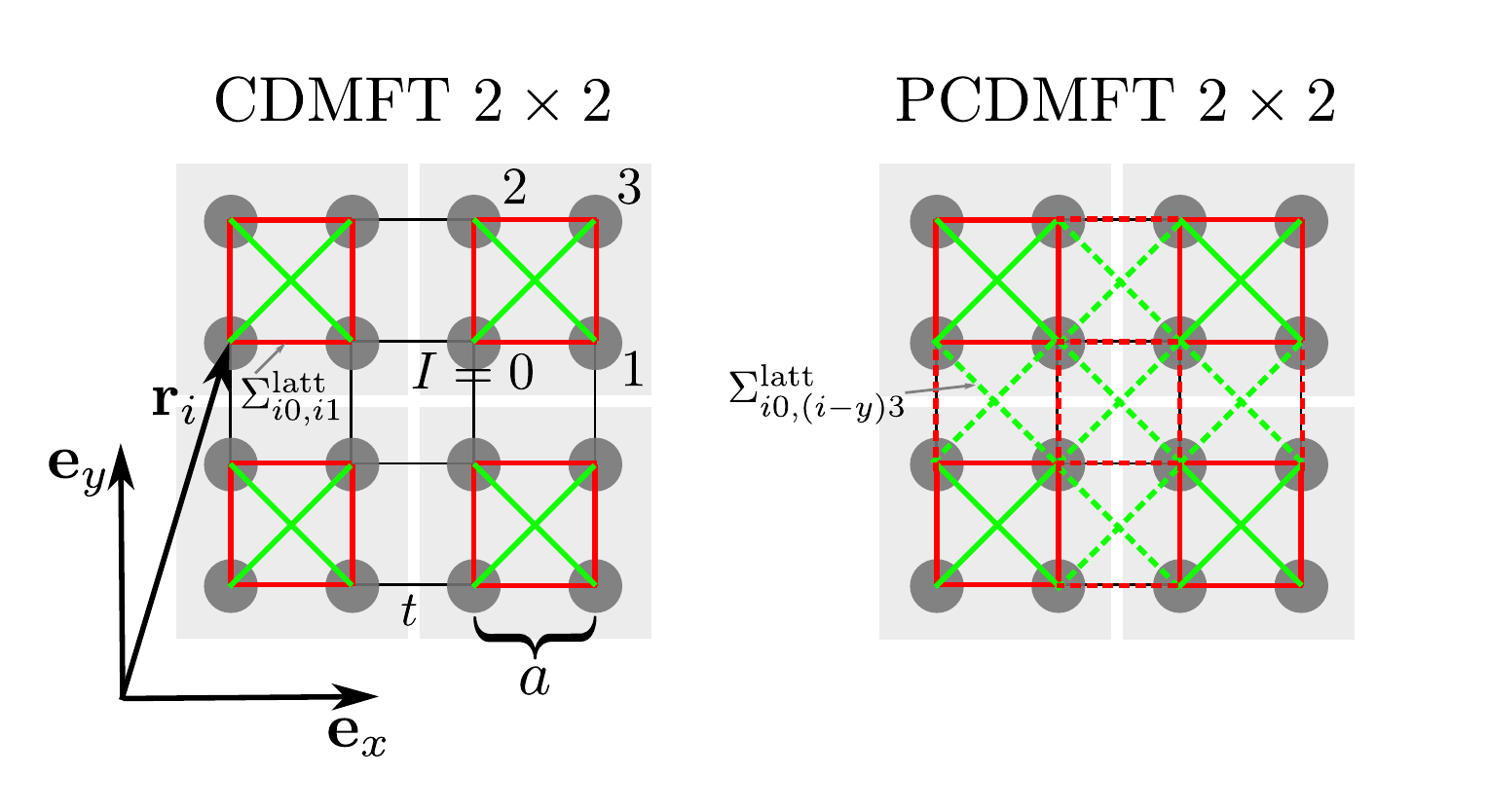} 
 \caption{ Gray circles are lattice sites. Light gray squares are supercells. A site is denoted by the supercell index $i$ and its index within the supercell $I$. In CDMFT the self-energy is non-zero only between sites within a single supercell (green and red lines). In PCDMFT, it is copied by hand onto inter-cell bonds (dashed green and red lines). $t$ is the hopping amplitude, $a$ the lattice spacing, $\mathbf{e}_{x,y}$ are the super-lattice vectors.
 }
 \label{fig:cdmft_and_pcdmft} 
\end{figure*}

\subsection{Cellular DMFT (CDMFT)} \label{sec:cdmft}

Cellular DMFT rewrites the lattice problem in terms of supercells\cite{Civelli2009,Civelli2008,Civelli2005,deLeo2008,TremblayPRB2008,Kotliar2001,Kyung2006a,Kyung2006,Okamoto2010,Osolin2017,Parcollet2004,Park2008,CivelliImadaPRB2016,Sordi2011,Sordi2010,Sordi2013,Tanaskovic2011,Zhang2007}. The lattice-site index is replaced by a double index - index of the supercell and the index of the site within the supercell.
$$i\rightarrow (i,I),\;\;\;\; G_{ij} \rightarrow G_{iI,jJ}$$
We denote with $i,j,...$ the index of the supercell, and with $I,J,...$ the index of the site within the supercell. From here, the derivation proceeds just as in single-site DMFT - one may view CDMFT as ``single-super-cell DMFT''. The approximated LW functional then depends only on Green's function components within a single super cell
\begin{equation}
 \Phi[\{\hat{G}_{ij}\}_{\forall i,j}] \approx \Phi[\{\hat{G}_{ii}\}_{\forall i}] = \sum_{i}\Phi[\hat{G}_{ii}]
\end{equation}
where with ``hat'' we denote matrix objects - $\hat{G}_{ij}$ is a matrix in the space of the $I,J$ indices.
Impurity action is given by
\begin{eqnarray} \nonumber
 S &=& \sum_{IJ,\sigma} \iint d\tau d\tau' c^+_{\sigma,I}(\tau) [-\hat{\cal G}^{-1}_\sigma]_{IJ}(\tau-\tau') c_{\sigma,J} (\tau') \\ \label{eq:cdmft_action}
   &&  + U \sum_I \int d\tau c^+_{\up,I}(\tau) c^+_{\dn,I}(\tau) c_{\dn,I}(\tau) c_{\up,I}(\tau)
\end{eqnarray}
and the self-consistency condition reads
\begin{equation}
 \hat{G}^\mathrm{imp}(i\omega_n) = \hat{G}^\mathrm{latt}_{ii}(i\omega_n) \equiv \sum_{\mathbf{k} \in \mathrm{RBZ}} \hat{G}_\mathbf{k}^\mathrm{latt}(i\omega_n)
\end{equation}
where RBZ stands for ``reduced Brilloun zone''. Note however, that in the derivation below we also rescale the lattice constant $a\rightarrow a/2$ so that no extra prefactors appear in the expressions, and the RBZ extends from 0 to $2\pi$ along both axes. The lattice Dyson equation now involves a matrix inversion
\begin{equation}
 \hat{G}_\mathbf{k}^\mathrm{latt}(i\omega_n) = \Big( \hat{G}_{0,\mathbf{k}}^{-1}(i\omega_n) - \hat{\Sigma}^{\mathrm{latt}}_{\mathbf{k}}(i\omega_n)\Big)^{-1}
\end{equation}
The self-energy approximation is simply
\begin{equation}
 \hat{\Sigma}^{\mathrm{latt}}_{\mathbf{k}}(i\omega_n) \approx \hat{\Sigma}^{\mathrm{imp}}(i\omega_n)
\end{equation}
Note that, physically, the self-energy is put exclusively on bonds \emph{within} a super cell, and \emph{not} on bonds between supercells. This artificially breaks the translational symmetry of the lattice.

The bare propagator and the dispersion need to be rewritten in the supercell language. Here we present the expressions in the simple $2\times 2$ tiling ($\mathbf{e}_{x,y}/a \rightarrow 2\mathbf{e}_{x,y}/a$)
\begin{equation}
 \hat{G}_{0,\mathbf{k}}(i\omega_n) = [(i\omega_n+\mu)\hat{I} - \hat{\varepsilon}_\mathbf{k}]^{-1}
\end{equation}
\begin{equation}
 \hat{\varepsilon}_\mathbf{k} = t\cdot\hat{u}_\mathbf{k}
\end{equation}
\begin{equation}
 \hat{u}_\mathbf{k} =\left[\begin{array}{cccc}
                       &u(k_x)&u(k_y)& \\
                       u^*(k_x)&&&u(k_y) \\
                       u^*(k_y)&&&u(k_x) \\
                       &u^*(k_y)&u^*(k_x)& \\
                      \end{array}\right]
\end{equation}
\begin{equation}
 u(k) = 1+e^{-ik}
\end{equation}

The drawback of this approach is that no simple interpretation of the result in terms of the original, translationally invariant lattice is possible.
To obtain a translationally invariant self-energy which can be plotted in the original BZ requires a post-processing step, or ``periodization``. In the present case
\begin{subequations} \label{eq:cdmft_periodization}
\begin{eqnarray}
 \Sigma^\per_{\mathbf{r}=(0,0)} &=& \Sigma^\imp_{00} \\
 \Sigma^\per_{\mathbf{r}=(0,1)} &=& \Sigma^\imp_{01} \\
 \Sigma^\per_{\mathbf{r}=(1,1)} &=& \Sigma^\imp_{03} 
\end{eqnarray}
The real-space vectors $\mathbf{r}$ are given in the basis of the original lattice-vectors.
The rest of the real-space vectors can be filled in by symmetry
\begin{equation}
 \Sigma^\per_{\mathbf{r}=(x,y)} = \Sigma^\per_{\mathbf{r}=(\pm x,\pm y)} = \Sigma^\per_{\mathbf{r}=(\pm y,\pm x)}
\end{equation}
and then we can Fourier transform to $\mathbf{k}$-space
\begin{equation}
\Sigma^\per_\mathbf{k} = \sum_\mathbf{r} e^{i\mathbf{k}\cdot\mathbf{r}}\Sigma^\per_\mathbf{r}
\end{equation}
\end{subequations}
Note that periodization is an ad-hoc procedure that does not have a clear physical interpretation in terms of the LW approximation. Also, the physical quantity that is being periodized can be chosen arbitrarily, and different choices will in general lead to different results.

\subsection{Periodized CDMFT (PCDMFT)}

The idea of PCDMFT\cite{Biroli2004,Kotliar2006,Li2015,Lichtenstein2000,Stanescu2006} is that the periodization should be performed in each DMFT iteration, and that the self-consistency should be closed using the translationally-invariant Green's function, rather than the super-lattice one. This scheme cannot be simply derived from an approximation of the LW functional.
The impurity action remains the same as in CDMFT, Eq.~\ref{eq:cdmft_action}.

The idea of PCDMFT can be achieved either by placing the missing self-energies on the super-lattice
\begin{eqnarray} \nonumber
 \hat{\Sigma}^{\mathrm{latt}}_{\mathbf{k}}(i\omega_n) &\approx& \Sigma^{\mathrm{imp}}_{00}(i\omega_n)\hat{I} 
                                                               + \Sigma^{\mathrm{imp}}_{01}(i\omega_n)\hat{u}_\mathbf{k}
                                                               + \Sigma^{\mathrm{imp}}_{03}(i\omega_n)\hat{w}_\mathbf{k} \\ 
                                                            &=& \hat{\Sigma}^{\mathrm{imp}}(i\omega_n)\circ\Big( \hat{I} 
                                                               + \hat{u}_\mathbf{k}
                                                               + \hat{w}_\mathbf{k} \Big)
\end{eqnarray}
where $\circ$ denotes element-wise product, and
\begin{equation}
 \hat{w}_\mathbf{k} =\left[\begin{array}{cccc}
                       &&& w_1(\mathbf{k}) \\
                       && w_2(\mathbf{k})& \\
                       &w^*_2(\mathbf{k})&& \\
                       w^*_1(\mathbf{k})&&& \\
                      \end{array}\right]
\end{equation}
\begin{eqnarray} 
 w_1(\mathbf{k}) &=& 1+e^{-ik_x}+e^{-ik_y}+e^{-i(k_x+k_y)} \\ 
 w_2(\mathbf{k}) &=& 1+e^{-ik_x}+e^{-ik_y}+e^{-i(k_x-k_y)}
\end{eqnarray}
or, equivalently, by periodizing the self-energy with Eq.~\ref{eq:cdmft_periodization} and rewriting the self-consistency condition with
\begin{equation} \label{eq:pcdmft_sc}
 \hat{G}^\imp = \left[\begin{array}{cccc}
                       G^\per_{\mathbf{r}=(0,0)}&G^\per_{\mathbf{r}=(0,1)}&G^\per_{\mathbf{r}=(0,1)}&G^\per_{\mathbf{r}=(1,1)} \\
                       G^\per_{\mathbf{r}=(0,1)}&G^\per_{\mathbf{r}=(0,0)}&G^\per_{\mathbf{r}=(1,1)}&G^\per_{\mathbf{r}=(0,1)} \\
                       G^\per_{\mathbf{r}=(0,1)}&G^\per_{\mathbf{r}=(1,1)}&G^\per_{\mathbf{r}=(0,0)}&G^\per_{\mathbf{r}=(0,1)} \\
                       G^\per_{\mathbf{r}=(1,1)}&G^\per_{\mathbf{r}=(0,1)}&G^\per_{\mathbf{r}=(0,1)}&G^\per_{\mathbf{r}=(0,0)}
                      \end{array}\right]
\end{equation}
where
\begin{equation}
G^\per_{\mathbf{r}} = \sum_{\mathbf{k}\in BZ} e^{-i\mathbf{k}\cdot\mathbf{r}} G^\per_\mathbf{k} =  \sum_{\mathbf{k}\in BZ}  \frac{e^{-i\mathbf{k}\cdot\mathbf{r}}}{G_{0,\mathbf{k}}^{-1}(i\omega_n) - \Sigma^\per_\mathbf{k}(i\omega_n)}
\end{equation}
The final result is the translationally invariant self-energy which solves Eq.~\ref{eq:pcdmft_sc}.

Note there is another variant of PCDMFT method (proposed in Ref.\onlinecite{Biroli2002}) where the self-energy is periodized with additional coefficients,
so that it is rigorously causal. In the present case, this method would correspond to restoring translational invariance on the lattice the following way
\begin{eqnarray} \nonumber
 \hat{\Sigma}^{\mathrm{latt}}_{\mathbf{k}}(i\omega_n) &\approx& \hat{\Sigma}^{\mathrm{imp}}(i\omega_n)\circ\Big( \hat{I} 
                                                               + \frac{1}{2}\hat{u}_\mathbf{k}
                                                               + \frac{1}{4}\hat{w}_\mathbf{k} \Big)
\end{eqnarray}
We observe that this method corrects the local part of self-energy in the difficult regime compared to regular PCDMFT, but the non-local part is strongly underestimated throughout the phase diagram (results not shown).

\begin{figure}[!ht]
 %\centering{}
 \includegraphics[width=2.9in,trim=0cm 0cm 0cm 0cm]{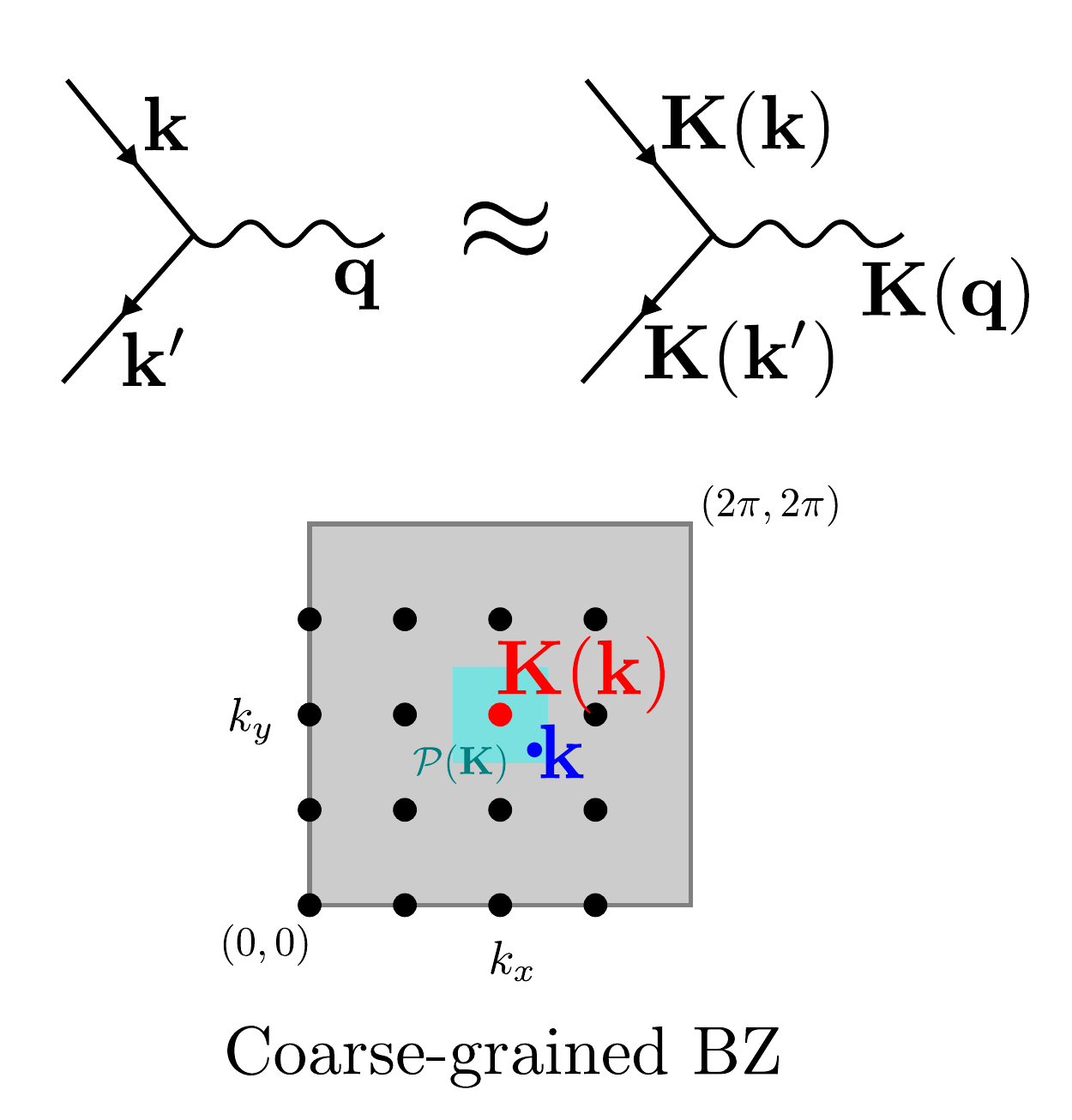} 
 \caption{ DCA approximates momentum-conservation. Lattice self-energy at wave-vector $\mathbf{k}$ is obtained from the impurity self-energy at the closest coarse grained wave-vector $\mathbf{K}$. Therefore, it is constant within each Voronoi patch ${\cal P}(\mathbf{K})$. The example presented is the $4\times 4$ scheme.
 }
 \label{fig:dca} 
\end{figure}

\subsection{Dynamical cluster approximation (DCA)} \label{sec:dca}

In DCA
\cite{Aryanpour2002,Dang2015,Ferrero2009,Ferrero2008,Ferrero2010,Gull2010,Hettler2000,Hettler1998,Huscroft2001,Jarrell2001,Kent2005,Leblanc2015,Macridin2004,Macridin2006,Macridin2007,Maier2004,MaierPRL2005,MaierPRL2006,Maier2011,MaierArXiv2015}
method, the conservation of momentum in LW diagrams is approximated by\cite{Aryanpour2002,Hettler2000}
\begin{equation}
 \mathbf{k}'-\mathbf{k} = \mathbf{q} \longrightarrow \mathbf{K}(\mathbf{k}')-\mathbf{K}(\mathbf{k}) = \mathbf{K}(\mathbf{q})
\end{equation}
where $\mathbf{K}$ represents the ''coarse-grained`` BZ points. The coarse-grained BZ contains only a certain finite and discrete subset of wave-vectors. The notation $\mathbf{K}(\mathbf{k})$ means ''the coarse grained wave-vector closest to the wave-vector $\mathbf{k}$``.
Because of the relaxation of momentum conservation, the diagrams factorize: the LW functional still depends on all $G$ components, but only through their sums
\begin{equation}
 \Phi[\{G_\mathbf{k}\}_{\forall \mathbf{k}}] \approx \Phi\Bigg[\bigg\{\sum_{\mathbf{k}\in{\cal P}(\mathbf{K})}G_\mathbf{k}\bigg\}_{\forall \mathbf{K}}\Bigg] \equiv 
 \Phi[\{G_\mathbf{K}\}_{\forall \mathbf{K}}]
\end{equation}
Here ${\cal P}(\mathbf{K})$ is the set of fine-grain wave-vectors $\mathbf{k}$ that are closest to the coarse-grained wave-vector $\mathbf{K}$ (Voronoi patch\cite{Aurenhammer1991} around $\mathbf{K}$).
This approximation leads to a piecewise-constant self-energy in $\mathbf{k}$-space, because of
\begin{equation}
 \frac{\partial \Phi[G_\mathbf{K}]}{\partial G_\mathbf{k}} =  \frac{\partial \Phi[G_\mathbf{K}]}{\partial G_\mathbf{K}}\frac{\partial G_\mathbf{K}}{\partial G_\mathbf{k}} = \frac{\partial \Phi[G_\mathbf{K}]}{\partial G_\mathbf{K}}\delta_{\mathbf{k}\in{\cal P}(\mathbf{K})}
\end{equation}

The impurity action is given by
\begin{eqnarray} \label{eq:dca_action}
 S &=& \sum_{\mathbf{K},\sigma} \iint d\tau d\tau' c^+_{\sigma,\mathbf{K}}(\tau) [-{\cal G}^{-1}_{\sigma,\mathbf{K}}(\tau-\tau')] c_{\sigma,\mathbf{K}} (\tau') \\ \nonumber
   &&  + U \sum_{\mathbf{K},\mathbf{K}',\mathbf{Q}} \int d\tau c^+_{\up,\mathbf{K}+\mathbf{Q}}(\tau) c^+_{\dn,\mathbf{K}'-\mathbf{Q}}(\tau) c_{\dn,\mathbf{K}'}(\tau) c_{\up,\mathbf{K}}(\tau)
\end{eqnarray}
and it corresponds to a finite cyclic cluster in real space $\mathbf{R}$.

Self consistency condition reads
\begin{equation}
 G_{\mathbf{K}}^\mathrm{imp}(i\omega_n) = G_\mathbf{K}^\mathrm{latt}(i\omega_n) \equiv \sum_{\mathbf{k} \in {\cal P}(\mathbf{K})} G_\mathbf{k}^\mathrm{latt}(i\omega_n)
\end{equation}
As already mentioned, the self-energy on the lattice is simply
\begin{equation}
 %\Sigma^{\mathrm{latt}}_{\mathbf{k}\in  {\cal P}(\mathbf{K})}(i\omega_n) = \Sigma^{\mathrm{imp}}_\mathbf{K}(i\omega_n)
 \Sigma^{\mathrm{latt}}_{\mathbf{k}}(i\omega_n) = \Sigma^{\mathrm{imp}}_{\mathbf{K}(\mathbf{k})}(i\omega_n)
\end{equation}
Note that more general coarse-graining schemes exist,  and that ${\cal P}(\mathbf{K})$ does not necessarily present a Voronoi patch around the wave-vector $\mathbf{K}$. Patches may have different shapes\cite{Ferrero2009}, and may even be interlaced\cite{Staar2016}. In the present paper, we only use the simplest scheme where patches are Voronoi patches, and all have the same shape.

\begin{figure}[!ht]
 %\centering{}
 \includegraphics[width=2.9in,trim=0cm 0cm 0cm 0cm]{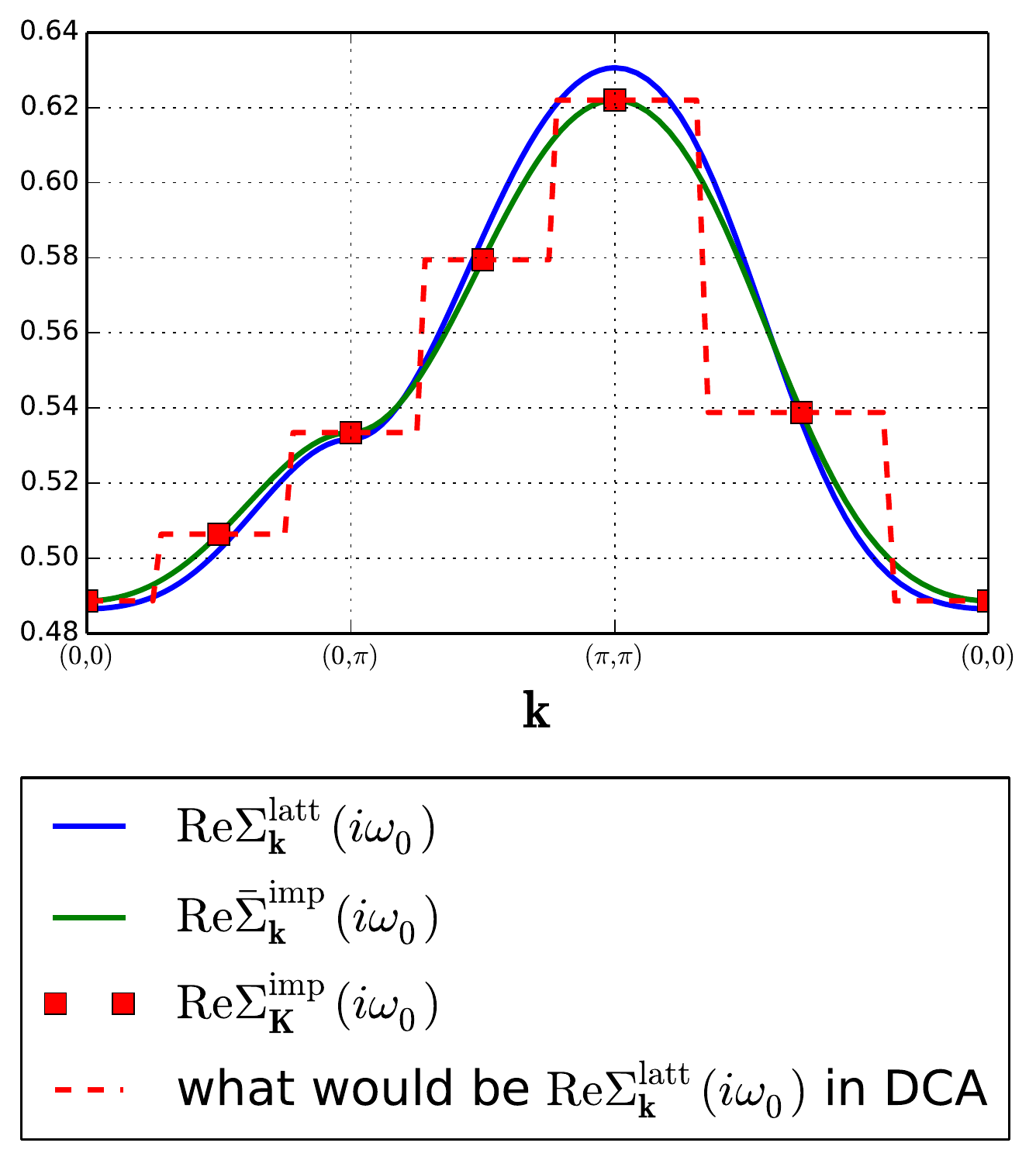} 
 \caption{ Example of various quantities appearing in DCA$^+$. Parameters of the calculation: $U/D=1.4$, $\delta=8\%$, $T/D=0.125$.
 }
 \label{fig:dca_plus_illustration} 
\end{figure}

\subsection{Continuous self-energy DCA (DCA$^+$)}

DCA$^+$\cite{Staar2016,MaierPRB2014,Staar2013} aims at improving the interpretation of the impurity self-energy in terms of the self-energy on the lattice. A piecewise constant self-energy is strongly counter intuitive and hard to compare to other methods. It is a natural step to try and interpolate the coarse-grained self-energy to obtain a smooth self-energy on the lattice. However, this scheme does not have a clear derivation as a LW function approximation. Furthermore, the interpolation can be done in various ways, and the method is not uniquelly defined. In this paper we implement (and present here) the version of the method as proposed in the original paper, Ref.\onlinecite{Staar2013}.

The impurity action is the same as in DCA, Eq.~\ref{eq:dca_action}.

The difference from DCA is the addition of a self-consistency condition that needs to be satisfied
\begin{equation} \label{eq:dca_plus_condition}
 \Sigma_{\mathbf{K}}^\mathrm{imp}(i\omega_n) = \Sigma_\mathbf{K}^\mathrm{latt}(i\omega_n) \equiv \sum_{\mathbf{k} \in  {\cal P}(\mathbf{K})} \Sigma_\mathbf{k}^\mathrm{latt}(i\omega_n)
\end{equation}
Here $\Sigma^\latt$ is a smooth function of $\mathbf{k}$. Note that for a given $\Sigma^\imp_\mathbf{K}$, $\Sigma^\latt$ is not uniquely defined.
This self-consistency condition imposes
\begin{equation}
 \Sigma^\latt_\loc = \Sigma^\imp_\loc
\end{equation}
but in general
\begin{equation}
 \Sigma^\latt_{\mathbf{k}=\mathbf{K}} \neq \Sigma^\imp_\mathbf{K}
\end{equation}

While a general interpolation of $\Sigma^\imp_\mathbf{K}$ is unlikely to satisfy the condition \eqref{eq:dca_plus_condition}, a Bayesian approach can be employed to find the most probable interpolation that does satisfy it. The method used is Richardson-Lucy deconvolution, and it is performed with respect to an interpolation of $\Sigma^\imp_\mathbf{K} \rightarrow \bar{\Sigma}^\imp_\mathbf{k}$ such that 
\begin{equation}
 \bar{\Sigma}^\imp_{\mathbf{k}=\mathbf{K}} = \Sigma^\imp_\mathbf{K}
\end{equation}
One starts from an initial guess for $\Sigma^\latt$ (say, $\Sigma^\latt_\mathbf{k}=\bar\Sigma^\imp_{\mathbf{k}}$), and iterates
\begin{equation}
 \Sigma^\latt_\mathbf{k} \leftarrow \Sigma^\latt_\mathbf{k} \sum_{\mathbf{k}'\in{\cal P}(\mathbf{k})}\frac{ \bar\Sigma^\imp_{\mathbf{k}'}}{\sum_{\mathbf{k}''\in{\cal P}(\mathbf{k'})}  \Sigma^\latt_{\mathbf{k}''} }
\end{equation}
until convergence is reached. Here ${\cal P}(\mathbf{k})$ denotes a patch of the same shape/size as the Voronoi patches of the coarse-grained BZ, but centered at the fine-grain wave-vector $\mathbf{k}$. The final result has the property
\begin{equation}
 \bar{\Sigma}_{\mathbf{k}}^\mathrm{imp} = \sum_{\mathbf{k}' \in {\cal P}(\mathbf{k})} \Sigma_{\mathbf{k}'}^\mathrm{latt},\;\;\forall\mathbf{k}
\end{equation}
which satisfies a stronger requirement than necessary.

Note also that the actual interpolation is performed not on $\Sigma^\imp$, but on an auxiliary quantity $\Xi$ which is by construction more local than the self-energy.
The method of interpolation proposed is the Wannier interpolation
\begin{eqnarray}
 \Xi_\mathbf{K}(i\omega_n) &=& (\Sigma^\imp_\mathbf{K}(i\omega_n) - \sgn(\omega_n)i\alpha)^{-1} \\
 \Xi_\mathbf{R} &=& \sum_\mathbf{K} e^{-i\mathbf{K}\cdot\mathbf{R}} \Xi_\mathbf{K} \\ 
 \bar{\Xi}_\mathbf{k} &=& \sum_\mathbf{R} e^{i\mathbf{k}\cdot\mathbf{R}} \Xi_\mathbf{R} \\
 \bar{\Sigma}^\imp_\mathbf{k}(i\omega_n) &=& \bar{\Xi}^{-1}_\mathbf{k}(i\omega_n) + \sgn(\omega_n)i\alpha,\;\; \alpha>0  
\end{eqnarray}
Note that $\bar{\Xi}_\mathbf{k}$ does not necessarily satisfy all the lattice symmetries. One way to restore lattice symmetries is to calculate it as
\begin{equation}
 \bar{\Xi}_\mathbf{k} = \frac{1}{N_M}\sum_{\hat{M}} \sum_\mathbf{R} e^{i\mathbf{k}\cdot\hat{M}\mathbf{R}} \Xi_\mathbf{R}
\end{equation}
where $\hat{M}$ runs over all the symmetry operations on the lattice, of which there are $N_M$. On the square lattice there are $N_M=8$ operations ($\mathbf{R}_{x,y}\rightarrow \pm \mathbf{R}_{(x,y),(y,x)}$), which restore the 8-fold symmetry in  $\bar{\Xi}_\mathbf{k}$.

\bibliography{refs}
\bibliographystyle{apsrev4-1}

\end{document}